\documentclass[times,onecolumn,final,longtitle]{elsarticle}

\usepackage{smhl}
\usepackage{framed,multirow}

\usepackage{amssymb}
\usepackage{latexsym}

\usepackage{url}
\usepackage{xcolor}
\definecolor{newcolor}{rgb}{.8,.349,.1}

\usepackage[hidelinks]{hyperref}
\hypersetup{
    colorlinks=true,
    linkcolor=blue,
    urlcolor=copperRed,
    linktoc=page,
    citecolor=blue!80
}
\newcommand*{\doi}[1]{\href{http://dx.doi.org/#1}{{\color{black}DOI:} #1}}
\usepackage{breakcites}
\usepackage{makecell}
\usepackage{adjustbox}
\usepackage{footnote}
\usepackage{tablefootnote}
\usepackage{footmisc}
\usepackage{float}
\usepackage{tasks}
\usepackage{marvosym}
\usepackage{stackengine}
\usepackage{paralist}
\usepackage{enumitem}
\usepackage{bm}
\usepackage{graphicx}
\usepackage{colortbl}
\usepackage{tikz}
\usepackage{pgfplotstable}
\usepackage{pgfplots}
\usetikzlibrary{shapes.geometric, arrows}
\usetikzlibrary{calc}
\usepackage{tabularx}

\definecolor{notes}{rgb}{0,0,1}
\definecolor{comments}{rgb}{1,0,0}
\definecolor{comfyblue}{RGB}{52,143,235}
\definecolor{belblue}{RGB}{54, 111, 186 }
\definecolor{copperRed}{RGB}{127,38,11} 
\definecolor{papayYollow}{RGB}{255, 230, 128}
\definecolor{dayanaYollow}{RGB}{255, 240, 179}
\definecolor{dragonberry}{RGB}{	102, 33, 70}
\definecolor{leveldorGray}{RGB}{239, 239, 242}
\definecolor{eggplant}{HTML}{3531FF}
\definecolor{ruby}{RGB}{88, 24, 69} 
 \definecolor{brick}{HTML}{711D00}
 \definecolor{kosternil}{HTML}{154F85}

\newcommand{\q}[1]{``#1"}

\journal{Smart Health}

\begin{document}

\verso{Mansura A Khan \textit{etal}}

\begin{frontmatter}
\title{Investigating Health-Aware Smart-Nudging with Machine Learning to Help People Pursue  Healthier Eating-Habits}%

\author[1]{Mansura A \snm{Khan}\corref{cor1}}
\cortext[cor1]{Corresponding author: 
  Email: mansura.a.khan@gmail.com}  
\author[1]{Khalil \snm{Muhammad}}
\author[1]{Barry \snm{Smyth}}
\author[1]{David \snm{Coyle}}
\address[1]{School of Computer Science, University College Dublin}


\begin{abstract}
Food-choices and eating-habits directly contribute to our long-term health. This makes the food recommender system a potential tool to address the global crisis of obesity and malnutrition. Over the past decade, artificial-intelligence and medical researchers became more invested in researching tools that can guide and help people make healthy and thoughtful decisions around food and diet.  In many typical (Recommender System) RS domains, \textbf{smart nudges} have been proven effective in shaping users' consumption patterns. In recent years, knowledgeable \textbf{nudging} and \textbf{incentifying choices} started getting attention in the food domain as well.  To develop smart nudging for promoting healthier food choices, we combined  Machine Learning and RS technology with food-healthiness guidelines from recognized health organizations, such as the World Health Organization, Food Standards Agency, and the National Health Service United Kingdom. In this paper, we discuss our research on, \emph{persuasive visualization for making users aware of the healthiness of the recommended recipes}. Here, we propose three novel nudging technology, the \textbf{WHO-BubbleSlider}, the \textbf{FSA-ColorCoading}, and the \textbf{DRCI-MLCP}, that encourage users to choose healthier recipes. We also propose a Topic Modeling based portion-size recommendation algorithm. To evaluate our proposed smart-nudges, we conducted an online user study with 96 participants and 92250 recipes. Results showed that, during the food decision-making process, appropriate healthiness cues make users more likely to click, browse, and choose healthier recipes over less healthy ones.
\end{abstract}

\end{frontmatter}

\section{Introduction}

Healthiness, nutritional balance, and sensible diets collectively contribute to a vast volume of knowledge, and it is often challenging for people to comprehend this and make an appropriate judgments for their food choices. Building healthier eating habits and continuously following a sensible diet plan requires hard work and motivation. People regularly make conscious decisions outside of their comfort zone. However, people willing to make these difficult choices and looking to pursue healthier eating habits often struggle due to finding a lack of new, exciting and healthy food-ideas. In diverse domains, online RS has proved to be effective as a means to change users' behavior \cite{RecommenderSystemsHandbook}. By empowering users to overcome the information overload problem and assisting them with the decision-making process, (Food Recommender Systems) FRS can make a meaningful impact on users' eating-habit. As  FRS can be implemented in cross-technology platforms such as web, mobile, smart-watch, and ubiquitous mediums, they can become the ultimate tool to assist users in identifying healthy recipes and deciding on more nutritious options. While understanding users' food preference, subjected to what-when-where, is already a complex problem, recommending healthy-options aligned within users' taste and pursuing users to choose healthier options endure an even higher degree of complexity.

Previous studies discovered evidence of the effectiveness of \textbf{behavior change interventions} at the individual, community, and population levels \cite{Theoriesofbehaviourandbehaviourchange,reviewonTheeffectivenessofnudging2increaseFruitVegetable,behaviourIndividualCommunityAndPopulationLavel,nudgesAsystematicreview}. \textbf{Nudging}, a method of behavior change intervention, has proven to be an inexpensive approach to impact human behaviors positively 
\cite{smartNudgeCognitiveTecnology,SmartNudgeRecommendation,Theoriesofbehaviourandbehaviourchange}. It is a type of \textit{choice architecture} technique that favors some items among all those competing for users' attention. In (Recommender System) RS domain, nudging is \textit{incentifying} some options to help them stand out within the recommendation list. Despite the fact that nudging is a new concept in RS research, researchers have started to investigate \textit{gamification} and \textit{audiovisual incentives} for nudging users towards desired behavioral changes \cite{karlsen2019recommendations}. For promoting healthier food-choices, we have designed and developed three novel nudges \begin{inparaitem}
\item[\textbullet] \textbf{WHO-Bubbleslider}:  
A Bubble Slider Scale based on the (World Health Organization) WHO nutrient intake goal \cite{website:WHOintakeguide} \item[\textbullet] \textbf{FSA-ColorCoading}: A Color Code Scale base on the (Food Standard Agency) FSA nutrient intake guideline \cite{website:FSA:Nutrient} and \item[\textbullet] \textbf{DRCI-MLCP}: A machine-learned portion-size recommendation based on the WHO-BMI Risk factor \cite{WHOAdultBodyMassIndexBMIRiskfactorClassification}, the (Food and Agriculture Organization) FAO physical activity vector \cite{website:FAO_activitylevelClasses} the (National Collaborative on Childhood Obesity Research) NCCOR calorie adjustment guideline \cite{Obesity3}, and the (National Health  Service United Kingdom) NHS-UK eat-well guideline \cite{NHSoneyou:eat-better}\end{inparaitem}.

All three nudges are carefully designed following nutrient-intake guidelines proposed by leading health organizations  \cite{website:WHO-BMI,website:Who_healthyDiet,website:Who_generalStrategyForObesityPrevention,website:FAO_activitylevelClasses,FSA:NutrientDetailed,website:FSA:Nutrient,WorldHealthOrganizationTechnicalReportSeries724,Obesity3}. Each of the nudging techniques consists of assessing recipe healthiness and conveying the information produced from the assessment to the user. For the DRCI-MLCP nudge, we developed a personalized portion-size recommender algorithm. We used (Ensemble Topic Modeling) \(E_{ns}TM\) \cite{MarkBelford} to determine the food-type of a recipe. Based on the food-type and various health features (e.g., weight and physical-activity) of a user, our proposed method determines health-aware portion-size for each recipe. We developed \textbf{nutri-visualization} contents (web-contents composed of text and graphics) to visualize the healthiness information. To evaluate our proposed nudging techniques, we conducted a user study with 96 participants and 92250 recipes, comparing three recommendation scenarios with each of the three proposed smart nudges and a recommendation scenarios without nudging. Unlike other popular RS domains, (Food Recommendations) FR deals with recommending items that are highly likely to be consumed physically \cite{consumptionTheory}, implying the prerequisite of personalizing the computation of food healthiness to each user. For generating personalized recommendations, we applied \(F\_T\_R\) recommender, a hybrid feature and topic based algorithm \cite{khan2021addressing}. In one of our previous work \cite{khan2021addressing}, \(F\_T\_R\) has proven to be the best performing personalization algorithm regarding prediction accuracy.  We used food-feature based open user-modeling \cite{khan2021addressing} for modeling user' taste profiles. For health profiles, users reported their age, weight, height, and physical-activity-level. We created health profiles using The (Daily Recommended Calorie Intake) DRCI algorithm proposed in section \ref{sec:healthprofileCal}. Our proposed equation takes into account various malnutrition and obesity guidelines proposed by leading health organizations \cite{website:WHO-BMI,website:Who_healthyDiet,website:Who_generalStrategyForObesityPrevention,website:FAO_activitylevelClasses}. This enabled us to personalize the proposed nudges to users' health requirements. The personalized and health-aware characteristics make our proposed nudging strategies smart-nudges. A demo of our proposed smart-nudges can be found at the YouTube link \cite{Thedemo}.

Few previously existing FRS works, including Gary Sacks et al.'s work in \cite{Impact_of_front_of_pack_dap032}, Simon Howard et al. in  \cite{NutritionalContentSsupermarketReadyMeals} and Christoph Trattner et al. \cite{Trattner:2017:RCI:3099023.3099072,InvestigatingtheHealthinessofInternetSourcedRecipes}, have evaluated large-scale recipe copra using FSA \cite{FSATrafficLightSystemFoodLabelling} and WHO \cite{website:WHOintakeguide} nutritional guidelines. These works have only proposed mechanism to asses recipe healthiness. However, not many of these works produced nudging technologies to communicate the healthiness information to users and attract their attention to healthier recipes.  Alain Starke et al. in \cite{DifferentEatingGoals} have investigates goal centric personalised health aware recommendation. In this paper, we propose visual nudging contents that not only aims to inform people of a recipe's healthiness but also to encourage users to choose the healthier option. The user study assessed the nudging contents for \textit{understandability} and \textit{persuasiveness}. Each nudging strategy is evaluated on its ability to \textit{attract users' attention to healthier recipes} and \textit{successfully makes users choose healthier options}. The results showed nutri-visualization techniques help healthy recipes attract users' attention over other options. We observed significant bias towards healthier recipes in users' browsing-time under the recommendation scenarios incorporating smart-nudging. Users reported that the nutri-visualization contents made it easier to find healthier recipes.

\section{Related Literature}

Food being directly connected to our physical and physiological health \cite{article74HowPeopleInterpretHealthyEating} made \textit{\textbf{nutrition}} both active and passive constraints for FRS research \cite{AnOverviewofRecommenderSystemsinTheHealthyFoodDomain,RecommenderSystemsandTheirEthicalChallenges}. The  (Nutrition aware Food Recommender System ) nutri-FRS possess the potential for becoming the ultimate \textbf{personal health guide} application that can assist people in defining their \textbf{health goals} and guide them plan diet, menu, grocery, and physical activity to achieve those goals. Which fueled the enthusiasm among  RS and medical researchers to investigate, design, and develop for nutri-FRS\cite{article87NutritionalInformationandUserPreferences,article83HealthAwareFoodRecommenderSystem,ASystematicReviewofNutritionRecommendationSystems:WithFocusonTechnicalAspects,AnOverviewofRecommenderSystemsinTheHealthyFoodDomain,article65surveyAINutritionRec}. Over the past few decades, researchers have produced seminal contributions towards nutri-FRS to ensure user-preference, diversity, novelty, and nutritional development in diet decisions, such as \cite{DBLP:journals/corr/abs-1711-02760,Jiis1,article47reviewRSinHealhyFood,SmartNudgeRecommendation,nudgesAsystematicreview,article74HowPeopleInterpretHealthyEating,article101FoodComputingSurvey}. In the following section, we briefly discuss the state-of-the-art nutri-FRS research addressing diverse challenges in generating valid health-aware FR. We categorized the research under various sub-categories, each representing a significant area of nutri-FRS research.

\subsection*{ \small Health Bias}
The most prevailing strategy for generating health-aware recommendations has been \textit{introducing health bias in the recommendation process}. Between two recipes with the same \textit{prediction scores}, health bias favors the healthier one. Health bias can be generated  by considering \textit{health properties} (e.g. macro-nutrients and association to diseases or cures) within the prediction calculation \cite{Ge:2015:HFR:2792838.2796554}. Tsuguya Ueta et al. in \cite{Ueta:2011:RRS:2186633.2186642}  identified the macro-nutrients and minerals that best satisfy user's health requirements and computed the recommendations only considering those.   Mouzhi et al. in \cite{Ge:2015:HFR:2792838.2796554}  considered both \textit{health properties} and ingredients for computing preferences. However, they assigned greater weights on \textit{health properties} over ingredients. Which ultimately yielded a higher preference score for healthier recipes.  While being very promising and effective, both approaches worked with few core health conditions and limited recipes corpus. Health bias can provide direct access to healthier recipes; however, it possesses the susceptibility of deviating from users' \textit{taste} preference.

\subsection*{ \small Calorie Requirement Based Filtering}
Another most common nutri-FRS strategy is Daily Calorie Intake (DCI) based filtering \cite{Chi:2015:CDD:2953211.2953554,Kim:2009:DDR:1674656.1676485,Buisson:2008:NNS:1346360.1346617}. DCI-based filtering algorithms retrieve food-items based on the match between \textit{users' DCI} and \textit{the number of calories} in the food-items. This strategy is beneficial for systems working with obesity or weight-watch purposes. Yu-Liang Chi et al. in \cite{Chi:2015:CDD:2953211.2953554}, proposed a semantic-rule based approach, where they employed semantic rules to determine meal-plans that best matches the user's DCI. They determine users DCI from their (Basal Metabolic Rate) BRM. Harris-Benedict equations \cite{HarrisBenedictequation} has been the most popular BMR calculation strategy and adopted by many, including \cite{Article23TowardsAutomaticMealPlan,Chi:2015:CDD:2953211.2953554,article43Fityou,article44MonitoringAwarenessHealthyEating,Harvey:2015:ARH:2792838.2796551,article87NutritionalInformationandUserPreferences,syahputra2017scheduling}. Jong-Hun Kim et al. in \cite{Kim:2009:DDR:1674656.1676485} developed a calorie-to-diet table. The table stored \textit{calorie ranges} and corresponding age, height, and weight as cases. Under each \textit{calorie range}, the table stored a set of diet-plans recommended by nutritionists. When a new user arrives, their recommender retrieves similar cases from the table. Depending on the \textit{calorie range} the user falls into, their recommender suggests a  diet-plan. While very interesting their approach worked with an even smaller numbers of recipes.  In \cite{TorreMhealthAppMonitoringHealthyeating} Isabel de la et al. followed a similar approach as \cite{Chi:2015:CDD:2953211.2953554,Kim:2009:DDR:1674656.1676485}  to identify users DCI; They also considered a fixed value of \emph{2400 Cal} as default DCI or basic energy-need if no user information was available. 
 
 Manuel Garcia \cite{article82Plan-Cook-Eat} proposed a robust approach for calculating DCI. Their proposed approach calculated users' DCI as Total Daily Energy Expenditure(TDEE) from users' Basal Metabolic Rate (BMR), Thermic
Effect of Food (TEF), Non-Exercise Activity Thermogenesis
(NEAT), and Thermic Effect of Activity (TEA). Based on users' TDEE, using the Acceptable Macronutrient Distribution Range (AMDR) \cite{AMDR}, their recommender proposed balanced diet-plans that ensure the total energy to be approximately equal to users' TDEE. Contributing to the robust knowledge on predicting user's DCI, these seminal research works became the pioneers in personalized Nutri-FRS technology. However, most of these DCI-based filtering approaches generate recommendations based on the number of calories in one serving size of recipes. Majority of the existing works do not provide any portion-size guidelines. 

Few advanced nutri-FRS works looked into further informative recommendations, such as calorie adjustment for weight gain/loss. Jean-Christophe Buisson et al. in \cite{Buisson:2008:NNS:1346360.1346617} used fuzzy arithmetic and reasoning to balance the portions within a meal-plan according to users' DCI.  When a retrieved meal-plan has a total calorie close to users' DCI, they employed a heuristic search algorithm to find replacements of member-recipes within the meal-plan. This replacement is done to force the meal-plan's total calorie to be the same as the predicted DCI. David Elsweiler et al. in \cite{Article23TowardsAutomaticMealPlan}, implemented a variation of the Harris-Benedict equation to predict users' BMR, followed by the addition or removal of 500 calories from the BRM to support weight gain/loss plans. They recommend meal-plans in acclimation with this DCI. Rung-Ching Chen et al. in \cite{Chen:2013:CDR:3104813.3104878} employed fuzzy ontological rule and JENA rules \cite{ApacheJena} to establish the most effective relationships between user-entities, \{\textit{height, weight, kidney function, hypertension, and hyperlipidemia}\} and food-entities\{\textit{diet-plans, recipes, ingredients}\}. Greedy Knapsack methods \cite{10.1016/j.cor.2011.02.010} were used to combine the fuzzy results and JENA results to recommend a better group of recipes. Hannah Forster et al. in \cite{article45ADietaryFeedbackSystem} implemented a nutrient supplement intake recommender where they incorporated DCI based filtering with their more complex nutrition requirement prediction algorithm.  Isabel de la et al. in \cite{TorreMhealthAppMonitoringHealthyeating}, adopted the WHO guidelines on carbohydrate, Lipid and Protein balance in daily consumption \cite{website:WHOportionCarbLipidProtaine} to define \textit{nutritional balance} in their meal-plan recommendations.

  There remains much interest on \textit{watching calorie intake} among greater population. Which has made BMR and DCI based filtering the most practiced nutri-FRS strategy. However, the emphasis on the calorie range can reduce the degree of personalization. This can make users lose interest and drop-out of the healthy \textit{diet plan}.

\subsection*{  \small Health Goal}

Goal-based algorithms have proven to be a great success in education 
\cite{GoalBasedCourseRecommendation,GoalBasedinELearningRS,GoalBasedHybridFiltering}
and were first investigated in the nutri-FRS domain by Tsuguya  Ueta et al. in \cite{Ueta:2011:RRS:2186633.2186642,article93GoalOrientednutriRec}. In recent years, Goal-based hybrid filtering approaches have seen increasing research interest among health technology developers \cite{Chi:2015:CDD:2953211.2953554,Jung:2016:KDN:2898233.2898316,article24HealthPromotionTools,Lee:2010:TFO:1821731.1821743,Chen:2016:DIR:2964284.2964315,article54lightweightfooddieres,article69NutritionAssistanceSystems}. Goal-based algorithms comprise hybrid filtering approaches inclined to achieve a cumulative result known as \textbf{the Goal}. While generating recommendations, Goal-based models prioritize items that will fulfill users' \textit{long-term} and \textit{short-term} Goals over items that are more close to users' taste-preferences. For example, if users set their Goal as \textit{pentonic acid consumption}, the Goal-based model will prioritize recipes containing \textit{pentonic acid} over recipes with high preference scores. Such models are designed to compromise the personalization performance in the present for future gain.

Tsuguya Ueta et al. in \cite{Ueta:2011:RRS:2186633.2186642}, proposed a Goal-based nutri-FRS that aims to deal with a list of health issues, e.g., acne, fatigue, insomnia, and stress. They generated two co-occurrence tables, such as a \textbf{health-issue to macro-nutrient} table and a \textit{ macro-nutrient to recipe} table.  Using the co-occurrence tables, the FRS recommended recipes that can help cure the health-issues mentioned in users' Goals. Hanna Schafer et al., in  \cite{article69NutritionAssistanceSystems}, proposed a Rasch-scale and Tailored-Goal based nutri-FRS to assist users to increase daily consumption of certain macro-nutrients. They adopted food-logging \cite{Food4MeFoodFrequencyQuestionnaire} to observe users' nutrition intake. Their nutri-FRS recommended recipes comparing uses' Goal and their daily food-log. Hoill Jung et al. in \cite{Jung:2016:KDN:2898233.2898316}, and  David Elsweiler et al. in \cite{Article23TowardsAutomaticMealPlan},   combined Goal-based methods with their more advanced hybrid personalization algorithms to assist users in gaining/losing weight. Yu-Liang Chi et al. \cite{Chi:2015:CDD:2953211.2953554} combined a variation of goal-based approach with their OWL-based Ontologies and Semantic Rules based personalization algorithms. The adaptation of goal-based ensured that all recommendations are aligned with the health guideline for CKD patients.

The robustness of Goal-based methods encouraged researchers to investigate applications that can bring behavioral change among users.  Andrea G. Parker et al. in \cite{article24HealthPromotionTools} adopted a Goal-based approach to foster a community around healthy \textit{food habits} and \textit{physical activity}.  Chia-Fang  Chung et al. in \cite{article54lightweightfooddieres}, adopted Goal-based models in their health-expert system, enabling a community to share \textit{health knowledge} and focus on individual health Goals.  Alain Starke et al. in \cite{DifferentEatingGoals} proposed a novel multi-list food recommender approach to accommodate multiple goals. While Goal-based filtering can enable users to communicate with the FRS in natural language, the recommendations are often strictly controlled by the Goals. Users are recommended very specific items to satisfy a goal; however, these recommendations often come without any explanations.  These no explanation recommendations and the lack of control on the recommendation process can reduce interest and incur user dropout.

\subsection*{\small Human in the Loop nutri-FRS}

The \textit{Human in the Loop (HitL)} model incorporates human moderators to examine the practical applicability of generated recommendations. In such models, \textit{human experts} are involved in one or many steps of developing a nutri-FRS, such as  \begin{inparaitem}
\item[\textbullet] knowledge discovery \item[\textbullet] defining diet-plans considering specific health issues \item[\textbullet] fine-tuning the weights on different nutrient contents in the filtering algorithms and \item[\textbullet] checking the generated recommendations' validity\end{inparaitem}. The most practiced HitL model incorporates a service called \emph{intervention-access}. Such systems make a user's RecList available to the user's nutritionists; using the intervention-access, the nutritionist can approve, update or reject food-items in the RecList. Few HiiT projects commissioned nutritionists to generate various health-mapping tables. Later, to determine whether a recipe or ingredient is healthy for a user, they analyzed food-items according to those mapping-tables.

Significant projects investigating HitL models includes: \textbf{\textbf{grocery recommendation} }\textbf{ \{} Hanshen Gu et al., \cite{Gu:2009:CFB:1701835.1701851}; and Prashanti Angara et al. \cite{article58conversationalRecforsmartkitchen}\textbf{\}}, \textbf{food-package recommendation for older adults at care facility} \textbf{\{}Cristina Bianca Pop et al. \cite{article26HealthyMenusforOlderAdults} \textbf{\}}, \textbf{menu-plan recommendation} \textbf{\{} Jong-Hun Kim et al.\cite{Kim:2009:DDR:1674656.1676485};  Rung-Ching  Chen, et al. \cite{Chen:2013:CDR:3104813.3104878}; Vanesa Espin et al. \cite{espin2016nutrition}; and  Jean-Christophe Buisson \cite{Buisson:2008:NNS:1346360.1346617} \textbf{\}},  \textbf{diet-plan for patients with Hyperglycemia}  \{Chang-Shing Lee et al. \cite{Lee:2010:TFO:1821731.1821743} \textbf{\}}, \textbf{diet-plan for at home older adults} \textbf{\{}Adel  Taweel et. al. \cite{Taweel:2014:DSS:2939574.2939584} \textbf{\}}, and \textbf{nutrition consultation} \textbf{\{} Vanesa Espin et al. \cite{espin2016nutrition};  IsabelTorre Diez et al.  \cite{article44MonitoringAwarenessHealthyEating}; Hannah Forster et al. \cite{article45ADietaryFeedbackSystem}; and Hanna Schafer et al.   \cite{article69NutritionAssistanceSystems} \textbf{\}}.

Many of these HitL works are the pioneers of healthy recommendations in their corresponding domain. Health experts' involvement ensured practical recommendations, which helped to gain users' trust and acceptance.   However, HitL FRS is often limited to working with a distinct target user group and smaller food corpus.

\subsection*{  \small Embedded System and Internet of Things Platforms}

In recent years, researchers have introduced a few multimedia nutri-FRS involving advanced technologies, such as Embedded System,  Multi-agent Architecture, Internet of Things (IoT) and Smart-health. Some of the significant works in this category include: \textbf{Ontology-driven Personalized FRS for IoT-based Healthcare System } \textbf{\{}Adel Taweel et al.  \cite{article60ontologyDivenPersonalizedFoodRec} \textbf{\}},\textbf{ Multi-agent Architecture based disease-driven nutri-FRS }  \textbf{\{} Todor Ivascu et al.  \cite{article59DiseaseDrivenfoodRec} \textbf{\}},  \textbf{ Smart-nudging  based FRS to guide eating behavior of the older adult who has recently been diagnosed with type II diabetes}  \{ Wen-Yu Chao et al.\cite{FoodRecommenderSystemforNudgingEating},\textbf{ SMASH usability heuristics and emotion detection based FRS} \textbf{\{} Tsaihsuan  Tsai et al. \cite{article92MoodCanteen} \textbf{\}}, \textbf{central administration based cloud system}  \textbf{\{ }Shreya B. Ahire et al. \cite{article34APersonalizedFramework} \textbf{\}}, \textbf{Oracle Data Miner based treatment model prediction based FRS}\textbf{\{} Abdullah A Aljumah et al. \cite{DiabetesHealthCareYoungPatients} \textbf{\} } and \textbf{adaptive diet monitoring based FRS} \textbf{\{ } Giuseppe Agapito et al. \cite{article85DIETOS} \textbf{\}}. While impressive for combining electronic and RS technology, many of the works mentioned here only support fundamental FR strategies, certain target user groups, and fixed sets of food-items.

\subsection*{  \small Knowledge Discovery for nutri-FRS}

One of the significant outcomes of nutri-FRS research over the years is domain knowledge. nutri-FRS researchers have contributed to acquiring knowledge on direct relations between particular food preferences and corresponding health-consequences and strategies to explain to the user why a recipe is particularly healthy for the individual.  Further knowledge discovery included methods to introduce change in food decisions and strategies to map health guidelines on available food-data to filter for personalized, healthy recommendations. Few literature-review works looked into the success and failure of fundamental and domain-specific RS algorithms in nutri-FRS, contributing to knowledge on \emph{what works and what doesn't in the nutr-FRS domain}.

To design and develop tools to assist the user in food activities, researchers need to understand \emph{how people make their food decisions}. Knowledge of which variables, to what extent, and under which circumstances impact these decisions is significant. Different decision-variables are prioritized and negotiated in various ways during the food decision-making process, depending on the circumstances. Over time people develop their own rules and strategies for simplifying the processes involved in making food decisions \cite{FoodChoiceProcessesofOlderAdults}. Jeffery Sobal et al. in \cite{ article:FoodChoiceIsMultifacetedContextualDynamicMultilevelIntegratedandDiverse,Aconceptualmodelofthefoodchoice,article89ConstructingHumanFoodChoice} and Marketa Dolejsova  et al. in \cite{Dolejsova:2019:CTI:3301019.3319994} investigated \textit{human eating-habits} and summarized which attributes influence food decisions and how these attributes change over time. Mariya Vizireanu et al in \cite{LayPerceptionsofHealthyEatingStyles}, Carole Bisogni et al. in \cite{article74HowPeopleInterpretHealthyEating} and Jeffery Sobal et al. in \cite{ManagingHealthyEating:DefinitionsClassificationsStrategies,ManagingValueInPersonalFoodSystems, FoodChoiceProcessesofOlderAdults} outlined how individuals develop an understanding of healthy food. Ristovski Slijepcevic et al. in \cite{DietaryGovernmentalityintheFamilyFoodPractice} looked into how healthy eating concepts are passed on through generations in families. Jeffery Sobal  et al. \cite{ManagingValueInPersonalFoodSystems} discussed how individuals adopt preexisting sound advice on healthy food and develop their own strategies to make better food decisions. 
Significant research has been done to investigate the impact of different 
stimulus factors on \textit{human food choice}, such as \textbf{family} \textbf{\{} Ristovski Slijepcevic et al. \cite{DietaryGovernmentalityintheFamilyFoodPractice} and Charles Abraham et al. \cite{behaviourIndividualCommunityAndPopulationLavel} \textbf{\}}, \textbf{Society and Communities} \textbf{\{} Suzanne Higgs  et al. \cite{SocialInfluencesOnEating};  Tegan Cruwys  at al.\cite{SocialInfluenceAffectsFoodIntake} and Ewelina Swierad et al. \cite{communityInflunceonEatingHabit} \textbf{\}}  \textbf{Social-media} \textbf{\{} Markus Rokicki et al.  \cite{article78ImpactofRecipeFeatures};  Andrew Arnold et al. \cite{SocialMediaCanImpactYourConsumption}  and   Mariarosaria Simeone  et al. \cite{SocialMediaAffectFoodChoices}\textbf{\}} and \textbf{Scientific findings} \textbf{ \{} Ristovski Slijepcevic et al. \cite{EngagingWithHealthyEatingDiscourse}; Lynn Spitzer et al.  \cite{HumanEatingBbehavior} and  Susan Michie et al el.   \cite{EatingBehaviourChangeTechniques} \textbf{\} }.

Stefanie Mika in \cite{article86Challenges4NutritionRS}, Cristoph Trattner et al. in \cite{DBLP:journals/corr/abs-1711-02760}, Felicia Cordeiro et al. in \cite{article95challangesinFoodJournelling}  and Kerry Shih-Ping in \cite{UsingPublicDisplays2encourageHealthyEating} outline various challenges and research problems in the domain of nutri-FRS.  Shahabeddin Abhari et al. in \cite{ASystematicReviewofNutritionRecommendationSystems:WithFocusonTechnicalAspects}, Thi Ngoc  et al. in \cite{article47reviewRSinHealhyFood} ,  Weiqing Min et al.  \cite{article101FoodComputingSurvey}, Thomas Theodoridis et al in \cite{article65surveyAINutritionRec} and Christoph Trattner et al. in \cite{DBLP:journals/corr/abs-1711-02760,trattnerevaluatonofFoodrecommendersASurvey} studied the performance of various fundamental as well as food-domain-specific RS algorithms and reviewed which algorithmic strategies are the best for addressing specific nutri-FRS challenges. In many of their seminal research works, such as \cite{article73predictabilityofthepopularityofonlinerecipes,InvestigatingtheHealthinessofInternetSourcedRecipes,trattnerevaluatonofFoodrecommendersASurvey,Trattner:2017:RCI:3099023.3099072,article80Whatonlinedatasayabouteatinghabits,article75foodRecipeUploadBehavior}, Christoph Trattner et al. looked into the potential of crowdsourced online recipes. Along with the healthiness of recipes available online,  their research also looked at the hidden patterns within peoples' uploaded recipes. Markus Rokicki et al. in  \cite{GenderDifferenceOnlineCooking,RelationshipBetweentCookingInterestandHobbies,article78ImpactofRecipeFeatures}, investigated large online recipe corpuses to identify various relationships between lifestyle attributes and food preferences. Their research produced knowledge on possible correlations between certain food types and users-clusters, classified based on lifestyle attributes. They also provide various suggestions on which lifestyle variables are more likely to associate with healthier lifestyles. David Elsweiler et al., in many of their significant research works, such as \cite{Harvey:2013:YYE:2651320.2651339,article76LearningUserTastes,article79VisualFoodTasteswithOnlineRecipe,Elsweiler:EFC:3077136:3080826}, investigated strategies to identify various taste-biases within online recipes and exploit the discovery for generating healthy recommendations. 

The nutri-FRS research community's significant contribution has generated a versatile and extensive knowledge base relevant to nutri-FRS, enabling the more recent, optimistic research on machine-learned personalized healthy FR.

\subsection*{  \small Nudging in nutri-FRS}

Smart nudge is a choice architecture strategy that includes displaying users the recommendations, followed by information regarding each item to motivate and help the user choose toward suggested behavior \cite{RecommendationWithNudge,SmartNudgeRecommendation,reviewonTheeffectivenessofnudging2increaseFruitVegetable}. A Nudge aims to influence an individual's behavior towards decisions that are beneficial for their long-term interest. In the physical environment, nudging has been proven effective in incentifying choices and gaining a positive impact on peoples' food decisions \cite{nudgesAsystematicreview,SmartNudgeRecommendation,EatingBehaviourChangeTechniques}. In an extensive literature review \cite{SmartNudgeRecommendation}, Tamara Bucher et al. found that 88\% of the reviewed works achieved change in participants' food choices in the same direction as the corresponding nudge.

Wen-Yu  Chao et al. in 
\cite{FoodRecommenderSystemforNudgingEating} proposed a user-interface design inspired by the concept of nudging.  Carl Anderson in \cite{pointsBasedNudgingatWW}  discussed the numeric-goal based nudging, introduced by the Weight Watchers (WW) \cite{WW}. Felicia Cordeiro et al. in \cite{article95challangesinFoodJournelling} investigated how negative nudging can work against motivation to plan and practice a new diet. David Elsweiler et al. in \cite{Elsweiler:EFC:3077136:3080826} and Carlos Celis-Morales et al. \cite{article108EffectPersonalizedNutrition} discussed the concept of information driven behaviour change in food domain. Mashfiqui Rabbi et al. in \cite{article25Mybehaviour} propose an implicit nudging approach, where their FRS learn user's behaviors from tracking their activity and recommends minor variations in those observed activities. Research on behavior change theory and nudging in nutri-FRS is at its early stage; however, it can become the ultimate technology to educate the mass population on nutrient guidelines and healthy eating.   

\subsection*{  \small Limitation and Research Scope} 

While previous research on nutri-FRS has seen significant success and produced numerous tools to assist people in finding healthy food, a big proportion of the current works are targeted to specific user groups. For example, diabetic patients \{ Ritika Bateja et al.  \cite{article62PaitentcentricRecommendation}; Chang-Shing Lee et al \cite{Lee:2010:TFO:1821731.1821743} and  Wen-Yu Chao et al. \cite{FoodRecommenderSystemforNudgingEating}, patients with chronic kidney disease (CKD){ Chi Yu-Liang et al. \cite{Chi:2015:CDD:2953211.2953554}; Giuseppe Agapito et al.  \cite{article49DIETOS}; Jei-Fuu  Chen et al. \cite{article30.5EffectsofjournalingDietaryintake} and Vladimir Villarreal et al. \cite{Villarreal:2014:MUA:2588895.2588938} }, patients with obesity \{ Hoill Jung et al. \cite{jung2016knowledge} , older adult in care facility \{  Vanesa spin et al. \cite{espin2016nutrition};  Adel Taweel et al. \cite{Taweel:2014:DSS:2939574.2939584}; Cristina Bianca Pop et al. \cite{article26HealthyMenusforOlderAdults} and Wen-Yu Chaoet al \cite{FoodRecommenderSystemforNudgingEating},  toddlers   Yiu-Kai Ng et al. \cite{article46personalizedFRSforTodller} and Carol Boushey et al.  adolescentscite \cite{MobileFoodRecord} Hannah Forster et al \cite{article45ADietaryFeedbackSystem,article108EffectPersonalizedNutrition}. In most cases, such nutri-FRS deal with carefully selected small sets of recipes and often compromise personalization and diversity to achieve a higher match on calorie range and other nutrient criteria.  Users are displayed recommendations that are less inclined with their preference and more constrained by the health requirements, often not backed-up by any explanation. The lack of explanation on \emph{why a recipe is considered healthy} or \emph{how healthy a recipe is} can make users feel less in control of their food decision, contributing to low user satisfaction and increasing user drop-outs.

In this paper, we propose three nutrition-aware recommendation strategies that give back more control to users and try to educate users about healthy eating instead of restricting their options to a confined set of recipes. This work aims to produce advanced tools that can guide the greater population on healthy eating-habits. Providing visual explanations on \textit{how healthy each item within the RecList is}, our proposed FRS strategies allow users to make their own food decisions.  Studies have proven that users gain more trust in the system under such conditions, leading to greater user-satisfaction and long-term user engagement\cite{OpenUserProfiles,UserModelingforAdaptiveNewsAccess}.

\section{Dataset}

 Our dataset consist of 92,250 recipes, collected from the recipe portal food dot com \cite{website:fooddotcom}. Along with \textit{recipe contents} and various meta-data, each recipe also contained information on eight nutrient contents, such as protein, carbohydrate, sugar, sodium, total fat, saturated fat,  dietary fibre, and cholesterol. Each recipe also contained a valid image. Table \ref{tab:healtFooddotcom} illustrates the distribution of the our  over  WHOHealthScale\footnote{discussed in section \ref{WhoHealthscore}} and FSAHealthScale \footnote{discussed in section \ref{FSAHealthscore}}. \vspace{-1cm}

\noindent\begin{table}[!h]
    \centering
    \scalebox{.8}{
\begingroup
   \setlength\tabcolsep{7pt}
\def\arraystretch{1.2}
\fontsize{9pt}{9pt}\selectfont
\scalebox{1}{
\noindent\begin{tabular}{p{.46\textwidth}p{.01\textwidth}p{.45\textwidth}}
\def\arraystretch{1.65}
       \noindent\begin{tabular}{p{.12\textwidth} p{.28\textwidth}}

           & Recipes             \\\cline{2-2}
    WHO score &   Count (Percentage)\\
    \hline
    0   &	33 (0.04)	      \\\hline
    1   &	945 (1.03)   \\\hline
    2   &	5492 (6.00)   \\\hline
    3   &	24811 (26.90)  \\\hline
    4   &	51100 (55.40)  \\\hline
    5   &	8820 (9.6)    \\\hline
    6   &	1003 (1.90)      \\\hline
    7   &	46   (0.05)     \\\hline
    8   &   0    (0.00)          \\\hline
        & n = 92,250 

\end{tabular}  & &  \noindent\begin{tabular}{p{.1\textwidth} p{.28\textwidth}}
      & Recipes \\[1ex] \cline{2-2}
    
   FSA score &Count (Percentage)\\
    \hline
  4   &    16886(18.30)\\\hline
    5   &	13284(14.40)\\\hline
    6   &	17385(18.85)\\\hline
    7   &	14356(15.56)\\\hline
    8   &	7627(8.27)\\\hline
    9   &	6560(7.11)\\\hline
    10  &	5165(5.60)\\\hline
    11  &	5495(5.96)\\\hline
    12  &	2400(2.60)\\\hline
    13  &	2024(2.205)\\\hline
    14  &	935(1.02)\\\hline
    15  &	27(.029)\\\hline
    16  &	106(0.15)\\\hline
        & n =92,250
\end{tabular}
     \end{tabular}
}
\endgroup}
\caption{Overall healthiness of the recipes in our dataset.}
\label{tab:healtFooddotcom}
\end{table}

\section{Personalized Portion-Size Recommender \label{sec:portionsize}}

Existing literature shows, BMR/DCI based filtering algorithms has been adopted by many significant works including \cite{Chi:2015:CDD:2953211.2953554,Kim:2009:DDR:1674656.1676485,Buisson:2008:NNS:1346360.1346617,Chi:2015:CDD:2953211.2953554,Kim:2009:DDR:1674656.1676485,Buisson:2008:NNS:1346360.1346617}. While BMR and DCI based filtering has seen the most attention in the nutri-FRS domain, the degree of personalization achieved by these approaches is relatively deficient. Such nutri-FRS try to fit the recipes, meal-plans, or diet-plans within users' DCI while compromising users' content and context preferences.  Which can make users frustrated, and consequently, can reduce the acceptance rate of recommendations. Guiding users with the appropriate portion-size for recipes to maintain healthy calorie intake can be a better solution. In this section, we propose a health-aware personalized portion-size recommender. Here, we investigated \(E_{ns}TP\) to predict the food-type (e.g., main-course, snack, side and drink) of a recipe. And based on the food-type and the user's health factors (e.g., gender, weight, physical activity and obesity index) the portion-size predictor determines the quantity of any recipe that is healthy for the user. We incorporated a wide range of internationally accepted nutritional guidelines including (National Health Service United Kingdom) NHS-UK one-you-eat-well Meal Over The Day guideline \cite{NHSoneyou:eat-better}, (World Health Organization) WHO adult body mass index  risk factor classes \cite{WHOAdultBodyMassIndexBMIRiskfactorClassification}, (National Collaborative on Childhood Obesity Research) NCCOR quantification of the effect of energy imbalance on body-weight \cite{Obesity3},  WHO Technical Report Series 724 \cite{WorldHealthOrganizationTechnicalReportSeries724}, (Food and Agriculture Organization) FAO energy requirement guideline \cite{website:FAO_activitylevelClasses},  Report of a Joint FAO/WHO/UNU Expert Consultation, 1991 \cite{JointReport_FAO_WHO_UNU_Expert_Consultation} and (National Research Council United States) NRC-US recommended dietary allowances guideline \cite{Recommendeddietaryallowances}. Unlike existing approaches,  our proposed portion-size recommender checks for obesity and underweight risks and adopts the recommendation accordingly.

\begin{itemize}[label={}]
\item The problem of personalized portion-size recommendation consists of two main challenges.
\begin{enumerate}
\small 
\fontfamily{cmtt}\selectfont
     \setlength\itemsep{.05em}
    \item Determining (Daily Recommended Calorie Intake) DRCI  that is healthy for an individual.
    \item Identifying how this DRCI should be distributed among the (Meals Over the Day) MOD.
    \end{enumerate}
\end{itemize}

\subsection{Calculating User's Health-aware DRCI \label{sec:healthprofileCal}}
The strategy to calculate DRCI accurately is the basis for our personalized portion-size recommendation.  Along with an in-depth understanding of human calorie needs and expenditures, the DRCI calculation also requires knowledge of how different \textit{biological} and \textit{lifestyle} variables contribute to those needs and expenditures. Our DRCI  prediction algorithm consists of 6 distinct steps.

\begin{tabular}{p{.9\textwidth}}
\cellcolor{leveldorGray}     
\begin{itemize}\setlength\itemsep{.05em}
  \setlength\itemsep{.05em}
\item \textbf{Step 1} {\fontfamily{cmss}\selectfont Calculate users' BMR from their height, weight, gender, and age.}

\item  \textbf{Step 2} {\fontfamily{cmss}\selectfont Calculate users' Daily Calorie Intake (DCI) based on \emph{how active they are in everyday life?}} 
\item  \textbf{Step 3} {\fontfamily{cmss}\selectfont Calculate user} BMI.

\item \textbf{Step 4} {\fontfamily{cmss}\selectfont Compering the user's BMI and WHO-BMI risk factor \cite{obesityRiskUN,WHOAdultBodyMassIndexBMIRiskfactorClassification} identify \emph{if the user needs adjustment on their DCI}? }
\item  \textbf{Step5} {\fontfamily{cmss}\selectfont If step 4 is true, identify the  energy/calorie adjustment  required within user's DCI to obtain a healthier Life.
\item  \textbf{Step 6} Using step 2,4 and 5 calculate users health-aware DRCI.}
\end{itemize}
\end{tabular}\newline\vspace{1cm}

 Following is a brief description of each step of the DRCI prediction algorithm.

\subparagraph*{ \textbf{\textit{ Step 1}}} 

The BMR is the minimum energy or calorie expended by an individual each day to comprehend the physical activity of independent living. The state-of-the-art of BMR calculation is the novel Harris-Benedict equation \cite{HarrisBenedictequation}. However, to achieve a higher degree of personalization in BMR calculation, we adopted a set of equations proposed FAO ENERGY REQUIREMENTS OF ADULTS \cite{FAOENERGYREQUIREMENTSOFADULTS}. The equations calculate BMR using the age, weight, height, and gender coefficients proposed by WHO in technical report Series 724, 1985 \cite{WorldHealthOrganizationTechnicalReportSeries724}. The coefficients make the calculation fine-tuned to corresponding age, weight, and gender groups. Table \ref{tab:BMR calculation} organizes the equations under \textit{gender} and \textit{age} categories.

\begin{table}[!h]
\centering
\small
\setlength\tabcolsep{10pt}
\def\arraystretch{1.2}
\begin{tabular}{ccc}
\rowcolor[HTML]{f3faff}
 &
  \cellcolor[HTML]{FFFFFF}{\color[HTML]{000000} \textbf{\begin{tabular}[c]{@{}c@{}}Age range\\ (years)\end{tabular}}} &
  \multicolumn{1}{c}{\cellcolor[HTML]{FFFFFF}{\color[HTML]{000000} \textbf{\begin{tabular}[c]{@{}c@{}}Equation for BMR (cal) calculation \\   \(w\tablefootnote{WHO weight coefficient \label{wc}} \times  W\tablefootnote{user's weight\label{uw}} + h\tablefootnote{WHO height coefficient\label{wh}}  \times H\tablefootnote{user's height\label{uh}} +a\tablefootnote{WHO age coefficient\label{wa}}\) \end{tabular}}}} \\ \cline{1-3} 
\multicolumn{3}{l}{\cellcolor[HTML]{ECF4FF}Male}                                      \\ \cline{2-3} 
\cellcolor[HTML]{e7f5fe}                   & 10 -18           & \(16.6\times W + 77 \times H + 572\)    \\ \cline{2-3} 
\cellcolor[HTML]{e7f5fe}                   & 18-30           & \(15.4 \times W - 27 \times H + 717\)    \\ \cline{2-3} 
\cellcolor[HTML]{e7f5fe}                   & 30-60           & \(11.3 \times W + 16 \times H + 901\)    \\ \cline{2-3} 
\multirow{-4}{12pt}{\cellcolor[HTML]{e7f5fe }} & \textgreater 60 & \(8.8 \times W + 112 \times H - 1 071\) \\ \cline{1-3} 
\multicolumn{3}{l}{\cellcolor[HTML]{DAE8FC}Female}                                    \\ \cline{2-3} 
\cellcolor[HTML]{DAE8FC}                   & 10-18           & \(7.4\times W + 482 \times H + 217\)    \\ \cline{2-3} 
\cellcolor[HTML]{DAE8FC}                   & 18-30           & \(13.3 \times W + 334 \times H + 35\)    \\ \cline{2-3} 
\cellcolor[HTML]{DAE8FC}                   & 30-60           & \(8.7\times W - 25 \times H + 865\)     \\ \cline{2-3} 
\multirow{-4}{12pt}{\cellcolor[HTML]{DAE8FC}} & \textgreater 60 & \(9.2 \times W + 637 \times H - 302\)   \\ \cline{2-3} 
\end{tabular}
\caption{List of BMR equations proposed by WHO Technical Report Series 724 \cite{WorldHealthOrganizationTechnicalReportSeries724}. \label{tab:BMR calculation}}
\end{table} 

\subparagraph*{ \textbf{\textit{ Step 2}} }

BMR only estimates the number of calories required to execute only the activities corresponding to independents living.  The actual amount of energy exhausted by one is defined by their level of physical activity per day \cite{ENERGYREQUIREMENTSANDDIETARYENERGYRECOMMENDATIONS}. We calculated a user \(u_a\)'s DCI, \(DCI_{u_a}\), using their BMR, \(BMR_{u_a}\), as shown in equation \ref{eq:1}.

\noindent\begin{equation} \label{eq:1}
\fontsize{14pt}{13pt}\selectfont
 DCI_{u_a} =  BMR_{u_a} \times PhysicalActivityFactor
\end{equation}

The \(PhysicalActivityFactor\) is the \textbf{activity coefficient} that determines the final quantity of calories essential for \({u_a}\) to consume everyday for maintaining the level of activity involved in their everyday life. Unlike much significant health-aware FRS, we intend to address the difference in energy requirement for a similar level of physical activity by individuals from different genders. Hence, we adopted the activity-coefficients \cite{Recommendeddietaryallowances} proposed by the NRC-US, summarized in table \ref{tab:physicalActivityFactor}.

\subparagraph*{ \textbf{\textit{ Step 3}} } 

To guide users in healthier decision-making on calorie consumption, it is essential to learn \textit{whether a user is currently suffering from \textit{obesity} and \textit{underweight} risk or not.} To determine obesity status, we considered BMI, also known as Quetelet Index, a WHO-defined measure for indicating nutritional status in adults \cite{WHOAdultBodyMassIndexBMIRiskfactorClassification}. We calculated the user's BMI using \emph{WHO-BMI equation} \cite{website:WHO-BMI}.

\begin{table}[!h]

\centering
\small
\setlength\tabcolsep{8pt} 
\def\arraystretch{1.3}
\begin{tabular}{|p{1\textwidth}p{.07\textwidth}|}
\hline
\rowcolor[HTML]{000000} 
\multicolumn{1}{|c|}{\cellcolor[HTML]{000000}{\color[HTML]{FFFFFF} \textbf{Level of Activity}}} &
  \multicolumn{1}{c|}{\cellcolor[HTML]{000000}{\color[HTML]{FFFFFF} \textbf{Activity Coefficient}}} \\ [1.5ex] \hline
\rowcolor[HTML]{EFEFEF} 
\multicolumn{2}{|l|}{\cellcolor[HTML]{EFEFEF}Sedentary} \\ \hline 
\multicolumn{1}{|m{.75\textwidth}}{ \color[HTML]{3531FF}{A sedentary lifestyle includes only the physical activity required for independent living.}} & { \begin{tabular}{p{.05\textwidth}p{.025\textwidth}}
    {\cellcolor[HTML]{ECF4FF} Male }& {\cellcolor[HTML]{ECF4FF}1.3 } \tabularnewline
     {\cellcolor[HTML]{DAE8FC} Female} & {\cellcolor[HTML]{DAE8FC} 1.3}
     
\end{tabular}} 
\\ \hline

\rowcolor[HTML]{EFEFEF} 
\multicolumn{2}{|l|}{\cellcolor[HTML]{EFEFEF}Moderately  Active} \\ \hline 
\multicolumn{1}{|m{.75\textwidth}}{ \color[HTML]{3531FF}{ Along with independent living activities, a moderately active lifestyle includes physical activities equivalent to walking approximately \textit{1.5 to 3} miles per day at a speed of \textit{3 to 4} miles per hour.}} & { \begin{tabular}{p{.05\textwidth}p{.025\textwidth}}
    {\cellcolor[HTML]{ECF4FF} Male }& {\cellcolor[HTML]{ECF4FF}1.7 } \tabularnewline
     {\cellcolor[HTML]{DAE8FC} Female} & {\cellcolor[HTML]{DAE8FC} 1.6}
     
\end{tabular}} 
\\ \hline
\rowcolor[HTML]{EFEFEF} 
\multicolumn{2}{|l|}{\cellcolor[HTML]{EFEFEF}Very  Active} \\ \hline 
\multicolumn{1}{|m{.75\textwidth}}{ \color[HTML]{3531FF}{ A lifestyle that includes independent living activities and intense sports or exercises, 6-7 days per week.}} & { \begin{tabular}{p{.05\textwidth}p{.025\textwidth}}
    {\cellcolor[HTML]{ECF4FF} Male }& {\cellcolor[HTML]{ECF4FF}2.1 } \tabularnewline
     {\cellcolor[HTML]{DAE8FC} Female} & {\cellcolor[HTML]{DAE8FC} 2.9}
     
\end{tabular}} 
\\ \hline
\rowcolor[HTML]{EFEFEF} 
\multicolumn{2}{|l|}{\cellcolor[HTML]{EFEFEF}Intensely  Active} \\ \hline 
\multicolumn{1}{|m{.75\textwidth}}{ \color[HTML]{3531FF}{ Along with independent living activities and intense daily exercises, the Intensely Active lifestyle includes a profession involving physical labor or two days per week, day-long, high-intensity sports, such as training,  marathon, and triathlon.}} & { \begin{tabular}{p{.05\textwidth}p{.025\textwidth}}
    {\cellcolor[HTML]{ECF4FF} Male }& {\cellcolor[HTML]{ECF4FF}2.4 } \tabularnewline
     {\cellcolor[HTML]{DAE8FC} Female} & {\cellcolor[HTML]{DAE8FC}2.2} \tabularnewline
\end{tabular}} 
\\ [1.5ex]\hline
\end{tabular}
\caption{Factors for estimating daily energy allowances for various levels of physical-activity. \cite{Recommendeddietaryallowances}. \label{tab:physicalActivityFactor}}
\end{table}

\subparagraph*{ \textbf{\textit{ Step 4}} }


Using the user's BMI and the WHO Adult BMI Risk Factor Matrix \cite{WHOAdultBodyMassIndexBMIRiskfactorClassification}, we predicted their \textbf{WHO BMI Risk class} \cite{WHOAdultBodyMassIndexBMIRiskfactorClassification}. The risk classes inform whether or not the user is susceptible to obesity and other comorbidities. Table \ref{tab:BMIRiskfactor} summarizes the WHO defined BMI range for different risk classes, such as \begin{inparaitem} \item[\textbullet] Underweight, \item[\textbullet]Normal Weight, \item[\textbullet] Overweight and \item[\textbullet] Obese \end{inparaitem}. For users outside the risk class \emph{Normal Weight}, Sept 4 recommends an energy adjustment on the user's previously calculated DCI.

\subparagraph*{ \textbf{\textit{ Step 5}} } 

The general rule of thumb for energy adjustment for healthier calorie intake has been, adding or subtracting \textit{2MJ or 500cals} from the DCI for individuals in Underweight and Overweight classes, respectively. Though this approach has been popular in nutri-FRS research \cite{Jung:2016:KDN:2898233.2898316,article44MonitoringAwarenessHealthyEating,Harvey:2015:ARH:2792838.2796551}, the generic 500cal change is not efficient in helping individuals from all three obese classes. According to health experts, to observe any significant change in BMI, an adult with a BMI of \(> 35\) needs at least a  change of \(>500 cal\) in their calorie consumption per day.  To calculate the required adjustment on a user's DCI, we adopted the \emph{\textbf{10 cal per pound per day}} \cite{Obesity3} energy adjustment strategy proposed by NCCOR.

\noindent\begin{equation} \label{eq:2}
\fontsize{14pt}{13pt}\selectfont
 E^{Adj}_{u_a} =  W^p_{u_a} \times 10
\end{equation}

Following the NCCOR guideline, equation \ref{eq:2} calculates the quantity of energy adjustment \({ E^{Adj}_{u_a}}\) for user \(u_a\) using \(W^p_{u_a} \). Here, \(W^p_{u_a} \) is  \(u_a\)'s weight in pounds (lbs). This robust energy adjustment strategy made it feasible to design and develop technology to assist users from any underweight or obesity class.

\begin{table}[!h]
\centering
\setlength\tabcolsep{8pt}
\def\arraystretch{1.5}
\small
\scalebox{.9}{
\begin{tabular}{llcc|}
\rowcolor[HTML]{000000} 
\multicolumn{2}{l}{\cellcolor[HTML]{000000}{\color[HTML]{FFFFFF} \textbf{ Classification}}} &
  {\color[HTML]{FFFFFF}\textbf{ BMI (\(\mathbf{kg/m^2}\) )}} &
  {\color[HTML]{FFFFFF}\textbf{ Risk of Comorbidities} }\\ \cline{3-4} 
\multicolumn{2}{l}{Underweight} &
  \textless{}18.5 &
Low Risk (risk of other clinical problems) \\\hline
\multicolumn{2}{l}{Normal Weight} &
  18.5-24.9 &
  \multicolumn{1}{l|}{Risk Free} \\ \hline
\multicolumn{2}{l}{Overweight (pre-obese)} &
  25.0-29.9 &
  \multicolumn{1}{l|}{Mildly Increased Risk} \\ \hline
\rowcolor[HTML]{f3faff} 
\multicolumn{2}{l}{\cellcolor[HTML]{f3faff}{\color[HTML]{000000}\textbf{ Obese}}} &
  {\color[HTML]{000000} \textbf{$\geq$ 30.0}} &
  \multicolumn{1}{l|}{\cellcolor[HTML]{f3faff}{\color[HTML]{000000} \textbf{Rink Range}}} \\ [1.5ex]\cline{2-4} 
 \rowcolor[HTML]{f3faff} &
  \cellcolor[HTML]{e7f5fe}Class III &
  \cellcolor[HTML]{e7f5fe}30.0-34.9 &
  \multicolumn{1}{l|}{\cellcolor[HTML]{e7f5fe}Moderate} \\ \cline{2-4} 
 \rowcolor[HTML]{f3faff} &
  \cellcolor[HTML]{dbf0fe}Class III &
  \cellcolor[HTML]{dbf0fe}35.0-39.9 &
  \multicolumn{1}{l|}{\cellcolor[HTML]{dbf0fe}Severe} \\ \cline{2-4} 
\rowcolor[HTML]{f3faff} \multirow{-3}{*}{} &
  \cellcolor[HTML]{DAE8FC}Class III &
  \cellcolor[HTML]{DAE8FC}$\geq$40.0 &
  \multicolumn{1}{l|}{\cellcolor[HTML]{DAE8FC}Very Severe} \\ \cline{1-4} 
\end{tabular}
}
\caption{WHO adult BMI risk factor classes  \cite{WHOAdultBodyMassIndexBMIRiskfactorClassification}.\label{tab:BMIRiskfactor}}
\end{table}

\subparagraph*{ \textbf{\textit{ Step 6}} } 

Finally, users' health aware DCI, \(DCI^h\), is calculated from the DCI determined in Step 2 and the energy adjustment determined in Step 5. For a user \(u_a\), equation \ref{eq:3} calculates \(DCI^h\). Here, \(DCI^h_{u_a}\) is \(u_a\)'s \(DCI^h\),  \(DCI_{u_a}\) is \(u_a\)'s DCI, and \(E^{Adj}_{u_a}\) is \(u_a\)'s energy  adjustment.

\noindent\begin{equation} \label{eq:3}
\fontsize{14pt}{13pt}\selectfont
 DCI^h_{u_a} =  DCI_{u_a} \pm E^{Adj}_{u_a}
 \end{equation}
 
 For underweight users, \( E^{Adj}_{u_a}\) is added to, and for overweight and obese users, \( E^{Adj}_{u_a}\)is taken from their DCI. This personalized and health-aware calorie intake recommendation, \(DCI^h_{u_a}\), is our proposed DRCI. The DRCI calculation approach aims to guide users towards achieving and maintaining a healthier BMI.

\subsection{ \(E_{ns}TM\) based Food-type Identification \label{sec:Enstmbasedfoodtypeidentification}}

The more challenging problem in recommending portion-size is \textit{ identifying the appropriate quantity/portion of each recipe, in a massive recipe corpus, for each user}. As a standard practice of food consumption, depending on the varying type of the food-item, such as main-course, snacks, sides, fruits, and drinks, people eat varying portions \cite{HumanEatingBbehavior,LayPerceptionsofHealthyEatingStyles}. Hence, while recommending portion-size, to determine the correct quantity, we first need to identify the food-type of the food-item in the recipe. Although over the online platforms, recipes often contain annotations on the dish-type (e.g., soup, pasta, and risotto), the dish-type information do not necessarily identify the food-type of a recipe. To predict the food-type of a recipe, we investigated \(E_{ns}TM\) based food-type identification approach.

 \begin{table}[!h]
\centering

\def\arraystretch{1.6}
\tabcolsep=6pt
\scalebox{.8}{%
\begin{tabular}{lllllll}
\rowcolor[HTML]{343434} 
\multicolumn{1}{c}{\cellcolor[HTML]{FFFFFF}} &
  \multicolumn{1}{c}{\cellcolor[HTML]{343434}{\color[HTML]{FFFFFF} \textbf{Topic Descriptor}}} &
  \multicolumn{1}{c}{\cellcolor[HTML]{343434}{\color[HTML]{FFFFFF} \textbf{Food-type}}} &
  \multicolumn{1}{c}{\cellcolor[HTML]{FFFFFF}{\color[HTML]{FFFFFF} \textbf{}}} &
  \multicolumn{1}{c}{\cellcolor[HTML]{FFFFFF}{\color[HTML]{FFFFFF} \textbf{}}} &
  \multicolumn{1}{c}{\cellcolor[HTML]{343434}{\color[HTML]{FFFFFF} \textbf{Topic Descriptor}}} &
  \multicolumn{1}{c}{\cellcolor[HTML]{343434}{\color[HTML]{FFFFFF} \textbf{Food-type}}} \\[1.7ex] \cline{1-3} \cline{5-7} 
\rowcolor[HTML]{FFFFFF} 
\multicolumn{1}{l}{\cellcolor[HTML]{FFFFFF}Topic 1} &
  \multicolumn{1}{l|}{\cellcolor[HTML]{FFFFFF}Quick and Easy Snack} &
  \multicolumn{1}{c|}{\cellcolor[HTML]{FFFFFF}snack} &
  \multicolumn{1}{l}{\cellcolor{white}} &
  \multicolumn{1}{l}{\cellcolor[HTML]{FFFFFF}Topic\ 16} &
  \multicolumn{1}{l|}{\cellcolor[HTML]{FFFFFF}Spicy and Umami curry } &
  \multicolumn{1}{c|}{\cellcolor[HTML]{FFFFFF}meal} \\ \cline{1-3} \cline{5-7} 
\rowcolor[HTML]{ DAE8FC} 
\multicolumn{1}{l}{\cellcolor[HTML]{ DAE8FC}Topic 2} &
  \multicolumn{1}{l|}{\cellcolor[HTML]{ DAE8FC}Easy Fish Mains} &
  \multicolumn{1}{c|}{\cellcolor[HTML]{ DAE8FC}meal} &
  \multicolumn{1}{l}{\cellcolor{white}} &
  \multicolumn{1}{l}{\cellcolor[HTML]{ DAE8FC}Topic\ 17} &
  \multicolumn{1}{l|}{\cellcolor[HTML]{ DAE8FC}Smoothies} &
  \multicolumn{1}{c|}{\cellcolor[HTML]{ DAE8FC}drink} \\ \cline{1-3} \cline{5-7} 
\rowcolor[HTML]{FFFFFF} 
\multicolumn{1}{l}{\cellcolor[HTML]{FFFFFF}Topic\ 3} &
  \multicolumn{1}{l|}{\cellcolor[HTML]{FFFFFF}Rice Dishes} &
  \multicolumn{1}{c|}{\cellcolor[HTML]{FFFFFF}side} &
  \multicolumn{1}{l}{\cellcolor{white}} &
  \multicolumn{1}{l}{\cellcolor[HTML]{FFFFFF}Topic\ 18} &
  \multicolumn{1}{l|}{\cellcolor[HTML]{FFFFFF}Turkey Mains} &
  \multicolumn{1}{c|}{\cellcolor[HTML]{FFFFFF}meal} \\ \cline{1-3} \cline{5-7} 
\rowcolor[HTML]{ DAE8FC} 
\multicolumn{1}{l}{\cellcolor[HTML]{ DAE8FC}Topic\ 4} &
  \multicolumn{1}{l|}{\cellcolor[HTML]{ DAE8FC}Tropical Juice and Desserts} &
  \multicolumn{1}{c|}{\cellcolor[HTML]{ DAE8FC}drink} &
  \multicolumn{1}{l}{\cellcolor{white}} &
  \multicolumn{1}{l}{\cellcolor[HTML]{ DAE8FC}Topic\ 19} &
  \multicolumn{1}{ p{.34\textwidth}|}{\cellcolor[HTML]{ DAE8FC}Citrus based Food-preserves, Drinks and Mains} &
  \multicolumn{1}{c|}{\cellcolor[HTML]{ DAE8FC}meal} \\ \cline{1-3} \cline{5-7} 
\rowcolor[HTML]{FFFFFF} 
\multicolumn{1}{l}{\cellcolor[HTML]{FFFFFF}Topic\ 5} &
  \multicolumn{1}{l|}{\cellcolor[HTML]{FFFFFF}Beef based Mains} &
  \multicolumn{1}{c|}{\cellcolor[HTML]{FFFFFF}meal} &
  \multicolumn{1}{l}{\cellcolor{white}} &
  \multicolumn{1}{l}{\cellcolor[HTML]{FFFFFF}Topic\ 20} &
  \multicolumn{1}{p{.3\textwidth}|}{\cellcolor[HTML]{FFFFFF}Holiday pies} &
  \multicolumn{1}{c|}{\cellcolor[HTML]{FFFFFF}snack} \\ \cline{1-3} \cline{5-7} 
\rowcolor[HTML]{ DAE8FC} 
\multicolumn{1}{l}{\cellcolor[HTML]{ DAE8FC}Topic\ 6} &
  \multicolumn{1}{l|}{\cellcolor[HTML]{ DAE8FC} Vegetable Dishes} &
  \multicolumn{1}{c|}{\cellcolor[HTML]{ DAE8FC}meal} &
  \multicolumn{1}{l}{\cellcolor{white}} &
  \multicolumn{1}{l}{\cellcolor[HTML]{ DAE8FC}Topic\ 21} &
  \multicolumn{1}{l|}{\cellcolor[HTML]{ DAE8FC}Pork Mains} &
  \multicolumn{1}{c|}{\cellcolor[HTML]{ DAE8FC}meal} \\ \cline{1-3} \cline{5-7} 
\rowcolor[HTML]{FFFFFF} 
\multicolumn{1}{l}{\cellcolor[HTML]{FFFFFF}Topic\ 7} &
  \multicolumn{1}{l|}{\cellcolor[HTML]{FFFFFF}Seafood Mains} &
  \multicolumn{1}{c|}{\cellcolor[HTML]{FFFFFF}meal} &
  \multicolumn{1}{l}{\cellcolor{white}} &
  \multicolumn{1}{l}{\cellcolor[HTML]{FFFFFF}Topic\ 22} &
  \multicolumn{1}{l|}{\cellcolor[HTML]{FFFFFF}Cheesy Dishes} &
  \multicolumn{1}{c|}{\cellcolor[HTML]{FFFFFF}side} \\ \cline{1-3} \cline{5-7} 
\rowcolor[HTML]{ DAE8FC} 
\multicolumn{1}{l}{\cellcolor[HTML]{ DAE8FC}Topic\ 8} &
  \multicolumn{1}{l|}{\cellcolor[HTML]{ DAE8FC}Chicken Mains} &
  \multicolumn{1}{c|}{\cellcolor[HTML]{ DAE8FC}meal} &
  \multicolumn{1}{l}{\cellcolor{white}} &
  \multicolumn{1}{l}{\cellcolor[HTML]{ DAE8FC}Topic\ 23} &
  \multicolumn{1}{l|}{\cellcolor[HTML]{ DAE8FC}Pasta Mains } &
  \multicolumn{1}{c|}{\cellcolor[HTML]{ DAE8FC}meal} \\ \cline{1-3} \cline{5-7} 
\rowcolor[HTML]{FFFFFF} 
\multicolumn{1}{l}{\cellcolor[HTML]{FFFFFF}Topic\ 9} &
  \multicolumn{1}{p{.3\textwidth}|}{\cellcolor[HTML]{FFFFFF}Health Conscious} &
  \multicolumn{1}{c|}{\cellcolor[HTML]{FFFFFF}side} &
  \multicolumn{1}{l}{\cellcolor{white}} &
  \multicolumn{1}{l}{\cellcolor[HTML]{FFFFFF}Topic\ 24} &
  \multicolumn{1}{l|}{\cellcolor[HTML]{FFFFFF}Soups and Stews } &
  \multicolumn{1}{c|}{\cellcolor[HTML]{FFFFFF}side} \\ \cline{1-3} \cline{5-7} 
\rowcolor[HTML]{ DAE8FC} 
\multicolumn{1}{l}{\cellcolor[HTML]{ DAE8FC}Topic\ 10} &
  \multicolumn{1}{l|}{\cellcolor[HTML]{ DAE8FC}Quick and Easy bread} &
  \multicolumn{1}{c|}{\cellcolor[HTML]{ DAE8FC}breakfast} &
  \multicolumn{1}{l}{\cellcolor{white}} &
  \multicolumn{1}{l}{\cellcolor[HTML]{ DAE8FC}Topic\ 25} &
  \multicolumn{1}{l|}{\cellcolor[HTML]{ DAE8FC}Floret Star-fries and Salads} &
  \multicolumn{1}{c|}{\cellcolor[HTML]{ DAE8FC}side} \\ \cline{1-3} \cline{5-7} 
\rowcolor[HTML]{FFFFFF} 
\multicolumn{1}{l}{\cellcolor[HTML]{FFFFFF}Topic\ 11} &
  \multicolumn{1}{l|}{\cellcolor[HTML]{FFFFFF}Banana based Desserts and Drinks} &
  \multicolumn{1}{c|}{\cellcolor[HTML]{FFFFFF}breakfast} &
  \multicolumn{1}{l}{\cellcolor{white}} &
  \multicolumn{1}{l}{\cellcolor[HTML]{FFFFFF}Topic\ 26} &
  \multicolumn{1}{l|}{\cellcolor[HTML]{FFFFFF}Chinese Desserts} &
  \multicolumn{1}{c|}{\cellcolor[HTML]{FFFFFF}snack} \\ \cline{1-3} \cline{5-7} 
\rowcolor[HTML]{ DAE8FC} 
\multicolumn{1}{l}{\cellcolor[HTML]{ DAE8FC}Topic\ 12} &
  \multicolumn{1}{l|}{\cellcolor[HTML]{ DAE8FC}Sweet Desserts for Holidays} &
  \multicolumn{1}{c|}{\cellcolor[HTML]{ DAE8FC}snack} &
  \multicolumn{1}{l}{\cellcolor{white}} &
  \multicolumn{1}{l}{\cellcolor[HTML]{ DAE8FC}Topic\ 27} &
  \multicolumn{1}{l|}{\cellcolor[HTML]{ DAE8FC}Quiches} &
  \multicolumn{1}{c|}{\cellcolor[HTML]{ DAE8FC}meal} \\ \cline{1-3} \cline{5-7} 
\rowcolor[HTML]{FFFFFF} 
\multicolumn{1}{l}{\cellcolor[HTML]{FFFFFF}Topic\ 13} &
  \multicolumn{1}{l|}{\cellcolor[HTML]{FFFFFF}Pies and Tarts} &
  \multicolumn{1}{c|}{\cellcolor[HTML]{FFFFFF}snack} &
  \multicolumn{1}{l}{\cellcolor{white}} &
  \multicolumn{1}{l}{\cellcolor[HTML]{FFFFFF}Topic\ 28} &
  \multicolumn{1}{l|}{\cellcolor[HTML]{FFFFFF}Corn based Mexican} &
  \multicolumn{1}{c|}{\cellcolor[HTML]{FFFFFF}side} \\ \cline{1-3} \cline{5-7} 
\rowcolor[HTML]{ DAE8FC} 
\multicolumn{1}{l}{\cellcolor[HTML]{ DAE8FC}Topic\ 14} &
  \multicolumn{1}{l|}{\cellcolor[HTML]{ DAE8FC} Savory Greek Mains} &
  \multicolumn{1}{c|}{\cellcolor[HTML]{ DAE8FC}meal} &
  \multicolumn{1}{l}{\cellcolor{white}} &
  \multicolumn{1}{l}{\cellcolor[HTML]{ DAE8FC}Topic\ 29} &
  \multicolumn{1}{l|}{\cellcolor[HTML]{ DAE8FC}Turkey Mains} &
  \multicolumn{1}{c|}{\cellcolor[HTML]{ DAE8FC}meal} \\ \cline{1-3} \cline{5-7} 
\rowcolor[HTML]{FFFFFF} 
\multicolumn{1}{l}{\cellcolor[HTML]{FFFFFF}Topic\ 15} &
  \multicolumn{1}{l|}{\cellcolor[HTML]{FFFFFF}Potato based Small Bites} &
  \multicolumn{1}{c|}{\cellcolor[HTML]{FFFFFF}side} &
  \multicolumn{1}{l}{\cellcolor{white}} &
  \multicolumn{1}{l}{\cellcolor[HTML]{FFFFFF}Topic\ 30} &
  \multicolumn{1}{l|}{\cellcolor[HTML]{FFFFFF}Roman} &
  \multicolumn{1}{c|}{\cellcolor[HTML]{FFFFFF}meal} \\ \cline{1-3} \cline{5-7} 
\end{tabular}%
}
\caption{Food-type label for each topic.}
\label{tab:foodtypetotopic}
\end{table}

For identifying food-type we extended our \(E_{ns}TM\) based food-topic identification approach, proposed in \cite{khan2021addressing}. We adopted the 30 food-topics {\{ \fontfamily{psv}\selectfont topic label, topic descriptors\}} and the recipe-to-topic association matrix discovered in \cite{khan2021addressing}. The topic descriptor is a set of 15 most significant food-features within a food-topic \cite{khan2021addressing}.  Although the food-topics represent the theme or the concept of corresponding food-items, they do not explicitly identify the food-type. However, the knowledge of the food theme can guide towards food-type identification. 


We classified the 30 topics into five \textit{food-types}, determined by NHS-UK \cite{NHSoneyou:eat-better}.  While determining the food-type or category of a food-topic, we considered both the \textit{topic label} and \textit{the topic descriptor}. Table \ref{tab:foodtypetotopic} shows the food-type assigned on each topic. The term \textbf{meal} is an umbrella-term for food-types lunch, dinner, and main course. The food-type \textbf{side} includes food those are usually consumed alongside foods of food-type \textit{meal}. And the food-type \textbf{snack} includes finger-foods, small-bites, fruits and vegetables, and other nibbles. We initially nominated both food-type \textit{side} and \textit{snack} for the food category breakfast. However, once identified as a side or snack, food-items are checked against a \textit{breakfast vocabulary dictionary} to separate breakfast food-items from other sides and snacks. Finally, the type \textbf{drinks} include juice, tea, coffee, smoothie, protein shake, and other liquid desserts.  Our five significant food-types are listed below.

\begin{tasks}(5)

\task meal
\task side
\task snack
\task drink
\task breakfast
  
\end{tasks}

For predicting the food-type of each recipe in a corpus, we use the recipe-to-topic association matrix that gives the association score, AS, for each  \(\{recipe, topic\}\) pair \cite{khan2021addressing}. The score states how strongly a topic describes a recipe. However, each recipe can have an association to multiple topics and consequently multiple food-types. To address this issue and determine which food-type is the most dominant for each recipe, we defined the food-type prediction approach. Our approach calculates the food-type of a recipe from the food-type labels of the top 7 topics that have the highest AS with the current recipe. The number of topics used in the prediction process, t=7, is defined as \textbf{6} ( {\color{belblue} 20\% of the total 30 topics}) + \textbf{1}( {\color{belblue}for dealing with ties}).

\begin{table}[!h]
    \centering
 \scalebox{.9}{
 \begin{small}
\setlength\tabcolsep{3.26pt}
\def\arraystretch{1.4}
\noindent\begin{tabular}{ |c|c|c|c|c|c|c|c|c|c|c|c|c|c|c|c|c| } 
 \cline{2-17}
 \rowcolor[HTML]{D9DBF1}\multicolumn{1}{c|}{\cellcolor[HTML]{FFFFFF} \(\)} & \(T_1\) & \(T_2\) &\(T_3\) & \(T_4\)&\(T_5\)& \(T_6\) & \(T_7\)&\(T_8\)& .&.& .& .& .& \(T_{28}\)& \(T_{29}\)& \(T_{30}\) \\ 
  \cline{1-17}
\cellcolor[HTML]{ECF4FF} \(R_1\) & \(0.79\)& \(0\)& 0.4&0.091& 0.312& 0.76& 0.231&0.2& .& .& .& .& .& 0& 0& \(0.031\)\\ 
    \hline
\cellcolor[HTML]{ECF4FF}  \(R_2\) & \(0.009\)& \cellcolor{papayYollow} \(0.341\)& \cellcolor{papayYollow}0.604& 0.105&\cellcolor{papayYollow}0.354& 0.003&\cellcolor{papayYollow} 0.4& \cellcolor{papayYollow}0.281& .& .& .& .& .& 0.108& \cellcolor{papayYollow} 0.201& \cellcolor{papayYollow}\(0.109\)\\ \hline
\cellcolor[HTML]{ECF4FF}  \(R_3\) & \(0.001\)& \(0.03\)& 0& 0&0.702& 0& 0.751&0.310& .& .& .& .& .&0.5& .001& \(.001\)\\ 
  \hline
 \cellcolor[HTML]{ECF4FF}     \(.\) & \(.\)& \(.\)& .& .&.& .& .&.& .& .& .& .& .& .& .& \(.\)\\ 
  \hline
\cellcolor[HTML]{ECF4FF}    \(.\) & \(.\)& \(.\)& .& .&.& .& .&.& .& .& .& .& .& .& .& \(.\)\\ 
  \hline
 \cellcolor[HTML]{ECF4FF}   \(R_{92248}\) & \(0\)& \(0.001\)& 0.304& 0.138&0.5& 0& 0.001&0.054&.& .& .& .& .& 0.341& 0.921& \(.4\)\\ 
    \hline
\cellcolor[HTML]{ECF4FF}   \(R_{92249}\) & \(0.001\)& \(0.034\)& 0.004& 0.513&0.45& 0.8& 0.091&0.018&.& .& .& .& .& 0.109& 0.0542& \(0.001\)\\ \hline
\cellcolor[HTML]{ECF4FF}  \(R_{92250}\) & \(0\)& \(0\)& 0.001& 0.061& 0.002& 0.719&0.321& 0.432& .& .& .& .& .&0.31&0.001 & \(0.41\)\\ \hline
\end{tabular}
 \end{small}
}

\caption{A pseudo topic-to-recipe matrix. \label{tab:recipeTopicpseudoTable}}
\end{table}

We estimate the possibility of a recipe being a food-type \(FoodType_x\) as the food-type sore, \( S(FoodType_x)\). The food-type sore \( S(FoodType_x)\)  for food-type \(FoodType_x\) in a recipe \(r_t\) is determined based on the presence or absence of topics corresponding to \(FoodType_x\). If in a recipe \(r_t\) one or more topics, in the top 7 topics, are of food-type \(FoodType_x\), the score \( S(FoodType_x)\) is calculated as the cumulative sum of the AS between \(r_t\) and the topics corresponding to \(FoodType_x\). For example, in table \ref{tab:recipeTopicpseudoTable}, Topic  \(T_2\),\(T_3\),\(T_5\),\(T_7\),\(T_8\) and \(T_29\) are the most dominant topics in recipe \(R_2\). The food-type of R2 is derived from the food-type of these seven topics. Equation \ref{eq:4} illustrates the food-type score estimation approach for \(r_t\). Here, T is a topic that is of food-type \(FoodType_x\), m is the number of topic those are of food-type \(FoodType_x\) and  \(AS_{(T,r_t)}\) is the AS between T and \(r_t\). The food-type with the highest score is predicted as the food-type for \(r_t\).

     \begin{equation} \label{eq:4}
     \fontsize{14pt}{13pt}\selectfont
 S(FoodType_x) =  \sum_{i=0}^{m}AS_{(T,r_t)}  \;\;\;\;\;\;\;\;     [m=0,1,2,..,7]
\end{equation}

\subsection{ Estimating Personalized  Portion-size}
 
The next step for predicting personalized portion-size recommendations is estimating the healthy portion based on the user's DRCI and the recipes' food-type. For deciding \textit{how the total calories per day to be distributed over different meals}, we extended the portion guidance proposed by the NHS-UK's \textit{OneYou:EatWell} project \footnote{NHS-UK governs the \textit{OneYou:EatWell}, providing information to mess population on physical activity and healthy food consumption to address the obesity crisis in the UK.} \cite{NHSoneyou:eat-better,Caloriereduction:Thescopeandambitionforaction}. Unlike our personalized DRCI, the \textit{OneYou:EatWell} gives a generic calorie guideline of 2000 or 2500 calories per day; and the distribution of \textit{\{400, 600, 600 and 400-to-900\}} calories over the meals \{\textit{breakfast, lunch, dinner and others(snacks and drinks)\}}, respectively. We adopted their meal to total calorie ratio and developed a distribution for users' personalized DRCI. Table \ref{tab:caloriedistribution} lists our proposed DRCI percentages to be allocated to different food-types.

 User's health-aware DRCI,\(DRCI_{u_a}\), is mapped on the \textit{DRCI to food-type} matrix based on the food-type of the current recipe \(cR_t\). Equation \ref{eq:foodtypeRatio2DCI} estimates the  \(P(cal, u_a)\), the proportion of DRCI the user \(u_a \) should be consuming from the recipe \(cR_t\). 
 
     \begin{equation} \label{eq:foodtypeRatio2DCI}
     \fontsize{14pt}{13pt}\selectfont
P \big(cal, u_a\big)=  DRCI_{u_a} \times  \frac{ \%(FoodType_x,{cR_t})}{100}    
\end{equation}
 
We determine how many portions of \(cR_t\) should be recommended to\(u_a \), \(\mathbb{Q} \big( cR_t\big)\),  using,
\begin{itemize}
\small
\setlength\itemsep{.02em}
    \item \(\mathbf{P(cal, u_a)}\), the predicted number of calories to be consumed from the recipe's food-type.
    \item \(\mathbf{Cal(cR_t)}\), the number of calories in one portion of \(cR_t\).
\end{itemize} 

Equation \ref{eq:2portioncal} generates the portion-size recommendation.

 \begin{equation} \label{eq:2portioncal}
 \fontsize{14pt}{13pt}\selectfont
\mathbb{Q} \big( cR_t\big) =  \frac{P\big(cal, u_a\big)}{ Cal(cR_t)}    
\end{equation}

\begin{table}[!h]
\centering
\small
\scalebox{.9}{
\setlength\tabcolsep{10pt}
\def\arraystretch{1.5}
\begin{tabular}{p{.12\textwidth}p{.01\textwidth}p{.2\textwidth}p{.005\textwidth}p{.12\textwidth}p{.2\textwidth}}
\rowcolor[HTML]{000000} 
\multicolumn{3}{c}{\cellcolor[HTML]{000000}{\color[HTML]{FFFFFF}\textbf{ 3 meals a day}}} &
 \multicolumn{1}{c} {\cellcolor[HTML]{FFFFFF}} &
  \multicolumn{2}{c}{\cellcolor[HTML]{000000}{\color[HTML]{FFFFFF} \textbf{2 meals a day}}} \\
\multicolumn{2}{c}{\cellcolor[HTML]{ECF4FF}Food-type} &  \cellcolor[HTML]{ECF4FF}{ \makecell{\% (approximate \\ percentage of calorie)}} &
   &
  \cellcolor[HTML]{ECF4FF}Food-type &
 \cellcolor[HTML]{ECF4FF}{ \makecell{\% (approximate \\ percentage of calorie)}} \\ \cline{1-3} \cline{5-6} 
\multicolumn{2}{c}{breakfast} &
 \multicolumn{1}{c}{ 20} &
   &
  \multicolumn{2}{c}{} \\ \cline{1-3}
\multicolumn{2}{c}{lunch} &
 \multicolumn{1}{c} {30} &
   &
  \multicolumn{2}{c}{} \\ \cline{1-3}
\multicolumn{2}{c}{dinner} &
  \multicolumn{1}{c}{30} &
   &
  \multicolumn{2}{c}{\multirow{-3}{*}{\setlength\tabcolsep{10pt}
  \begin{tabular}{p{0.15\textwidth}p{0.12\textwidth}}
    meal 1     &    \(\;\;\;\) 35-40\\ \hline
    meal 2        &  \(\;\;\;\) 40-45\end{tabular}}} \\ \cline{1-3} \cline{5-6} 
\multicolumn{2}{c}{drinks\tablefootnote{Drinks those contain calories e.g., smoothie, protein shakes,beers \label{ft:drinks}}} &
  \multicolumn{1}{c}{5-10} &
   &
  drinks \footref{ft:drinks} &
  \multicolumn{1}{c}{10-15}\\
\cline{1-3} \cline{5-6} 
\multicolumn{2}{c}{snack} &
  \multicolumn{1}{c}{10} &
   &
  snack &
  \multicolumn{1}{c}{10} \\ \cline{1-3} \cline{5-6} 
\multicolumn{2}{c}{side} &
  \multicolumn{1}{c}{10-15} &
  \multirow{-8}{*}{} &
  side &
  \multicolumn{1}{c}{10-15} \\ \cline{1-3} \cline{5-6} 
\end{tabular}}
\caption{ Calorie distribution over different meal and snacks during the day.}
\label{tab:caloriedistribution}
\end{table}
\section{ Proposed Health-Aware Smart-Nudging Strategies}

To lift the burden of processing complex and cross-domain nutrition information from users, the nutri-FRS needs to learn the healthiness of each recipe in its corpus. Furthermore, the nutri-FRS needs mediums that convey this knowledge to users without requiring much effort from them. In the context of the internet, a recipe is a similar content to news or articles, which encouraged us to investigate badge, keyword, and graphical visualization-based persuasion techniques that have seen significant success in persuasive news recommendation \cite{garrett2009echo,beam2014automating,messing2014selective}. To convey the health-information on recipes, we chose  two well established persuasive (User Interface) UI contents, \begin{inparaitem}\item[\textbullet]\textbf{an obvious-widget} \cite{wong2008transformation,buffet2013widget} \item[\textbullet] and\textbf{ a badge} \cite{chou2019actionable}\end{inparaitem}. 

An obvious-widget is an UI content that aims to attract users' attention during Multi-modal Graphical Interaction (MGI)\cite{cross2014supporting,MultimodalGraphicalInterFace,norris2004analyzing} between the user and an application (app). MGI is a competitive user-interaction scenario where diverse multi-modal contents strive to win attention, convey information, and receive responses. The typical structure, location on a web page, and chromaticity of an obvious-widget makes the widget indeed observed. To ensure maximum exposure of the obvious-widget, we adopted the cognitive theory on visual search in human visual-hemifields (left and right), proposed by Elena S Gorbunova et al. \cite{gorbunova2019visual}. According to Elena S Gorbunova et al., human brain tend to comprehend the words on the right side of the screen faster than those on the left. The results are the same across various platforms when graphical contents accompany texts. Hence, we deployed our obvious-widget on the right-most side of the screen/webPage. We also decided to adopt the design characteristic \textbf{\emph{fixed}} \cite{buffet2013widget,BootStrapWebpageComponents}  for our obvious-widget. The characteristic \textbf{\emph{fixed}} prevent a UI content from being affected, (e.g., clipped, partially or entirely hidden, and opacity reduction) when user scroll up-down or left-right.  Our final design decision on the visual contents, corresponding to our nudging strategies, is a rectangular widget, with the characteristics of \textit{obvious} and \textit{fixed} IU contents, to be implemented on the right side of the screen. This widget will provide users with various healthiness information generated by the nudging strategies.

 The widget displays additional information about each item in RecList, intriguing users' health paradigms when searching for recipes and leading them to healthier decisions.  However, the widget-based nudging can only be effective if users visit both healthy and less-healthy options. When the number of items in a RecList is not low (supposedly $\geq$ 8),  users give greater importance to the top items making them more likely to be selected \cite{halflife,Zhang2016}. RS researcher needs to design technology to spread user's attention over items that are not at the top of the list.  We developed a calorie-based badge to address this issue and incentify the healthier options RecList. A badge is  an \textit{UI  content} generally used for emphasizing numerical or categorical characteristics of objects it is attached to \cite{chou2019actionable}. Badges are usually numerical or graphical and have widely been used by popular applications, such as Facebook \cite{website:facebook}, LinkedIn \cite{website:linkedin} and Amazon \cite{website:amazon}. Social media applications most commonly use badges to emphasize aspects including verified profiles, number of clicks, number of purchases, user rank, and levels in games. For each of the three nudging strategy we designed corresponding badge to be implemented on the RecList.

  In the following section, we describe three novel visual smart-nudges that notify about the healthiness of a recipe when the user exercises a recipe retrieval action. Each of the visual nudges is designed to visualize domain-specific information and anticipate becoming the visual language of food healthiness in the online recommendation environment.

\subsection{ Daily Recommended Calorie Intake  based Machine Learned Consumption-Portion (DRCI-MLCP)\label{BMInudge}} 

  The choice architecture techniques adopted in DRCI-MLCP smart-nudge aim to provide users with personalized information on \begin{inparaitem} \item[\textbullet] \textit{if a food is suitable for them?} and \item[\textbullet] \textit{if yes, then how much?} \end{inparaitem}. The nudge attracts users' attention to compare their DRCI and the calorie contained within one portion of the recommended recipe. According to Public Health England (PHE), making specific information available to the user can help them make better choices regarding calorie intake \cite{phe2018calorie}. The DRCI-MLCP visual nudge is designed to guide users for healthy calorie intake.

The design and development of the DRCI-MLCP nudge can be divided into two major complex problems:

\begin{enumerate}\setlength\itemsep{.05em}
    \fontfamily{cmtt}\selectfont

    \item { Identifying the consumption portion-size, personalized to each user, for each food-item. }
    \item { Developing a visual nudge that can convey the food-healthiness information and help users identify healthy recipes.} 
 \end{enumerate}

We determined the healthy portion-size of each recipe for each user using the portion-size recommended proposed in section \ref{sec:portionsize}. 

 \subsubsection{Designing the Visual Content for DRCI-MLCP Obvious-Widget\label{sec:DRCI-MLCPbviouswidget}}
 
\begin{figure*}[!h]
  \centering
  \scalebox{.7}{
    \begin{tabular}{c c}
     \hskip.5cm\begin{tabular}{c}
     \begin{tikzpicture}
      \node[inner sep=0pt , rectangle] (russell) at (-4,0)
  {\includegraphics[width=.45\textwidth,height=!]{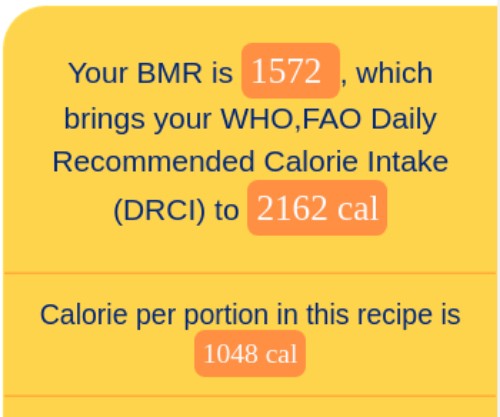}};
   
\node[inner sep=0pt , rectangle] (russell) at (-4.05,-6.1){
     \includegraphics[width=.453\textwidth,height=!]{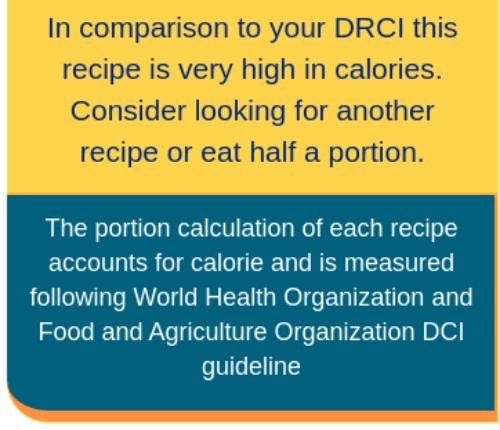}};
   
      \end{tikzpicture}
     \end{tabular}
    &  \hskip-.4cm\begin{tabular}{c}
         \includegraphics[width=.35\textwidth,height=2.8cm]{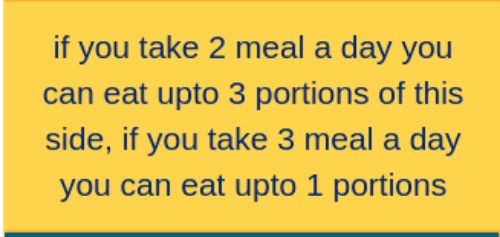} \\
         (b)\\
          \includegraphics[width=.35\textwidth,height=2.8cm]{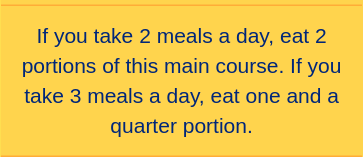} \\
         (c)\\
          \includegraphics[width=.35\textwidth,height=2.8cm]{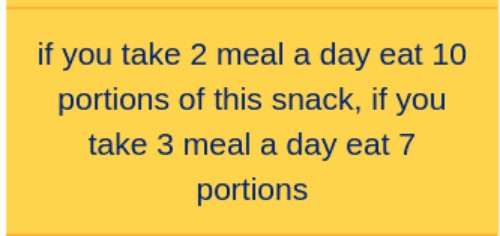}\\
          (d)
          \\
          \includegraphics[width=.35\textwidth,height=2.8cm]{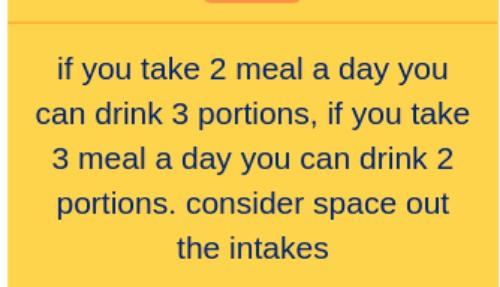}
         
    \end{tabular}
\\
(a)&(e)
    
    \end{tabular}
    }
\caption{The DRCI-MLCP Obvious-widget.}
    \label{fig:BMInudge}
\end{figure*}

To display the suitable portion-size recommendation for each recipe we designed DRCI-MLCP obvious-widget. The widget consisted of lexical and graphical contents, as shown in figure \ref{fig:BMInudge}(a). The widget is divided into four vertical sections, each providing distinct information. 

 \textbf{ \small Information Presented in Each of the  Four Vertical Sections of DRCI-MLPC Obvious-Widget.}
 \begin{itemize}\setlength\itemsep{.05em}
      \setlength\itemsep{.05em}
     \item \textbf{\small Top} Notify users of their BMR and health aware DRCI.
     \item \textbf{\small \(2^{nd}\) from  the Top}  Notify users of the number of calories in one portion of the current recipe.
     \item\textbf{\small \(3^{rd}\) from the Top} The Portion-size and explanations (optional). {\small \color[HTML]{3531FF}The optional explanation consists of explicit message on \emph{ why the current recipe is not a good fit for the user.}}
     \item \textbf{\small Bottom}  A short description of the source of the guidelines adopted in the portion-size calculation. This section aims at gaining subconscious-trust \cite{neal2011beyond} from user.
 \end{itemize}

As shown in figure \ref{fig:BMInudge}, the recommended portion-size, \(\mathbb{Q} \big( cR_t\big)\), is displayed at the widget's \textbf{ \(3^{rd}\)-from-the-Top} section. To help users decide on recipes by providing them more information regarding the recommendations, we check whether the recipe can avail the best satisfaction. Assuming \textit{users want to consume at least one whole portion of the recipe},  we compare \(P(cal, u_a)\) and \(Cal(cR_t)\), as shown in equation \ref{eq:fitcheck}. In the cases of one portion having greater calories than our predicted  \(P\big(cal, u_a\big)\), the DRCI-MLCP widget still recommends the suitable fractional proportion to the user. However, the widget provides an explanation on \textit{why the recipe is not suitable for them} and  \textit{why a portion-size of \emph{less-than-one} was recommended}.  We wrote a (Natural Language Processing) NLP script that generates explicit instructions on \textit{how much  of a food-item the user should consume}, as shown in figure \ref{BMInudge} (b), (c), (d) and (e).

  \begin{equation} \label{eq:fitcheck}
     \fontsize{14pt}{13pt}\selectfont
 Cal(cR_t)  \leq P\big(cal, u_a\big)    
\end{equation}

\begin{figure}[!h]
\centering
\begin{tikzpicture}
\node[inner sep=0pt , rectangle, draw] (russell) at (3,0)
    {  \includegraphics[width=.45\textwidth,height=!]{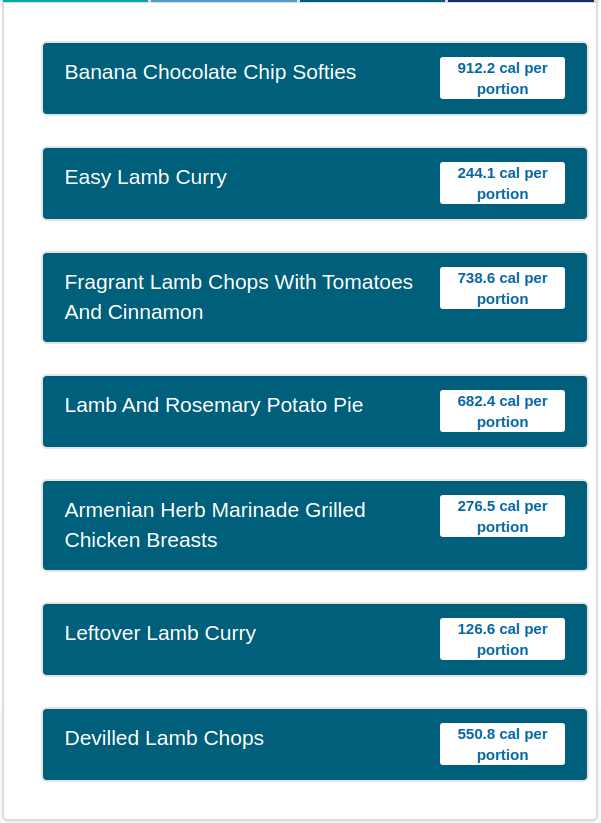}};

 \node[inner sep=0pt , rectangle,fill=white, minimum height=5pt,minimum width=1.3cm] at (5.4,-4) {\fontsize{6pt}{5pt}\selectfont \color{kosternil}1048 cal per};
 \end{tikzpicture}

\caption{The DRCI-MLCP Badge.}
    \label{fig:Calbadge}
         
\end{figure}

 For users who might want to eat larger portions, the DRCI-MLCP widget also suggests searching for similar recipes with fewer calories.  Unlike state-of-the-art nutri-FRS strategies, the recipes, unfit to \(u_a\)'s DRCI range, are still made available to \(u_a\). However, \(u_a\) is provided with information on \textit{ why the recipe is not a best match for \(u_a\)}. The widget also gives additional instructions, for example, \q{to divide the consumption into multiple occasions and keep time gaps between consummation event, when multiple portions are recommended for snacks and drinks}.

\subsubsection{Designing the Visual Content for DRCI-MLCP Badge}
 
 To guide users from the beginning of a recipe search event we developed DRCI-MLCP based badge, a CalBadge, as shown in figure \ref{fig:Calbadge}. Here, we adopted a numerical badge \cite{chou2019actionable}, notifying users about the number of calories in a single portion of a recommended recipe. If users have a pre-existing knowledge of their \textit{healthy calorie} ranges, this CalBadge will enable them to look out for suitable recipes since the early stage of the browsing process. For example, if a user is looking for a meal with high-calories (e.g., $\ge700cal$), they are likely to browse the \(1^{st}\), \(3^{rd}\) and \(7^{th}\) recipes in the RecList, shown in figure \ref{fig:Calbadge}. 
 
 Unlike most nutri-FRS, DRCI-MLCP widget dose not forbid options instead provides information on why a recipe may or may not be healthy for the user. Leaving the decisions entirely up to the users, the DRCI-MLCP widget encourages users towards healthier choices.  And the DRCI-MLCP badge guide users to recipes that are best match for them. It at aims reduce the number of clicks between the initiation of a \textit{recipe search} and a \textit{healthy recipe being found}.
 
 \begin{table}[!h]
\small
\centering
\def\arraystretch{1.2}
\scalebox{.8}{
\begin{tabular}{|m{.05\textwidth}|m{.4\textwidth}|m{.35\textwidth}|}
\hline
\rowcolor[HTML]{000000} 
\multicolumn{2}{|m{.425\textwidth}|}{\cellcolor[HTML]{000000}{\color[HTML]{FFFFFF} \textbf{Dietary-Factor /Macro-Nutrient}}} &
  {\color[HTML]{FFFFFF} \textbf{Goal (\% Macro-Nutrient unless otherwise stated)}} \\ \hline
\rowcolor[HTML]{EFEFEF} 
\multicolumn{2}{|m{.425\textwidth}|}{\cellcolor[HTML]{EFEFEF}Total fat}  & 15 - 30\%                                      \\ \hline
             \multicolumn{1}{|c} {}     & \multicolumn{1}{m{.4\textwidth}|}{\cellcolor[HTML]{ECF4FF}Saturated Fatty Acids }                              & \cellcolor[HTML]{ECF4FF}\textless{}10\%        \\ \cline{2-3} 
              \multicolumn{1}{|c} {}     & \multicolumn{1}{m{.4\textwidth}|}{\cellcolor[HTML]{ECF4FF}Polyunsaturated Fatty Acids }                        & \cellcolor[HTML]{ECF4FF}6 - 10\%               \\ \cline{2-3} 
                    \multicolumn{1}{|c} {}     & \multicolumn{1}{m{.4\textwidth}|}{\cellcolor[HTML]{ECF4FF}n-6 Polyunsaturated Fatty Acids    }                 & \cellcolor[HTML]{ECF4FF}5 - 8\%                \\ \cline{2-3} 
              \multicolumn{1}{|c} {}     & \multicolumn{1}{m{.4\textwidth}|}{\cellcolor[HTML]{ECF4FF}n-3 Polyunsaturated Fatty Acids }                    & \cellcolor[HTML]{ECF4FF}1 - 2\%                \\ \cline{2-3} 
                  \multicolumn{1}{|c} {}     & \multicolumn{1}{m{.4\textwidth}|}{\cellcolor[HTML]{ECF4FF}Trans fatty acids      }                             & \cellcolor[HTML]{ECF4FF}\textless{}1\%         \\ \cline{2-3} 
                                     \multicolumn{1}{|c} {}     & \multicolumn{1}{m{.4\textwidth}|}{ \cellcolor[HTML]{ECF4FF}
                              \textbf{Monounsaturated Fatty Acids }\newline {\begingroup (Saturated fatty acids + Polyunsaturated fatty acids + Trans fatty acids)\endgroup}  }         &  \cellcolor[HTML]{ECF4FF} \textless{}50\%        \\ \cline{2-3} 
\rowcolor[HTML]{C0C0C0} 
\multicolumn{2}{|m{.425\textwidth}|}{\cellcolor[HTML]{C0C0C0}\textbf{Total Carbohydrate} \newline (The percentage of total energy available after taking into account energy consumed as protein and fat. Hence the wide range.)} & 
  55 - 75\% \\ \hline
\rowcolor[HTML]{EFEFEF} 
\multicolumn{2}{|m{.425\textwidth}|}{\cellcolor[HTML]{EFEFEF}Sugar ( including Monosaccharides and Disaccharides)} & \textless{}10\%                                \\ \hline
\rowcolor[HTML]{EFEFEF} 
\multicolumn{2}{|m{.425\textwidth}|}{\cellcolor[HTML]{EFEFEF}Protein}                                              & 10 - 15\%                                      \\ \hline
\rowcolor[HTML]{EFEFEF} 
\multicolumn{2}{|m{.425\textwidth}|}{\cellcolor[HTML]{EFEFEF}Cholesterol}                                          & \textless{}300 mg per day                      \\ \hline
\rowcolor[HTML]{EFEFEF} 
\multicolumn{2}{|m{.425\textwidth}|}{\cellcolor[HTML]{EFEFEF}Sodium Chloride (sodium)}                             & \textless{}5g per day (\textless{}2 g per day) \\ \hline
\rowcolor[HTML]{EFEFEF} 
\multicolumn{2}{|m{.425\textwidth}|}{\cellcolor[HTML]{EFEFEF}Fruits and Vegetables}                                & $\geq$400 g per day                                 \\ \hline
\rowcolor[HTML]{EFEFEF} 
\multicolumn{2}{|m{.425\textwidth}|}{\cellcolor[HTML]{EFEFEF}Total Dietary Fibres}                                  & \textgreater{}25g                              \\ \hline
\rowcolor[HTML]{EFEFEF} 
\multicolumn{2}{|m{.425\textwidth}|}{\cellcolor[HTML]{EFEFEF}Non-starch Polysaccharides}                           & \textgreater{}3\%                              \\ \hline
\end{tabular}
}
\caption{ WHO nutrient intake goals for proportion of macro nutrient in a 100gm of food \cite{website:WHOintakeguide}.}
\label{tab:WhoGuideline}
\end{table}

\subsection{ The WHO-HealthScore based nudge: WHO-BubbleSlider \label{WhoHealthscore}}

In this nudging technology, we investigate food intake guidelines offered by WHO. In cooperation with FAO, WHO has issued guidelines on a balanced-diet. The guideline is described in terms of healthy ranges for various macro-nutrients that should be considered as intake goals. The list includes unique nutrients (e.g., protein and cholesterol) and secondary nutrients (e.g.,  n-6 Polyunsaturated fatty acids). Table \ref{tab:WhoGuideline} summarizes the WHO intake goals for the 15 more significant macro-nutrients. Though the guidelines on healthy-intake of various macro-nutrients, vitamins, and minerals are available, people often struggle to comprehend that information and make an aware judgment during an eating or cooking decision. We aim to develop a numeric healthiness scale, WHO-HealthScore, and corresponding visual contents to help people evaluate recipes on the health standers proposed by WHO. For developing the healthiness scale, we consider the seven major macro-nutrients, such as \begin{inparaitem} \item[\textbullet] protein, \item[\textbullet]carbohydrate, \item[\textbullet]sugar, \item[\textbullet]sodium, \item[\textbullet]total fat, \item[\textbullet]saturated fat, and \item[\textbullet]dietary fiber \end{inparaitem}. According WHO, these seven macro-nutrients are more crucial for nutritional deficiency risks and chronic non-deficiency diseases \cite{website:Who_healthyDiet}. Since for decades \textit{cholesterol} has been associated with chronic diseases (e.g. Coronary Artery Disease) \cite{Dairyproductsandplasmacholesterollevels}, we  also included cholesterol along with the seven major macro-nutrients.

Our novel WHO-HealthScore based nudge aims to help users evaluate their options on WHO defined health standers. The design and development of the WHO-HealthScore based nudge is divided into two significant problems.

\begin{enumerate}\setlength\itemsep{.05em}
\small
 \fontfamily{psv}\selectfont

     \item { Investigating approaches to define a standardized scale for recipe-healthiness  based on the WHO proposed guidelines.}
    \item {Developing  visual contents that help users understand the information given to them and consequently identify a healthy recipe.}
 \end{enumerate}

\subsubsection{Investigating approaches to define a standardized scale to assess the healthiness of recipes based on WHO proposed guideline}

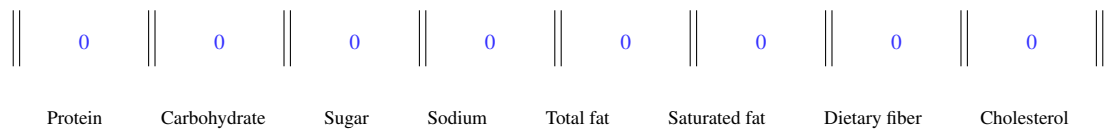
\begin{figure}[!h]
\centering
\scalebox{1}{
\begin{tikzpicture}
\node[inner sep=0pt , rectangle,minimum width=\textwidth] at (-4,0){
\setlength\tabcolsep{22pt} 
\def\arraystretch{2} \begin{tabular}{||c||c||c||c||c||c||c||c||}
\color[HTML]{3531FF} 0& \color[HTML]{3531FF}0&\color[HTML]{3531FF} 0&\color[HTML]{3531FF} 0&\color[HTML]{3531FF} 0& \color[HTML]{3531FF}0& \color[HTML]{3531FF}0&\color[HTML]{3531FF} 0 \\

\end{tabular}};
\node[inner sep=0pt , rectangle,minimum width=\textwidth] at (-4,-1){
\setlength\tabcolsep{11pt} 
\def\arraystretch{2} \small \begin{tabular}{||c||c||c||c||c||c||c||c||}

\multicolumn{1}{c}{Protein}& \multicolumn{1}{c}{Carbohydrate} &  \multicolumn{1}{c}{Sugar}&  \multicolumn{1}{c}{Sodium} &  \multicolumn{1}{c}{Total fat} &  \multicolumn{1}{c}{Saturated fat} & \multicolumn{1}{c}{Dietary fiber} & \multicolumn{1}{c}{Cholesterol}
\end{tabular}};
\end{tikzpicture}
}
\caption{The eight macro-nutrient bins for each recipe.}
\label{fig:nutrientbins}
\end{figure}

To convert the WHO-defined intake ranges into a continuous numeric scale, we adopted a variation of one-hot coding,  a similar approach investigated by Simon Howard et al. in  \cite{NutritionalContentSsupermarketReadyMeals}.  As shown in figure \ref{fig:nutrientbins}, in an array of eight bins, each macro-nutrient is assigned a bin with one bit set to zero. If the corresponding nutrient's proportion in a recipe falls within the WHO healthy range, the digit is set to one. Each recipe is compared against WHO healthy range for all eight macro-nutrients. Finally, the healthiness of the recipe is determined by the sum of all eight bins. For a recipe, the possible range for the sum of these eight bins is \textbf{0} ({ \color{brick} no nutrient goals are fulfilled}) to \textbf{8 }({\color{brick} all nutrient goals are fulfilled}). This 0-to-8 is our WHO guideline based food healthiness scale, the \textbf{WHO-HealthScale}. The healthiness score of a recipe on the WHO-HealthScale is their \textbf{WHO-HealthScore}.

To inform the user about the WHO-HealthScore of each item in the RecList, we developed an obvious-widget. For guiding users' attention towards healthier recipes, we designed and implemented the WHO-HealthScore badge on the RecList.

\subsubsection{ Designing the Visual Content for the WHO-HealthScore based Obvious-widget: the WHO-BubbleSlider Obvious-widget \label{sec:WHOBubbleSliderWidget based Nudge}}
 
\begin{figure}[!h]
\centering
\begin{tikzpicture}
\node[inner sep=0pt , rectangle] (russell) at (0,0)
    {  \includegraphics[width=.25\textwidth,height=!]{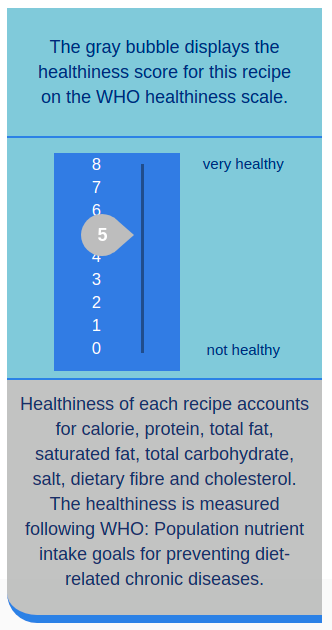}};
    
\end{tikzpicture}

\caption{The WHO-BubbleSlider Obvious-widget}
\label{fig:WHOHealthScaleWidget}
\end{figure}

 To transform this WHO-HealthScore into a visual nudge, we designed a graphical nudging content, a WHO-BubbleSlider, as shown in figure \ref{fig:WHOHealthScaleWidget}. The nudging strategy is a vertical scale accompanied by a bubble notifying a recipe's position on the scale.  Such scales are called bubble-slider-scale. The core of this nudging content is the vertical scale, with 0, at the bottom representing very unhealthy recipes, and 8, at the top representing very healthy recipes. The 0-8 range corresponds to WHO-HealthScale. Within this WHO-BubbleSlider, a gray bubble is dynamically placed along the vertical scale at the position of the WHO-HealthScore of a corresponding recipe. The gray bubble is designed along the vertical scale to draw users' attention and explicitly inform the corresponding recipe's healthiness. 
    
Along with the numeric declaration of the healthiness of a recipe, the BubbleSlider nudge strives to encourage a visual understanding of the recipe-healthiness guidance. For example, the higher the bubble is on the scale, the healthier a recipe. When browsing from one recipe to another, the change in the vertical position of the bubble gives a visual impression similar to sliding \cite{BootStrapWebpageComponents}. Which gives the additional advantage of visual comparison of healthiness of two consecutive recipes in a user's browsing flow. 

In this nudging strategy, to convey the recipe-healthiness information we implemented the WHO-BubbleSlider nudge as a \textbf{fixed} and \textbf{obvious} widget at the rightmost corner of the screen.  The WHO-BubbleSlider obvious-widget consists of three vertical sections, separated by horizontal-lines and varying backgrounds.

  \textbf{ \small Information Presented in Each of the  three Vertical Sections of WHO-BubbleSlider Widget.}
 \begin{itemize}\setlength\itemsep{.05em}
 \small
     \item \textbf{\small Top} A short note on the WHO-BubbleSlider scale to help first-time users understand the scale. ( {\small \color{eggplant} Long-term users are more likely to develop knowledge on different attributes of any smart application.})
     \item \textbf{\small \(2^{nd}\) from Top} The WHO-BubbleSlider scale to visualize the healthiness of the corresponding recipe. 
     \item \textbf{\small Bottom} A short description of the source of the healthiness calculation guidelines. This section aims at gaining subconscious-trust  \cite{neal2011beyond} from user.
 \end{itemize}

\subsubsection{ WHO-HealthScore Badge}

\begin{figure}[!h]
\centering
\begin{tikzpicture}
    \node[inner sep=0pt , rectangle, draw] (russell) at (0,0)
    {  \includegraphics[width=.45\textwidth,height=!]{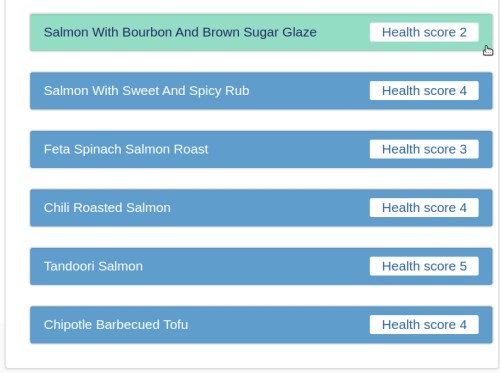}};

\end{tikzpicture}

\caption{The WHO-HealthBadge}
\label{fig:WHO-HealthScore}
\end{figure}

We also developed a numeric badge, WHO-HealthBadge, to attract users' attention towards healthier options within the RecLlist. The WHO-HealthBadges, attached to each item in the RecList, displays the WHO-HealthScore of the corresponding recipe, as shown in figure \ref{fig:WHO-HealthScore}.  The aim is to warn users about the healthiness of the recipe and assist them from the beginning of the recommendation process.

\subsection{The FSA-HealthScore based Nudge:  FSA-ColorCoding}\label{FSAHealthscore}

For the third and final nudging technology, the FSA-ColorCoding, we followed the FSA guidelines on \emph{nutritional balance} in healthy food \cite{website:FSA:Nutrient}. The FSA nutritional balance guideline corresponds to a relatively smaller number of macro-nutrients, such as  \begin{inparaitem}  \item [\textbullet] sugar, \item [\textbullet] sodium, \item [\textbullet] total fat, and \item [\textbullet] saturated fat \end{inparaitem}\cite{website:FSA:Nutrient}. In this work, we refer to the set of these four macro-nutrients as FSA-4. For each macro-nutrient in FSA-4, FSA proposed three different healthiness ranges, such as \textbf{healthy}: LOW, \textbf{moderate}: MEDIUM, and \textbf{unhealthy}: HIGH, as shown in table \ref{tab:FSA_guideline}. For example, according to FSA guidelines, as shown in table \ref{tab:FSA_guideline}, a recipe \(r_t\) with a total fat of $ \leq3\%$ is a healthy recipe concerning only \textit{total fat}. The FSA guideline also determined a traffic light food labeling scale for their healthiness ranges, such as  green (healthy), amber (moderate), and red (unhealthy) \cite{FSATrafficLightSystemFoodLabelling,website:FSA:Nutrient}. According to the FSA guideline, the recipe \(r_t\) is a green recipe with respect to \textit{total fat}. While the color-based explanation is much easier to comprehend and is used worldwide on food-packaging, it is still challenging for many to evaluate recipes explicitly for each macro-nutrient in FSA-4 and make a healthy judgment.

To determine a standardized healthiness assessment inspired by the FSA guideline and inform users of the overall healthiness of recipes, we developed \textbf{FSA-ColorCoding}. The FSA-ColorCoding informs the overall healthiness of a recipe based on all four macro-nutrients in FSA-4. We designed the nudging content for FSA-ColorCoding in the form of a colored disk. The design and development of the FSA-ColorCoding nudge can be divided in two major problems:

\begin{enumerate}\setlength\itemsep{.05em}
\small
 \fontfamily{psv}\selectfont
     \item {Investigating approaches to define a standardized scale for recipe-healthiness based on the FSA proposed guidelines.}
    \item {Developing  visual contents that help users understand the information given to them and consequently identify a healthy recipe.} 
 \end{enumerate}  

\subsubsection{Investigating approaches to define a standardized scale to assess the healthiness of recipes based on the FSA proposed guidelines }

\begin{itemize}\setlength\itemsep{.05em}
 \item {\textbf{The FSA-HealthScore}}

The FSA proposed three ranges (e.g., green, amber and red) for the FSA-4,  successfully cover all the recipes in our corpus. However, a bulk proportion of recipes falls in the FSA-defined \textit{red range} for one or more macro-nutrients, making a significantly large number of recipes unhealthy. This issue raised the possibility of having many red recipes in the RecList Which cam make users feel \emph{lacking of options}. Nudging technology does not restrict less healthy recipes; instead displays a notification on the lack of health in the recipe and leaves the decision up to users. However, as red is the least healthy range, showing many red recipes to users can cause depletion of their trust in the FRS. To differentiate between the \textit{unhealthy} and \textit{very-unhealthy} recipes and potentially find a least-healthy range much farther from the medium-range, we split the FSA-defined range \textit{red} into two. The splitting generated two new customized ranges, such as \textbf{High}: \textit{unhealthy} and \textbf{VERY-HIGH}: \textit{very-unhealthy}. Further classifying FSA red range,  can give users a better chance of differentiating between comparatively healthier recipes. We assigned the color \textbf{red} to the new unhealthy and the color \textbf{brown} to new very-unhealthy ranges. The span for the FSA-defined ranges green and amber remains unchanged.

\begin{table}[!h]
\centering
\small
\setlength\tabcolsep{22pt}
\scalebox{.9}{
\def\arraystretch{1.5}
\begingroup
\begin{tabular}{c
>{\columncolor[HTML]{C5E0B3}}c 
>{\columncolor[HTML]{FFE599}}c 
>{\columncolor[HTML]{FF9590}}c }
\cellcolor[HTML]{000000}{\color[HTML]{FFFFFF} \textbf{ Macro-Nutrient } } &
  \cellcolor[HTML]{000000}{\color[HTML]{FFFFFF} \textbf{LOW}} &
  \cellcolor[HTML]{000000}{\color[HTML]{FFFFFF} \textbf{MEDIUM}} &
  \cellcolor[HTML]{000000}{\color[HTML]{FFFFFF} \textbf{HIGH}} \\
    & \cellcolor{white} \color[HTML]{2E8B57} Green & \cellcolor{white} \color[HTML]{B8860B} Amber & \cellcolor{white} \color[HTML]{8B4513} Red \\
Total Fat            & $\leq$ 3 	    \% & \textgreater 3 \%      to $\leq$ 17.5 \% & \textgreater{}17.5 \% \\
Saturated Fat      & $\leq$ 1.5    \% & \textgreater 1.5 \%    to   $\leq$ 5.0 \% & \textgreater{}5.0 \%  \\
(Total) Sugars & $\leq$ 5.0    \% & \textgreater 5.0 \%   to   $\leq$ 22.5 \%  & \textgreater{}22.5 \% \\
Salt           & $\leq$ 0.3    \% & \textgreater 	0.3 \%   to   $\leq$ 1.5 \% & \textgreater{}1.5\%  
\end{tabular}
\endgroup
}
\caption{\small FSA guidelines on proportion of FSA-4 macro-nutrients in 100gm of food \cite{website:FSA:Nutrient}.}
\label{tab:FSA_guideline}
\end{table}

\begin{table}[!h]
\centering
\small
\scalebox{.7}{
\setlength\tabcolsep{11pt}
\def\arraystretch{1.8}
\begingroup
\begin{tabular}{c
>{\columncolor[HTML]{C5E0B3}}c 
>{\columncolor[HTML]{FFE599}}c 
>{\columncolor[HTML]{FF9590}}c 
>{\columncolor[HTML]{E6A060}}c}
\cellcolor[HTML]{000000}{\color[HTML]{FFFFFF} \textbf{ Macro-Nutrient } } &
  \cellcolor[HTML]{000000}{\color[HTML]{FFFFFF} \textbf{LOW}} &
  \cellcolor[HTML]{000000}{\color[HTML]{FFFFFF} \textbf{MEDIUM}} &
  \cellcolor[HTML]{000000}{\color[HTML]{FFFFFF} \textbf{HIGH}}  &
    \cellcolor[HTML]{000000}{\color[HTML]{FFFFFF} \textbf{VERY HIGH}}\\
    & \cellcolor{white} \color[HTML]{2E8B57} Green & \cellcolor{white} \color[HTML]{B8860B} Amber & \cellcolor{white} \color[HTML]{8B4513} Red & \cellcolor{white} \color[HTML]{8B4513} Brown \\
Total Fat            & $\leq$ 3 	    \% & \textgreater 3 \%      to $\leq$ 17.5 \% & \textgreater 17.5 \% to $\leq$ 26.25 \%& \textgreater 26.25 \%\\
Saturated Fat      & $\leq$ 1.5    \% & \textgreater 1.5 \%    to   $\leq$ 5.0 \% & \textgreater 5.0 \% to   $\leq$ 7.5 \% &\textgreater  7.5 \% \\
(Total) Sugars & $\leq$ 5.0    \% & \textgreater 5.0 \%   to   $\leq$ 22.5 \%  &\textgreater 22.5 \%  to  $\leq$   33.75 \%&\textgreater   33.75 \%\\
Salt           & $\leq$ 0.3    \% & \textgreater 	0.3 \%   to   $\leq$ 1.5 \% & \textgreater 	1.5 \%   to   $\leq$ 2.25 \%&\textgreater  2.25 \%
\end{tabular}
\endgroup
}
\caption{ The proposed optimization on nutrient-level classes.}
\label{tab:FSA_guidelineOptimize}
\end{table}

We identified the span for the \(4^{th}\) range \textit{brown} using equation \ref{eq:redefFSA}. In
equation \ref{eq:redefFSA}, for a macro-nutrient \(mc_i\), \(HIGH^{FSA}(mc_i)\) is the minimum value for the FSA defined least-healthy range \{high:unhealthy\} and \(VeryHIGH^{opt}(mc_i)\) is the minimum value for the new least-healthy range \{very high: very unhealthy\}).

    \begin{equation} \label{eq:redefFSA}
 VeryHIGH^{opt}(mc_i)  = HIGH^{FSA}(mc_i) \times  1.5    
\end{equation}

Table \ref{tab:FSA_guidelineOptimize} summarizes the new four healthiness ranges, such as green, amber, red and brown, for the FSA-4. To convert the FSA-inspired macro-nutrient guidelines ( healthiness ranges) into a standardized scale, we adopted Gary Sacks et al.'s work in \cite{Impact_of_front_of_pack_dap032}. We assigned an integer value to each range, such as healthy=1, medium-healthy=2, unhealthy=3, and very-unhealthy=4. The sum of each macro-nutrient's numeric score accumulates in a final range of 4 (very healthy recipe) to 16 (very unhealthy recipe). This \textbf{4-to-16} is our proposed FSA-HealthScale. Each recipe is checked against table \ref{tab:FSA_guidelineOptimize}, and scores are assigned based on the quantity of each of FSA-4  macro-nutrients.  A macro-nutrient \(mc_i\)'s score for a recipe \(r_t\) is determined based on the range that matches the quantity of \(mc_i\) in \(r_t\). The sum of the scores for each macro-nutrient in FSA-4 is the FSA-HealthScore of \(r_t\).  Table \ref{tab:healtFooddotcom} shows the spread of the our dataset over the FSA-HealthScore. 

The FSA guideline also suggests the healthy range for daily intake of \textit{dietary fiber}.  As part of our FSA-HealthScore, we generated a separate boolean scale for dietary fiber, the fibreScore. If the quantity of dietary fiber in a recipe satisfies the FSA healthy-range, the fibreScore is 1 and 0 otherwise.

\item {\textbf{ The FSA-ColorCode}}\label{FSAHealthScore}

We divided the numeric scale  4-to-16 into four epochs: healthy (4-6), moderately-healthy (7-9), unhealthy (10-12), and very-unhealthy (13-16).  The epochs were carefully designed to keep our \textit{healthy} and \textit{medium-healthy} ranges aligned with the FSA \textit{healthy} and \textit{moderate} ranges; only recipes in the FSA \textit{unhealthy-range} to be differentiated into \textit{unhealthy} and \textit{very unhealthy} recipes.  We assigned a color code to each epoch :

\hspace{.25\textwidth}\begin{tabular}{p{.4\textwidth}}
\cellcolor{leveldorGray}   
\begin{enumerate}\setlength\itemsep{.05em}
\item Healthy (4-7) = Green
\item Moderately-healthy (8-11) = Amber
\item Unhealthy (12-14) = Red
\item Very-unhealthy (15-16) = Brown
\end{enumerate}
\end{tabular}

This {\color{brick}\(\mathbf{| Green \rightarrow  Amber\rightarrow Red \rightarrow Brown |}\)} scale corresponding to the categorical scale  {\color{brick}\(\mathbf{| healthy \rightarrow  medium\mbox{-}healthy\rightarrow unhealthy  \rightarrow}\)\newline \(\mathbf{very\mbox{-}unhealthy |}\)} is our FSA-ColorScale. The  FSA-HealthScore of a recipe determines the color of the recipe on the  FSA-ColorScale. For example, any recipe with an FSA-HealthScore between 8 and 11 is given a FSA-ColorCode of Amber on the FSA-ColorScale.  To inform the user about the FSA-ColorCode of each item in the RecList, we developed an FSA-ColorCoading obvious-widget, as shown in figure \ref{fig:FSACOLORCOding} (a) and (b). For guiding users' attention towards healthier recipes, we designed and implemented the FSA-ColorBadge, as shown in figure \ref{fig:FSACOLORBadge} (a) and (b).

\end{itemize}

 \begin{figure}[!h]
\centering
\scalebox{.6}{
\begin{tikzpicture}
\node[inner sep=0pt , rectangle] (russell) at (-2,0)
    {  \includegraphics[width=.45\textwidth,height=!]{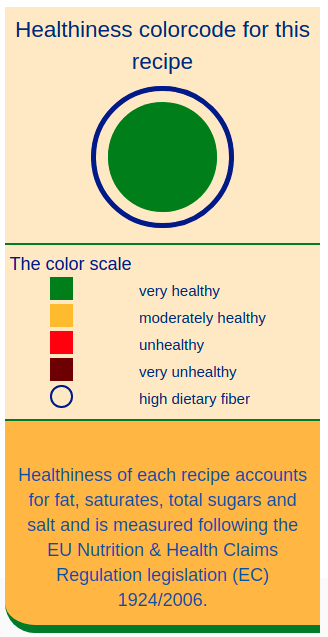}};
    \node[inner sep=0pt , rectangle] (russell) at (7,0)
    {  \includegraphics[width=.45\textwidth,height=!]{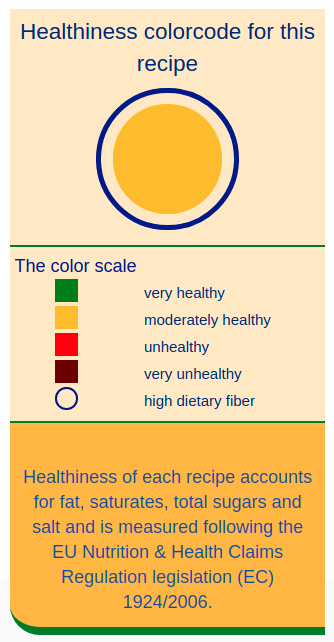}};
    
\node[] at (-2,-7.5) {(a)};
\node[] at (7,-7.5) { (b)};
\end{tikzpicture}
}
\caption{The FSA-ColorCoding Obvious-widget. }
\label{fig:FSACOLORCOding}
\end{figure}

\subsubsection{Designing the Visual Content for FSA-ColorCoading Obvious-Widget \label{sec:FSAcolorcoadinWidget}}

 To transform the FSA-ColorCode into a visual nudge, we designed a colored disk, the FSA-disk. As shown in figure \ref{fig:FSACOLORCOding}, FSA-disk takes the color of corresponding recipe's FSA-ColorCode.. To represent the boolean fiberScale, we designed a blue ribbon, fiberRibbon, around the FSA-disk. The ribbon is displayed if the dietary fiber score is 1 and hidden otherwise. This FSA-disk and the fiberRibbon is the core of FSA-ColorCoding nudge. The FSA-ColorCoding obvious-widget consists of three vertical sections, separated by horizontal-lines and varying backgrounds.

  \textbf{\small Information Presented in Each of the  Three Vertical Sections of FSA-ColorCoding Obvious-widget.}
 \begin{itemize}\setlength\itemsep{.05em}
     \item \textbf{\small Top} The FSA-disk (a color filled circle) with the dietary fiber ribbon around it.

    \item \textbf{\small \(2^{nd}\) from Top} Indications on the meaning of different colors in the ColorCoding scale to help first-time users understand the scale. ( {\small \color{eggplant} Long-term users are more likely to develop knowledge on different attributes of any smart application.})

     \item \textbf{\small Bottom} A short description of the source of the healthiness calculation guidelines. This section aims at gaining subconscious-trust  \cite{neal2011beyond} from user.
 \end{itemize}
 
 This nudging strategy aims to establish a color based recipe-healthiness language that describes recipes using five colors, e.g., green, amber, red, brown and blue, and help user make informed judgement.

\subsubsection{ Designing the Visual Content for The FSA-ColorBadge}
  
  \begin{figure}[!h]
\centering
\scalebox{.9}{
\begin{tikzpicture}
\node[inner sep=0pt , rectangle] (russell) at (-3.5,0)
    {  \includegraphics[width=.5\textwidth,height=9cm]{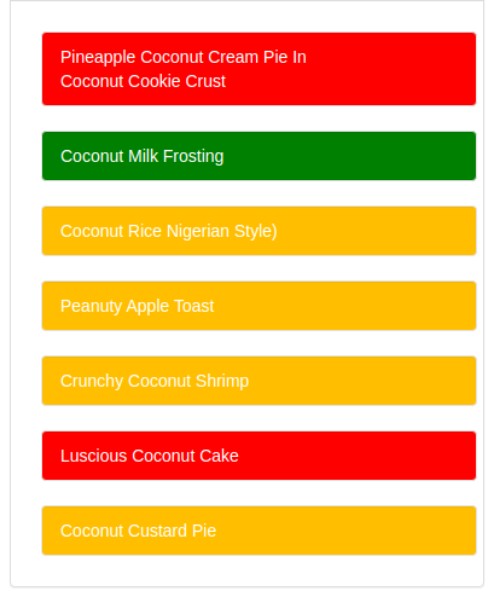}};
    
    \node[inner sep=0pt , rectangle] (russell) at (5,0)
    {  \includegraphics[width=.5\textwidth,height=9cm]{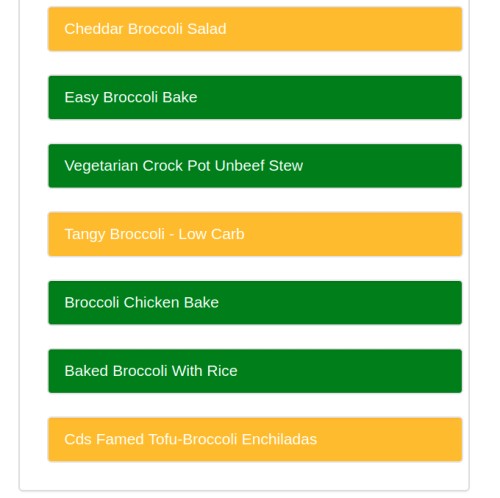}};
    
    \node[] at (-4,-5) {(a)};
\node[] at (5,-5) { (b)};
   
\end{tikzpicture}}

\caption{The FSA-ColorCoding healthiness badge. }
\label{fig:FSACOLORBadge}
\end{figure}

  We also developed FSA-ColorBadge, a color badge, to attract users' attention towards healthier options within the RecLlist. The FSA-ColorBadge, attached to each recipe in the RecList, displays the FSA-ColorCode of the corresponding recipe, as shown in figure \ref{fig:FSACOLORBadge} (a) and (b). The colored or badge-annotated RecList helps users to reach a suitable recipe in less time. The global association the color green  to the concept goodness \cite{naz2004color} creates high incentives for recipes presented with a green badge.

\section{Evaluation}

To analyze the performance of our proposed health-aware nudging techniques, we conducted an online user study. We developed a full-stack website called CookIT, with all necessary attributes for evaluating recommendation scenarios under the impression of our novel nudges. The system takes the user through the experiences of four recommendation scenarios:  one for each of the three health nudging strategies, along with one baseline scenario. The baseline scenario is a pure recommendation scenario with no nudging or choice architecture attribute. To encourage users in healthier food choices, recommending healthy recipes, which users also like, is essential. Taking this into account, we evaluated the nudging strategies in combination with a personalized recipe recommender. We applied  a hybrid Feature and Topic based algorithm \(F\_T\_R\)  \cite{khan2021addressing}. The user study aims to evaluate whether any of the three smart-nudges effectively motivate users to choose healthier options by making informed recommendations.

     

\subsection{The Experimental Setup: \label{sec:user-modeling}}

The course of action during the user study is divided into three major sections.

\subsubsection{Signing Up: User Modeling}

The user study requires users to create an account within the system to take part in the experiment. The process begins with displaying users information-sheets to inform them on the data \textit{collection} and \textit{storage} strategies. The information-sheets also informs about the possible future uses of the data collected during the experiment. Following up the information-sheets users are displayed with a consent-form. The consent questions are defined imposing UCD research regulation and data protection policies \cite{website:HRECGuideline,website:UCDDataProtectionPolicy,website:thepolicies1,website:healthResearchRegulation,website:HumanresearchEthicAssesment} and the GDPR \cite{website:TheGeneralDataProtectionRegulation,website:DataprotectionRsearchInhealth}.

\begin{figure}[!h]
\centering
\scalebox{.7}{
\begin{tikzpicture}
\node[inner sep=0pt , rectangle, draw] (russell) at (.1,0)
    {  \includegraphics[width=.95\textwidth,height=.6\textwidth]{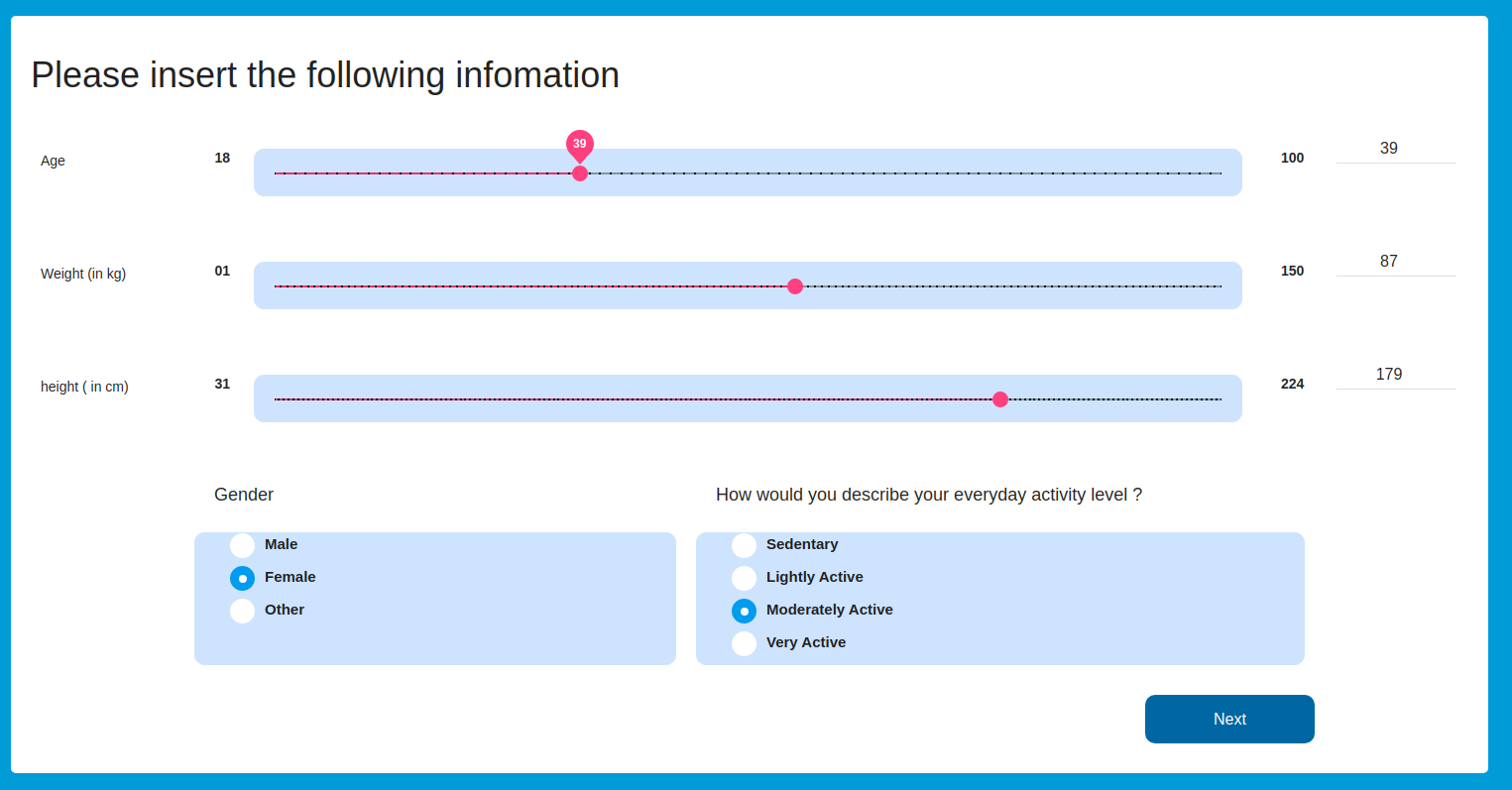}};
\node[fill=white, rectangle] at (-7,3) {  \textcolor{black}{\begingroup \fontsize{9pt}{9pt}\selectfont Age \endgroup} };
\node[fill=white, rectangle] at (-5.5,3) {  \textcolor{black}{\begingroup \fontsize{9pt}{9pt}\selectfont 18 \endgroup} };
\node[fill=white, rectangle] at (5.8,3) {  \textcolor{black}{\begingroup \fontsize{9pt}{9pt}\selectfont 100 \endgroup} };
\node[fill=white, rectangle] at (6.8,3.1) {  \textcolor{blue}{\begingroup \fontsize{8pt}{8pt}\selectfont 39 \endgroup} };

\node[fill=white, rectangle] at (-6.5,1.6) {  \textcolor{black}{\begingroup \fontsize{8pt}{8pt}\selectfont Weight(in Kg) \endgroup} };
\node[fill=white, rectangle] at (-5.4,1.6) {  \textcolor{black}{\begingroup \fontsize{8pt}{8pt}\selectfont 01 \endgroup} };
\node[fill=white, rectangle] at (5.8,1.6) {  \textcolor{black}{\begingroup \fontsize{9pt}{9pt}\selectfont 150 \endgroup} };

\node[fill=white, rectangle] at (6.8,1.7) {  \textcolor{blue}{\begingroup \fontsize{8pt}{8pt}\selectfont 87 \endgroup} };

\node[fill=white, rectangle] at (-6.5,.2) {  \textcolor{black}{\begingroup \fontsize{8pt}{8pt}\selectfont Height(in cm) \endgroup} };
\node[fill=white, rectangle] at (-5.4,.2) {  \textcolor{black}{\begingroup \fontsize{8pt}{8pt}\selectfont 01 \endgroup} };
\node[fill=white, rectangle] at (5.8,.2) {  \textcolor{black}{\begingroup \fontsize{9pt}{9pt}\selectfont 224 \endgroup} };
\node[fill=white, rectangle] at (6.8,1.6) {  \textcolor{black}{\begingroup \fontsize{8pt}{8pt}\selectfont 129 \endgroup} };

\node[fill={rgb,255:red,206; green,227; blue,254}, rectangle] at (-4.7,-2.7) {  \textcolor{black}{\begingroup \fontsize{7pt}{7pt}\selectfont Other \endgroup} };

\node[fill={rgb,255:red,206; green,227; blue,254}, rectangle] at (-4.5,-2.3) {  \textcolor{black}{\begingroup \fontsize{7pt}{7pt}\selectfont Female \endgroup} };

\node[fill={rgb,255:red,206; green,227; blue,254}, rectangle] at (-4.7,-1.9) {  \textcolor{black}{\begingroup \fontsize{7pt}{6pt}\selectfont Male \endgroup} };

\node[fill={rgb,255:red,206; green,227; blue,254}, rectangle] at (1.2,-2.3) {  \textcolor{black}{\begingroup \fontsize{7pt}{6pt}\selectfont Moderately Active \endgroup} };

\node[fill={rgb,255:red,206; green,227; blue,254}, rectangle] at (.9,-2.69) {  \textcolor{black}{\begingroup \fontsize{7pt}{6pt}\selectfont Very Active \endgroup}
};

\node[fill={rgb,255:red,206; green,227; blue,254}, rectangle] at (1.1,-3.05) {  \textcolor{black}{\begingroup \fontsize{7pt}{6pt}\selectfont Intensely Active \endgroup}};

\node[fill={rgb,255:red,0; green,103; blue,163}, rectangle] at (5,-3.9) {  \textcolor{white}{\begingroup \fontsize{9pt}{9pt}\selectfont Next \endgroup}};

rgb(0, 103, 163)
\node[fill={rgb,255:red,206; green,227; blue,254}, rectangle] at (.8,-1.95) {  \textcolor{black}{\begingroup \fontsize{7pt}{6pt}\selectfont Sedentary \endgroup} };

\node[fill=white, rectangle] at (-5,-1.2) {  \textcolor{black}{\begingroup \fontsize{9pt}{9pt}\selectfont Gender \endgroup}};

\node[fill=white, rectangle] at (3,-1.2) {  \textcolor{black}{\begingroup \fontsize{9pt}{9pt}\selectfont How would you describe your daily activity level? \endgroup}};user health data acquisition 

\end{tikzpicture}}
    \caption{User interface for the health profiling. \label{fig:thehealthprofileImage} 
    }
\end{figure}

As the proposed smart nudges are personalized to users' health variables (e.g. age and obesity risk class), the evaluation requires capturing information regarding these variables. We implemented an interactive \textit{input pane} to capture users' health information. The input pane supports both the type-in and slider input options, as shown in figure \ref{fig:thehealthprofileImage}.  Users are requested to provide their age, weight, height, and gender. They also had to identify \textit{how active they are in their everyday life.} Table \ref{tab:physicalActivityFactor} summarizes a brief description of the four levels of being physically active: \begin{inparaitem}  \item[\textbullet] Sedentary  \item[\textbullet]  Moderately Active  \item[\textbullet]  Vary Active and \item[\textbullet] Intensely active\end{inparaitem}. After capturing users' health information CookIT generates users' health-profile including BMR, BMI, DRCI, and Risk class, following section \ref{sec:healthprofileCal}.

To capture user's taste-preference, we adopted our food feature-based intelligent user-modelling, proposed in \cite{khan2021addressing}. The approach display's users a list of food features in the order of their significance-score and users are asked to select features that the likes and dislikes. Following this preference reading, users taste-profile, user-to-feature and user to-topic matrices, are determined. After capturing a user's consent, health-info and feature preference, the system generates a \textbf{user-id} and a \textbf{participant-number} for every participant. To implement de-identification,  all preference and response data are stored under this \(\{user\mbox{-}id, participant\mbox{-}number\}\) identification pair. The system informs participants of their participant-numbers through email. To log into the system and take part in the experiment, the participant needs to use their participant number and password.

\begin{figure}[!h]
\centering
\scalebox{1}{
\begin{tikzpicture}
\node[inner sep=0pt , rectangle, draw] (russell) at (-5,0)
    {  \includegraphics[width=.5\textwidth,height=.4\textwidth]{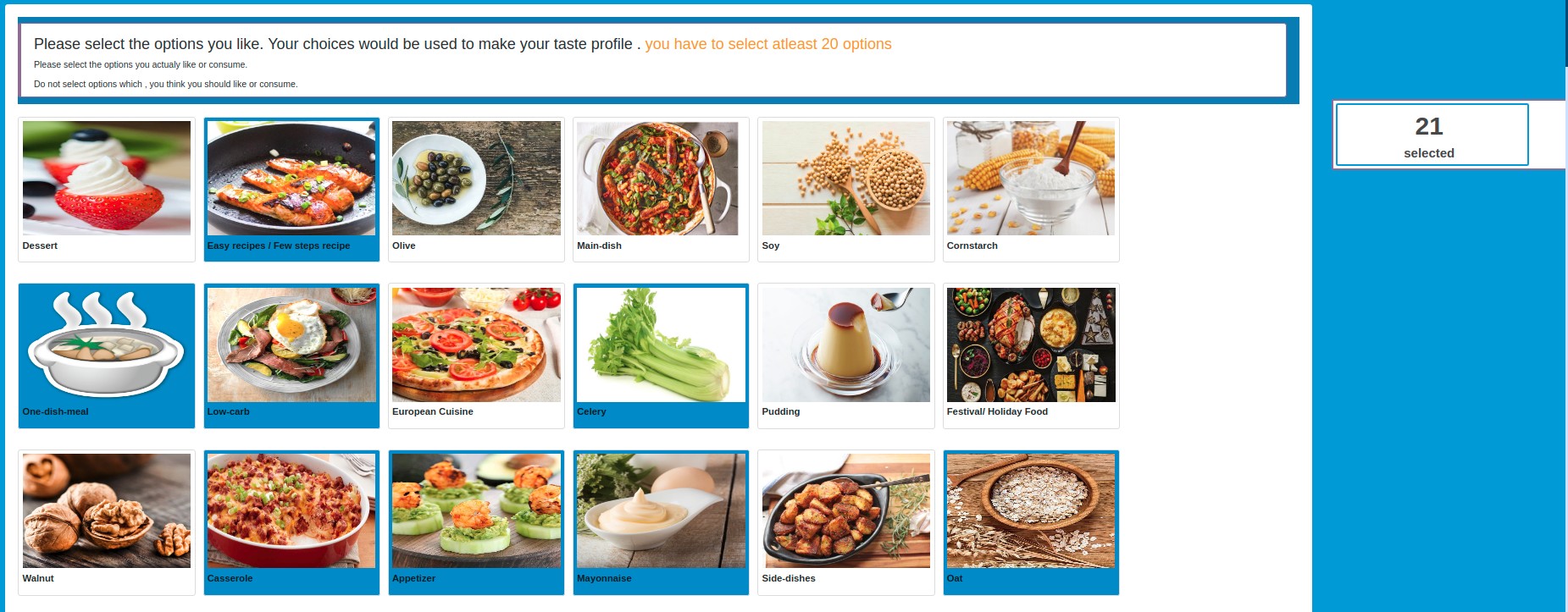}};
    
\node[inner sep=0pt, rectangle, draw] (whitehead) at (3.5
,0)
{\includegraphics[width=.5\textwidth,height=.4\textwidth]{usrLike.jpg}};

   \node[fill=white, rectangle] at (-1.5,1.7) {  \textcolor{black}{\begingroup \fontsize{7pt}{6pt}\selectfont selected \endgroup} };
    
     \node[fill=white, rectangle] at (-1.5,2) {  \textcolor{black}{\begingroup \fontsize{7pt}{7pt}\selectfont 21 \endgroup} };
    
    \node[inner sep=0] (image) at (2.7,-.5) {\includegraphics[width=0.39\textwidth, height=0.32\textwidth]{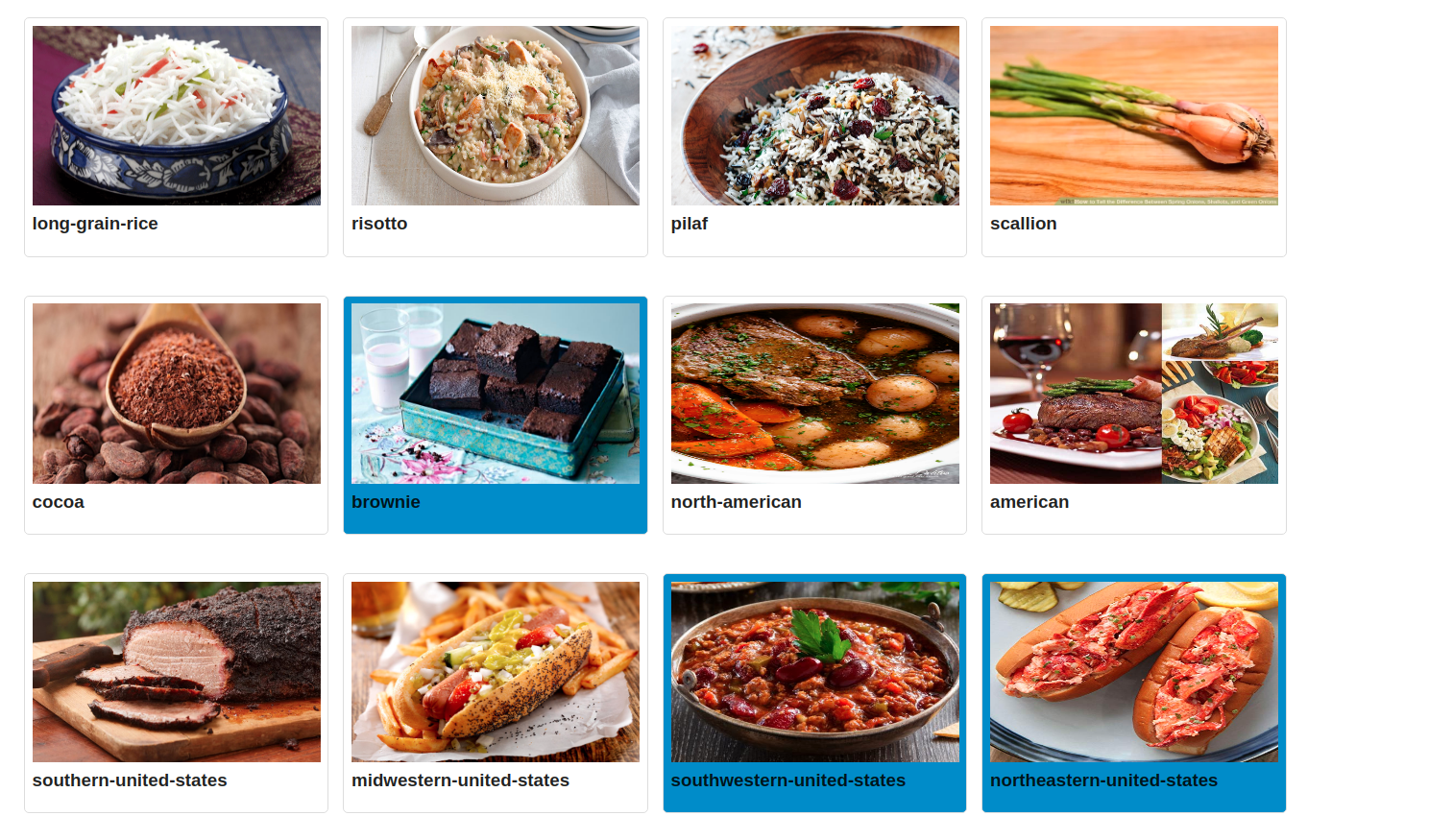}};

    \node[fill=white, rectangle] at (-6,2.8) {  \textcolor{black}{\begingroup \fontsize{7pt}{7pt}\selectfont Please browse all features and select the options you like. \endgroup} };
    
    \node[fill=white, rectangle] at (-7.6,2.5) { \textcolor{blue}{\begingroup \fontsize{7pt}{7pt}\selectfont Select at least 20 options \endgroup}};
    
    \node[fill=white, rectangle] at (6.9,1.7) {  \textcolor{black}{\begingroup \fontsize{7pt}{6pt}\selectfont selected \endgroup} };
    
     \node[fill=white, rectangle] at (7,2) {  \textcolor{black}{\begingroup \fontsize{7pt}{7pt}\selectfont 26 \endgroup} };
    
      \node[fill=white, rectangle] at (2.8,2.8) {  \textcolor{black}{\begingroup \fontsize{7pt}{7pt}\selectfont Please browse all features and select the options you don't like. \endgroup} };
      
      \node[fill=white, rectangle] at (.9,2.5) { \textcolor{blue}{\begingroup \fontsize{7pt}{7pt}\selectfont Select at least 20 options \endgroup}};

         \node[] at (-6,-3.7) {(a) Positive preference-acquisition};
         \node[] at (2,-3.7) {(b) Negative preference-acquisition};

\end{tikzpicture} 
}
\caption{User interface for user preference acquisition on the food-features. \label{fig:usermodelingTasteFeatureIndChapter} }
\end{figure}

\begin{figure}[!h]
     \centering
     \scalebox{1}{
     \begin{tikzpicture}
        \node[inner sep=0pt , rectangle] (russell) at (-10,0) { 
     \includegraphics[width=.5\textwidth,height=80pt]{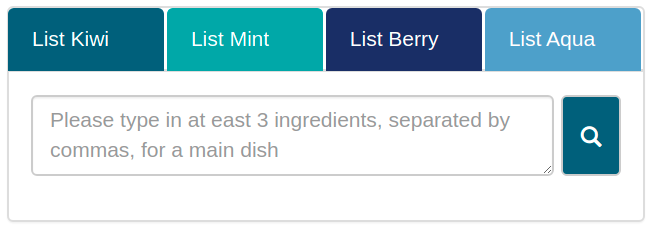}};
     
 \node[inner sep=0pt , rectangle] (russell) at (-2,0) { 
     \includegraphics[width=.48\textwidth,height=80pt]{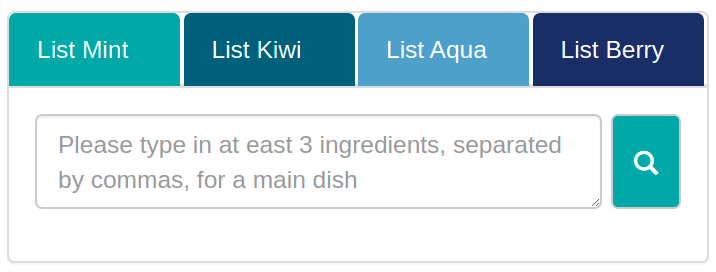}};
     
     \node[fill=white, rectangle] at (-10,-1.8) { (a) };
     \node[fill=white, rectangle] at (-4,-1) { (b) };
     \end{tikzpicture}}
     \caption{ Display sequence of recommendation scenarios for two different users.}
     \label{fig:tabsequence}
 \end{figure}

\subsubsection{The Recommendation Scenarios}

 We designed a \textit{Pill-based navigation control} \cite{BootStrapWebpageComponents} to accommodate four distinct recommendation scenarios. In the context of experimental designs, the most common nuisance factors to be counterbalanced are procedural variables (i.e., temporal or spatial position) that can create order and sequence effects. The complete counter balanced design ensure equal treatment to every hypothesis, in our cases recommendation scenarios, involved.  We designed Pills-controlled \textit{complete counter-balanced} user study.
Users are presented with a series of Pills, each associated with an individual recommendation scenario.  Each recommendation scenario is assigned a pseudo name to mitigate familiarity bias, such as \begin{inparaitem} \item[\textbullet] List Aqua,  \item[\textbullet]List  Mint,  \item[\textbullet]List Kiwi, and \item[\textbullet] List Berry\end{inparaitem}.   Clicking on a Pill loads the corresponding recommender and relevant user interface (recipe view, obvious widget and badge). For every \(n^{th}\) participant, CookIT presents the Pills in a unique sequence. In this study, we are evaluating four recommenders; hence for every the \(n^{th}\) participant CookIT organizes the four pills in one of the \(\mathbf{4!}\) possible combinations. During the sign-up process, each participant is assigned a \textit{display sequence} for the recommendation scenarios. Every time a participant logs into the system they experience the recommendation scenarios in the exact order as their \textit{display sequence}. Figure \ref{fig:tabsequence} illustrates the Pills-controlled \textit{display sequences} for two different participants. 

Users are instructed to navigate the pills in a \textit{left-to-right} manner. Clicking on each Pill allows users to access recommendations corresponding to the nudging strategy. When visiting a Pill (e.g., List Berry), participants are guided to conduct a recipe search,  as shown in figure \ref{fig:tabsequence}. In response to the user's query, CookIT generates a seven-item Reclist using the \(F_\_T\_r\) recommender.  Across all four recommendation scenarios, recipes are displayed in the order of their predicted preference score. However, based on which nudging technology is attached to the Pill, the corresponding badge is added to the RecList. For the no-nudge scenario, no badge is added.

      \begin{figure}[!h]
     \centering
     \scalebox{.95}{
      \begin{tikzpicture}
        \node[inner sep=0pt , rectangle, draw, line width=.02cm] (russell) at (-4,12) { 
     \includegraphics[width=1\textwidth,height=350pt]{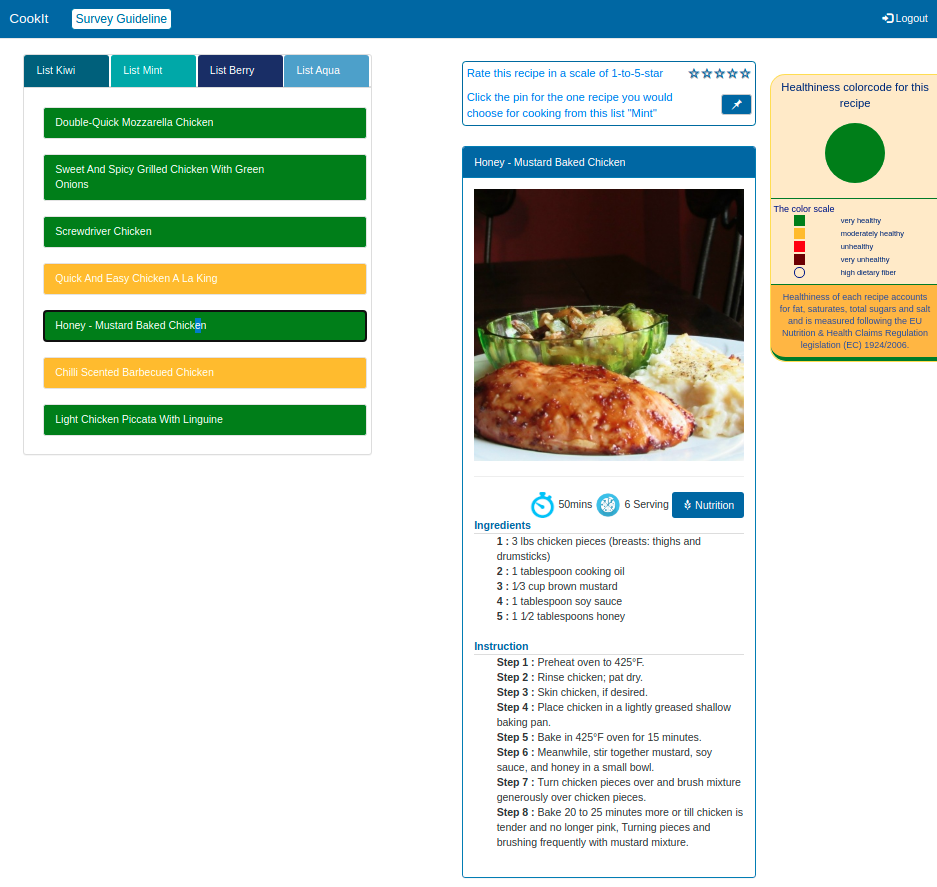}};
     \end{tikzpicture}
     }
     \caption{Visualizing customized nudging content on each recipe.}
     \label{fig:nudgingvisualization} 
 \end{figure}

Upon clicking each recipe, CookIT displays the recipe content, as shown in figure \ref{fig:nudgingvisualization}. For each recommendation scenario designated to a smart-nudge approach, along with the recipe contents (e.g., instructions and image) corresponding persuasive visual contents were also displayed.  For the no-nudge scenario only the recipe contents were displayed. Users were allowed to click and browse recipes in any order they wanted for as many times as they liked. Figure \ref{fig:nudgingvisualization} illustrate the recommendation scenario corresponding to the smart-nudge FSA-ColorCoding.


   \begin{figure}[!h]
     \centering
          \begin{tikzpicture}

      \node[inner sep=0pt , rectangle] (russell) at (-5.4,6) { 
     \includegraphics[width=.5\textwidth,height=!]{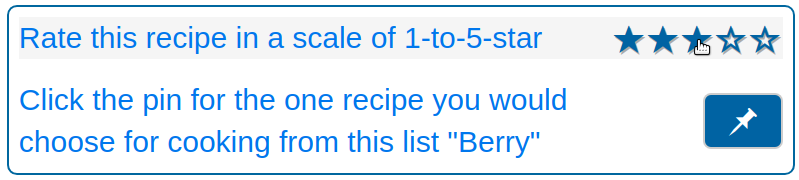}};
      
     \end{tikzpicture}
     \caption{ User interface for collecting users response on a recommended recipe.}
     \label{fig:rateArecipe}
 \end{figure}

 Within each list, participants were required to rate each individual recipe based on how much they liked the recipe, as shown in figure \ref{fig:rateArecipe}. The rating was taken on a \(5 \ star \ rating \ scale\), where \(\mathbf{0}\) and \(\mathbf{5}\) represented \( \small {did \ not\ like \ at \ all}\) and \( \small {liked \ very \ much}\), respectively. Participants are also required to pin one recipe for each list. Here, the activity \textbf{pin} is considered as intention to consume. To complete the section \textit{recommendation scenarios} and move on to the section  \textit{feedback questionnaire}, participants must rate all 28 recipes, seven in each recommendation scenario, and pin four recipes, one for each of the recommendation scenarios.  To support diverse evaluation strategies, CookIt watch and log user activities, such as click, visit, browse and pin. In cases where the participant missed rating or pinning recipes in one or more scenarios, CookIT notifies them \textit{what} and \textit{in which list} they missed.

 \begin{figure}[!h]
 \scalebox{.9}{
     \centering \begin{tikzpicture}
        \node[inner sep=0pt , rectangle] (russell) at (-4,0) { 
     \includegraphics[width=1\textwidth,height=!]{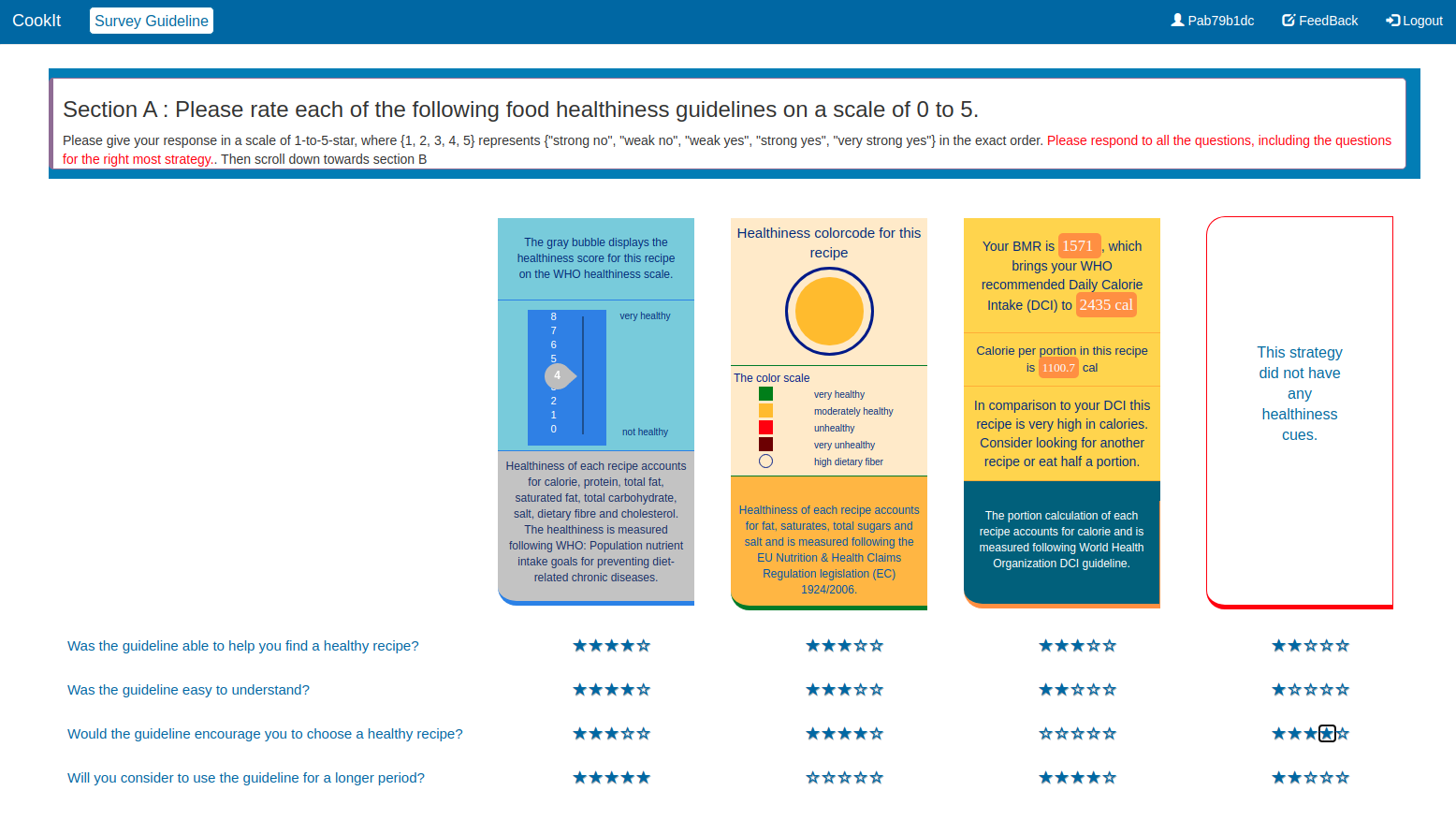}};
     
      \node() at (-4,-3.8)  [fill=white,minimum width=15cm,minimum height=3cm] {};
       
        \node[inner sep=0pt , rectangle] (russell) at (-1.4,0) { 
     \includegraphics[width=305pt,height=!]{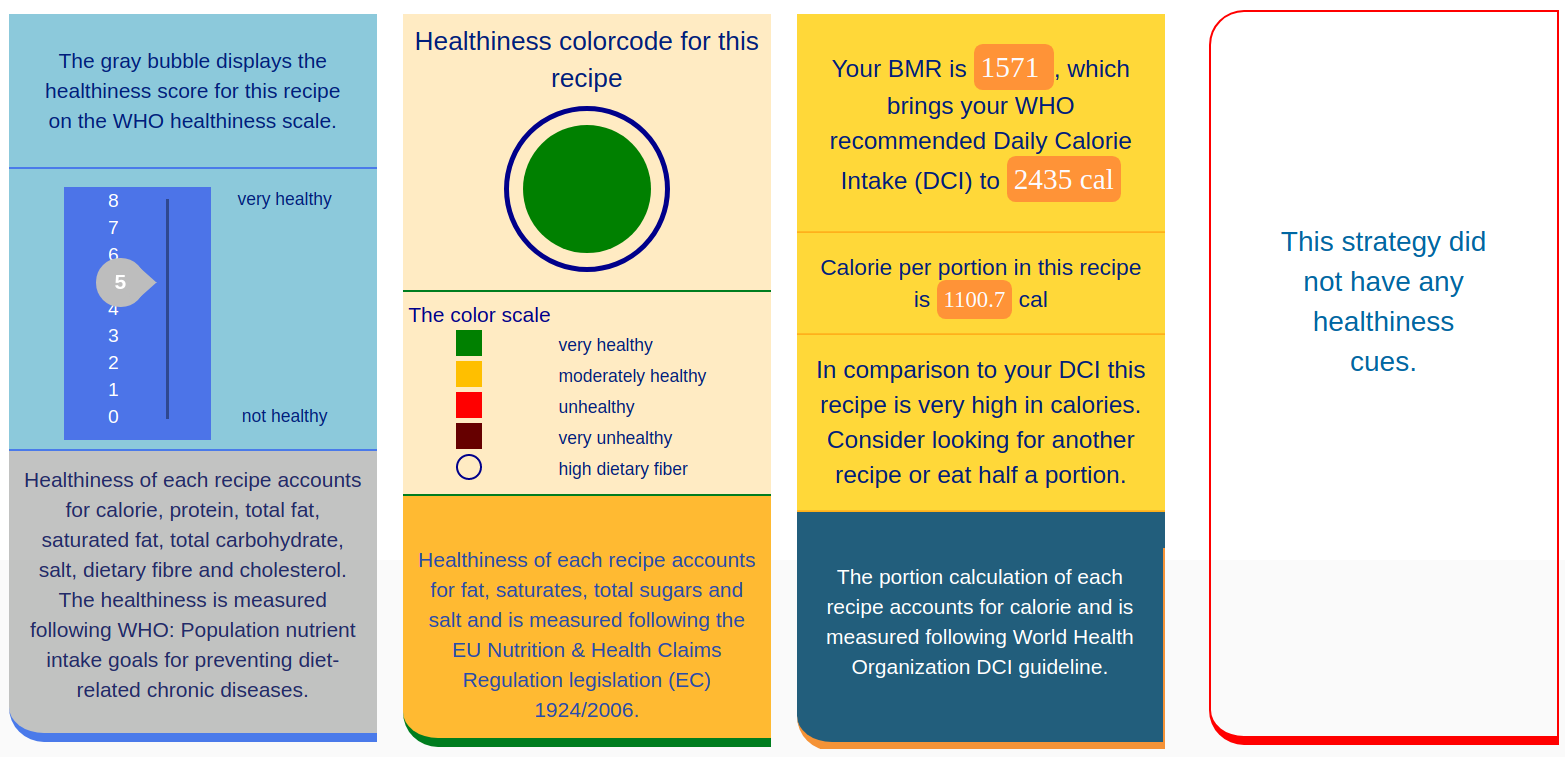}};

         \node[inner sep=0pt , rectangle] (russell) at (-1.2,-4.5) { 
     \includegraphics[width=300pt,height=110pt]{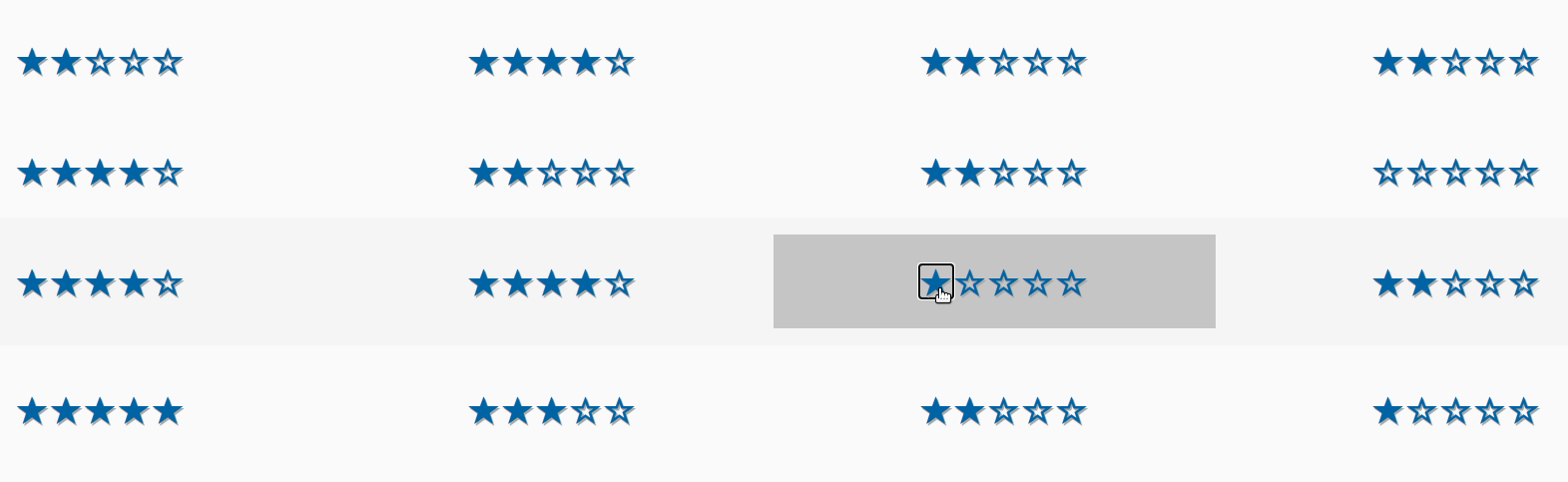}};
      \node[fill=white, rectangle] at (-5.4,3.55) { {\begingroup \fontsize{10pt}{9pt}\selectfont Please rate each of the following food-healthiness guidelines on a scale of 0-to-5 stars. \endgroup} };
         \node[fill=white, rectangle] at (-9.5,3) { {\begingroup \fontsize{7pt}{8pt}\selectfont  {\color{brick} corresponding to the right-most strategy} \endgroup} };
      \node[fill=white, rectangle, minimum width=.91\textwidth] at (-4,3.25) { {\begingroup \fontsize{7pt}{6pt}\selectfont The number of stars \{1,2,3,4,5\} represent \{strong no, no, maybe, yes, strong yes\}, respectively. {\color{brick} Please respond to every field, including the fields} \endgroup} };

      \node[fill=white, rectangle] at (-9.8,-2.8) {  \textcolor{belblue}{\begingroup \fontsize{9pt}{8pt}\selectfont Was the guideline able to help you find\endgroup} };
         \node[fill=white, rectangle] at (-11.1,-3.3) {  \textcolor{belblue}{\begingroup \fontsize{9pt}{8pt}\selectfont a healthy recipe? \endgroup} };
          \node[fill=white, rectangle] at (-9.8,-3.8) {  \textcolor{belblue}{\begingroup \fontsize{9pt}{8pt}\selectfont Was the guideline easy to understand? \endgroup} };
         \node[fill=white, rectangle] at (-9.8,-4.4) {  \textcolor{belblue}{\begingroup \fontsize{9pt}{8pt}\selectfont Would the guideline encourage you to\endgroup} };
         
         \node[fill=white, rectangle] at (-10.5,-4.8) {  \textcolor{belblue}{\begingroup \fontsize{9pt}{8pt}\selectfont choose a healthy recipe? \endgroup} };
         
              \node[fill=white, rectangle] at (-9.8,-5.3) {  \textcolor{belblue}{\begingroup \fontsize{9pt}{8pt}\selectfont Will you consider to use the guideline \endgroup} };
         
         \node[fill=white, rectangle] at (-10.85,-5.7) {  \textcolor{belblue}{\begingroup \fontsize{9pt}{8pt}\selectfont for a longer period? \endgroup} };

     \node[inner sep=0pt , rectangle] (russell) at (-9.75,4.6) { 
     \includegraphics[width=.3\textwidth,height=20pt]{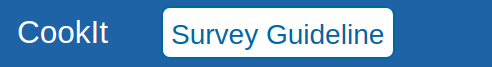}};
                      \node[inner sep=0pt , rectangle] (russell) at (-1.55,4.6) { 
     \includegraphics[width=.7\textwidth,height=20pt]{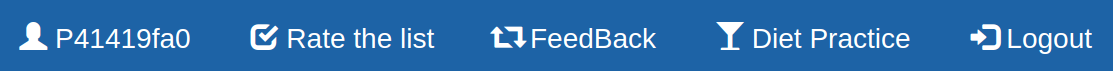}};
     
     \end{tikzpicture}}
     \caption{Questionnaire on multiple performance variables of recommender models. }
     \label{fig:performanceTabs1}
 \end{figure}
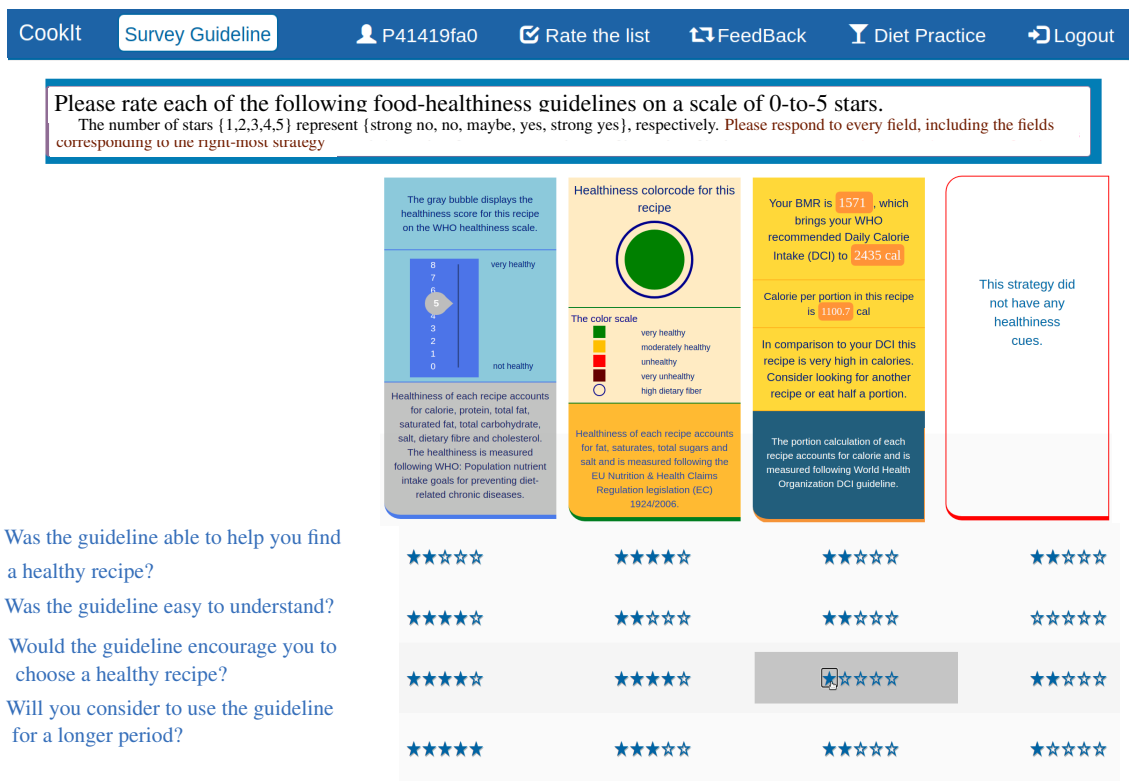

 \subsubsection{Feedback Questionnaire on Utility Gained in each Nudging Scenario: \label{sec:feedback}}

In this section, participants are required to complete a questionnaire on different performance criterion for all four recommendation scenarios. User are requested to rate each recommendation scenario on a \(5 \ star \ rating \ scale\), where \(\{\mathbf{1,2,3,4,5\}}\) represented \(\small\{very\ poor\), \(\small{poor}\), \(\small{moderate}\), \(\small{good }\), \(\small{very\ good}\}\), respectively.  We implemented categorical rating system to allow user assign negative rating expressing their level of distaste along with their level of satisfaction. The three nudging widgets and one empty-widget ( representing the no-nudging scenario) are displayed side by side to help participants compare while responding, as shown in figure \ref{fig:performanceTabs1}. If required, the participants can revisit the recommendation scenarios and browse the corresponding recipes for further comparison. Participants' search results for each scenario, along with their provided ratings, are stored as sessions. When the participant revisits a recommendation scenario,  CookIT loads the  corresponding session (RecList, rates and pins) from the database. During revisit, the user can not change the ratings or update the pinned recipe. Participants are required to respond to all 16 fields, four for each of the recommendation scenarios.

The four criteria on which participants evaluate each recommendation scenario are:
    \begin{enumerate}\setlength\itemsep{.05em}
    \small
        \item \textbf{Helpfulness or effectiveness}: Is the guideline able to help you find a healthy recipe?
        \item \textbf{Ease of understandability}: Was the guideline easy to understand? 
        \item \textbf{Persuasiveness}: Would the guideline encourage you to choose a healthy recipe?
        \item \textbf{Suitability for long-term use}: Will you consider to use the guideline for a longer period?
      \end{enumerate}

\subsection{ The User Group}

The web-system, CookIT, was hosted on the internet under the ucd.ie \cite{ucd} domain. We conducted a completely remote and online study. Participants were recruited via social media groups within the UCD \cite{ucd}, such as institution email, WhatsUp, and Slack groups. Participants were also recruited from online platforms, such as Facebook\cite{website:facebook}, Twitter \cite{website:twitter}, and  LinkedIn \cite{website:linkedin}. There were three requirements for being eligible to participate in the survey: 
\begin{itemize}\setlength\itemsep{.05em}
\small
  \fontfamily{psv}\selectfont

    \item Participants need to be 18 years old or older.
    \item Participants need to be actively involved in preparing their food (e.g., cooks twice a week or more)
    \item Participants need to be willing to give consent to the terms and conditions of the survey. 
    
\end{itemize}

In total 137 participants signed up on the website and took part in the survey; from which only 91 users provided responses to every required field. However, to ensure a total counterbalanced user study and equal exposure to each recommendation scenario, we only accepted 72 out of 91 valid participants. Every \(72/24 = 3\) participants were shown the four recommendation scenarios in a unique order. The final user group is spread over an age-range of \textbf{24-to-71} and includes students and professionals. 56\% of our participants identified themselves as female and 44\% as male. The participation was completely voluntary, and there was no remuneration for the participation.

\subsection{The Ethics Approval}
This study has been reviewed by and received ethics permission from the Human Research Ethics Committee, UCD \cite{website:UCD-office-of-research-ethics}. The ethics permission details are accessible with the  Research Ethics Exemption Reference Number: LS-E-20-41-Khan-Coyle.

\section{Results}

\subsection{ Are the Health Nudges Effective at All? }

The health aware smart-nudge technologies proposed in this paper aim to encourage users to prefer healthier options over others. We adopted Rank based Performance  Metrics (RPM) \cite{RecommendingEvaluatingChoices,Evaluatingrecommendationsystemsshani2011evaluating} to inspect the performance of these smart-nudges in \textit{leading users to choose healthier options when available}. RPM has been a popular performance metric for evaluating the distance between the predictions and users' true preferences \cite{RecommendingEvaluatingChoices,Evaluatingrecommendationsystemsshani2011evaluating,NPDMExamplemeyer2012recommender,NPDMExamplemusto2010enhanced}. RPM enable evaluating systems that impose encouragement on some items over others within the RecList.  We adopted the following two RPM to test the effectiveness of the proposed nudging technologies.

 \begin{itemize}\setlength\itemsep{.05em}
     \item  \textbf{RPM 1} The Rank Correlation.
     \item  \textbf{RPM 2} The Normalized Distance-based Performance Measure.
     
 \end{itemize}

As our proposed nudging aims to promote healthy options over relatively less healthy ones, we determined the system assigned ranks based on recipe healthiness. For user-assigned rank we considered participant's explicit rating on each recipe. Before investigating the competency of the nudging strategies based on these user-ratings, we look into the data to examine \textit{can the data support drawing reliable conclusions}\cite{EvaluationRS}.

\textbf{Shapiro Wilk's normality test} \cite{ShapiroNormalityTest} on the user-rating data, collected in each of the four recommendation scenarios, conferred the data to be \textit{platykurticly distributed} with \textit{negative kurtosis}. A pval of  \(\mathbf{\small( p \leq 1.88907e^{-11})}\) produced in each case indicates the low possibility of \textit{Type1 error} (e.g., rejecting a correct \(H_o\): the data is normally distributed). Considering this outcome, we applied a \textbf{non-parametric Friedman's test for repeated-measures\footnote{Friedman's test for repeated-measures is designed to compare the rank-ordering  \cite{EvaluatingCF,Howgoodyourrecommendersystemisilveira2019good} of data from three or more experiments where the same set of individuals (or matched subjects) participates in each experiment, which is the case in our user study.}} \cite{FriedmanTest} to test for differences amongst the groups. The result \( \mathbf{ \small ({X^2}_F(3) = 62.0628, \; pval<0.00001)}\) indicates significant differences among data, from the four scenarios, at a confidence level of \(p=0.001\). Follow-up two-tailed \textit{Mann Whitney Wilcoxon tests } \cite{MannWhitney}, a non-parametric multiple-comparisons test, were conducted between each pair among the four scenarios. In each case, the test has the same null hypothesis: \textit{no statistically significant difference between the user rates from the two different scenarios}. The results demonstrated statistically significant differences in participants' recipe ratings between each of the personalized recommendation scenarios with smart-nudging strategy and the baseline personalized recommendation scenario without any nudging, at a confidence level of \((p=0.001)\) in each case. We also observed a significant difference between the ratings in the WHO-Bubbleslider nudging scenario and both DRCI-MLCP and FSA-ColorCoding nudging scenarios, with a critical Z value \cite{criticalZscore} of \(\mathbf{\small(Z_c = 6.84834, \; pval < 0.00001)}\) and \(\mathbf{\small(Z_c = 7.33956, \; pval< 0.00001)}\), respectively, at a confidence level of \((p=  0.001)\). However, there was no significant difference between users' recipe ratings in  DRCI-MLCP and FSA-ColorCoding nudging scenarios, at a confidence level of \((p= 0.001)\). The result demonstrated statistical significance in the majority of the cases, hence considered capable of supporting reliable conclusions.

 \subsubsection{RPM 1: The Rank Correlation}

 A \textbf{rank correlation} \cite{CorrCofmukaka2012guide} measures the degree of similarity between two ranking approaches and can be used to assess the alignment of the relation between them. The rank correlation coefficient informs \textit{how positive or negatively two models are aligned}. Here we evaluated the performance of each recommendation scenario based on the correlation between the system-predicted rank for a recipe \(r_t\), \(P(R,r_t)\), and the user-assigned rank for \(r_t\), \(U(R,r_t)\). Here we adopted Pearson's Product-Moment Correlation Coefficient (PPMCC) \cite{CorrCofmukaka2012guide}. A perfect score of \(\mathbf{+1}\) represents a perfect linear relationship; that is, the system predicted every rank correctly for each user. A score of \(\mathbf{-1}\) represents, the system predicted every rank wrong and in the opposite direction. The score of 0 for a system is equivalent to random retrieval.

 The PPMCC measures a possible two-way linear association between two variables. In this case the variables are \begin{inparaitem} \item[\textbullet] how healthy is a recipe \(r_t\) \item[\textbullet] how much the user like \(r_t\) \end{inparaitem}. It is important to note that, here the system assigned rank is not the orders recipes are presented to the users, instead it is the order the system is trying to make the user notice and like the recipes. All four recommendation scenarios operate on the same recommendation algorithm and recipes in the RecList are displayed in the order of their prediction scores. This display decision was adopted to avoid \textit{positional bias} and give the baseline recommendation scenario (no-nudging) an equal chance to compete. The rank correlation investigates \textit{the effectiveness of the smart-nudges in implicitly pursuing users to like healthier recipes.}   


To test whether the smart-nudges are overfitted to any specific healthiness guideline, we investigated PPMCC evaluation with system-assigned ranks determined based on both the WHO-HealthScore and FSA-HealthScore.

\subsection*{Rank Correlation between User-assigned Ranks with WHO-HealthScore based System-assigned Ranks: \label{sec:whoHealthScalebasedSystemranking}}
    
    \begin{itemize}\setlength\itemsep{.05em}
        \item \textbf{System-assigned Rank} 
        
        The higher the WHO-HealthScore of a recipe, the higher the rank assigned by the system. The WHO-HealthScale ranges from 0-to-8, while the user rate is read on a 1-to-5 scale. Hence, we normalized the who-HealthScore of the seven recipes, in the RecList, into a 1-to-5 range. Recipes are sorted in descending order of their normalized WHO-HealthScore. The position of \(r_t\), in this new descending sorted RecList, is the system predicted rank, \(P(R,r_t)\), for  \(r_t\).
        
        \item \textbf{User-assigned Rank}
        
        The seven recipes, in the RecList corresponding to each recommendation scenarios, are sorted in the descending order of the user-rates on recipes. The position of \(r_t\) in the sorted list is the user-assigned rank \(U(R,r_t)\) for  \(r_t\). To preserve a user's ratings from being affected by the ranking process, we assigned the same rank for recipes, on which the user cast the same rate. We adopted a ceiling approach for deciding on the rank of the recipes with the same user-rating. 
   
    \end{itemize}
    
  \begin{figure}[!h]
        \centering
\scalebox{.7}{
\begin{tikzpicture}
\node[inner sep=0pt , rectangle] (russell) at (-3.5,0)
    { \includegraphics[width=1\textwidth,height=!]{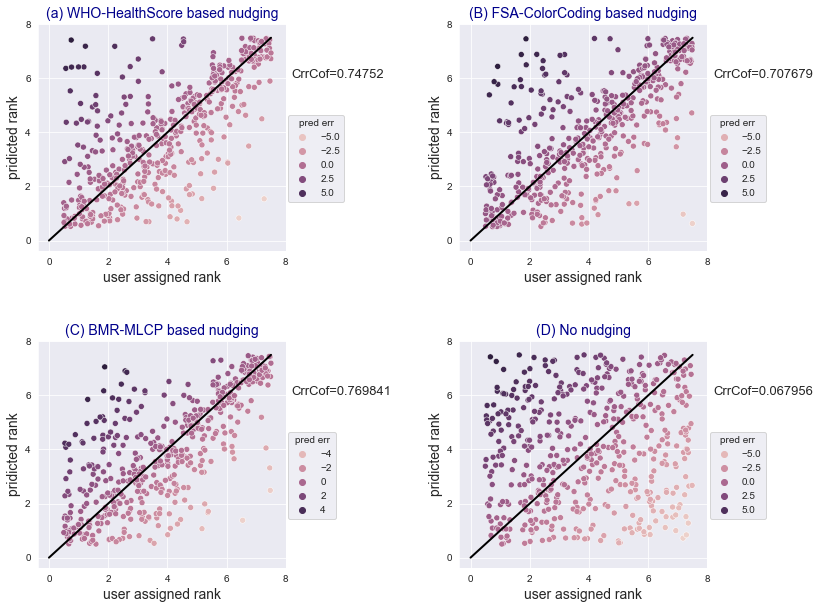}};

     \node[inner sep=0pt , rectangle,fill=white, minimum height=15pt,minimum width=5cm] at (-8.4,5.9) {\fontsize{9pt}{9pt}\selectfont \color[HTML]{090088} (a) WHO-BubbleSlider based nudging};
     
      \node[inner sep=0pt , rectangle,fill=white, minimum height=20pt,minimum width=2cm] at (-4.7,4.5) {\fontsize{9pt}{9pt}\selectfont \color[HTML]{000000} PPMCC=0.747520};
     
     \node[inner sep=0pt , rectangle,fill=white, minimum height=15pt,minimum width=5cm] at (0,5.8) {\fontsize{9pt}{9pt}\selectfont \color[HTML]{090088} (b) FSA-ColorCoding based nudging};
     
     \node[inner sep=0pt , rectangle,fill=white, minimum height=20pt,minimum width=2cm] at (3.6,4.5) {\fontsize{9pt}{9pt}\selectfont \color[HTML]{000000} PPMCC=0.707679};

         \node[inner sep=0pt , rectangle,fill=white, minimum height=15pt,minimum width=5cm] at (-8.05,-.5) {\fontsize{9pt}{9pt}\selectfont \color[HTML]{090088} (c) DRCI-MLCP based nudging};

             \node[inner sep=0pt , rectangle,fill=white, minimum height=20pt,minimum width=2cm] at (-4.7,-1.5) {\fontsize{9pt}{9pt}\selectfont \color[HTML]{000000} PPMCC=0.769841};
     
     \node[inner sep=0pt , rectangle,fill=white, minimum height=15pt,minimum width=5cm] at (0,-.5) {\fontsize{9pt}{9pt}\selectfont \color[HTML]{090088} (d) no nudging};
     
      \node[inner sep=0pt , rectangle,fill=white, minimum height=20pt,minimum width=2cm] at (3.6,-1.5) {\fontsize{9pt}{9pt}\selectfont \color[HTML]{000000} PPMCC=0.067956};

\end{tikzpicture}}

\caption{Correlation between the  WHO-HealthScore based system predicted ranks and the user-assigned ranks}
    \label{fig:rankpredErrWHOSCORING1}
\end{figure}
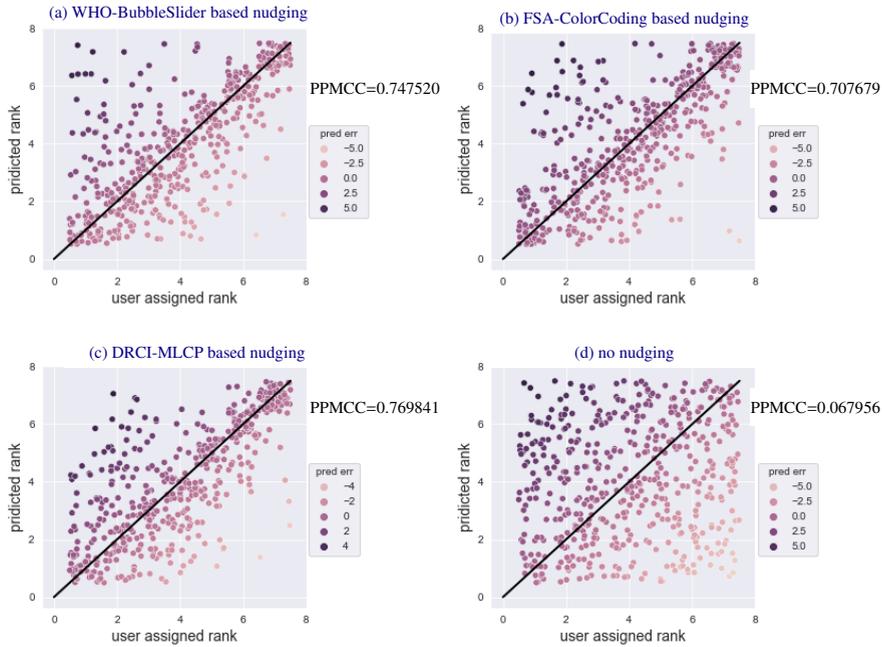
 
    Figure \ref{fig:rankpredErrWHOSCORING1} illustrates the correlation between the WHO-HealthScore based system-assigned ranking and the user-assigned ranking. The high positive PPMCC in recommendation scenarios, under the impression of smart-nudges, such as DRCI-MLCP nudge(+0.769841), WHO-BubbleSlider nudge (+0.74752), and FSA-ColorCoding nudge (+0.707679),  suggest that all three nudging scenarios produced a significant positive effect on users' preference towards healthier recipes. That is, under the influence of smart-nudges recipes with high WHO-HealthScore achieved higher preference from users. The DRCI-MLCP nudging strategy, with the highest positive PPMCC of +0.769841, is proven to be the most effective in motivating users to choose healthier recipes. In contrast, the no-nudging scenario, with the lowest PPMCC of +0.067956, demonstrates no positive or negative trend between the \textit{recipe healthiness} and \textit{user preference}.

As shown in figure \ref{fig:rankpredErrWHOSCORING1}(b), the high density of positive miss predictions in the recommendation scenario with the smart-nudge FSA-ColorCoding, suggests that the nudging strategy is more successful in de-incentfying less-healthy recipes than incentfying  healthy recipes. For the DRCI-MLCP  and WHO-BubbleSlider nudging, no such trends were observed in the missed predictions.

\subparagraph*{ Rank Correlation between User-assigned Ranks with FSA-HealthScore based System-assigned Ranks} \leavevmode

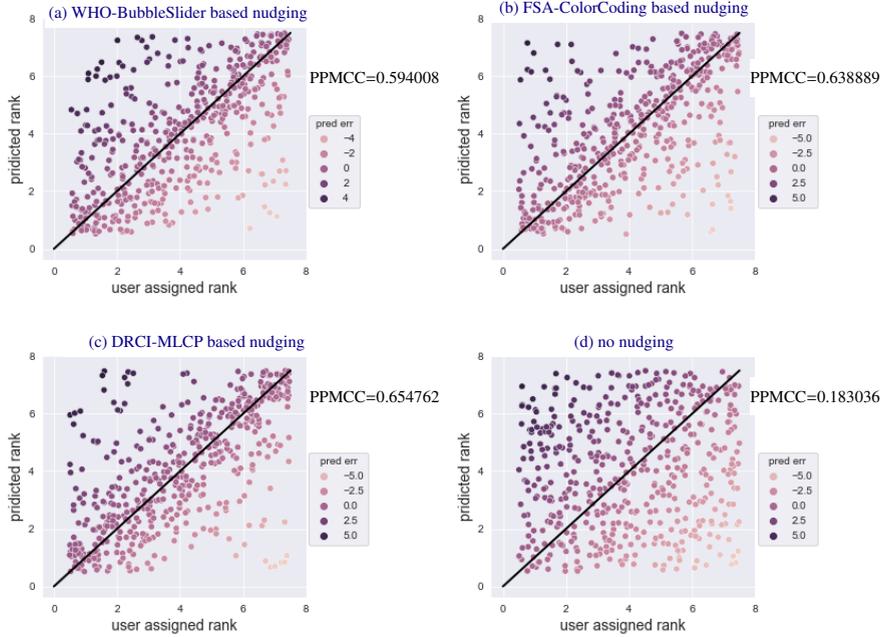
\begin{figure}[!h]
        \centering
\scalebox{.7}{
\begin{tikzpicture}
\node[inner sep=0pt , rectangle] (russell) at (-3.5,0)
    { \includegraphics[width=1\textwidth,height=!]{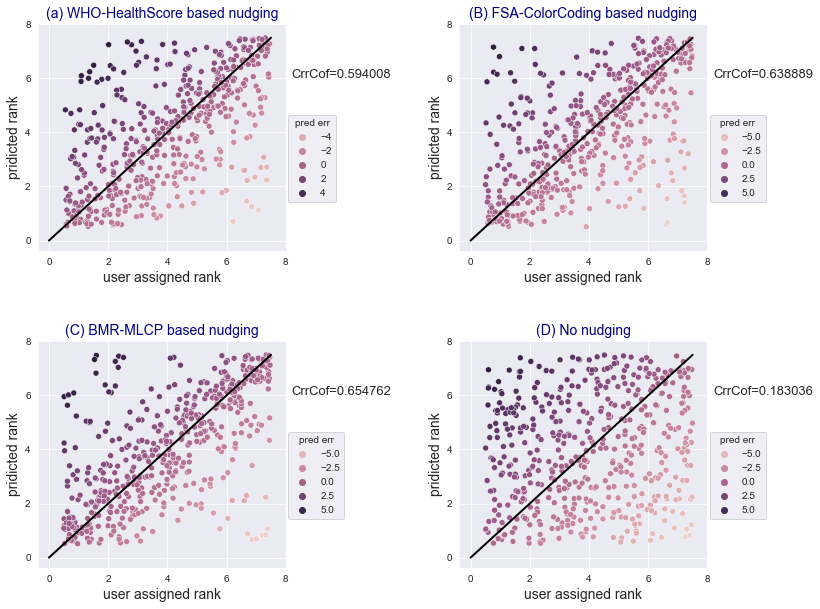}};
    \node[inner sep=0pt , rectangle,fill=white] at (-8.05,-.49) {\fontsize{8pt}{8pt}\selectfont \color[HTML]{090088}DRCI-MLCP based nudging};
     \node[inner sep=0pt , rectangle,fill=white, minimum height=15pt,minimum width=5cm] at (-8.4,5.7) {\fontsize{9pt}{9pt}\selectfont \color[HTML]{090088} (a) WHO-BubbleSlider based nudging};
     
      \node[inner sep=0pt , rectangle,fill=white, minimum height=20pt,minimum width=2cm] at (-4.7,4.5) {\fontsize{9pt}{9pt}\selectfont \color[HTML]{000000} PPMCC=0.594008};
     
     \node[inner sep=0pt , rectangle,fill=white, minimum height=15pt,minimum width=5cm] at (0,5.8) {\fontsize{9pt}{9pt}\selectfont \color[HTML]{090088} (b) FSA-ColorCoding based nudging};
     
     \node[inner sep=0pt , rectangle,fill=white, minimum height=20pt,minimum width=2cm] at (3.6,4.5) {\fontsize{9pt}{9pt}\selectfont \color[HTML]{000000} PPMCC=0.638889};

         \node[inner sep=0pt , rectangle,fill=white, minimum height=15pt,minimum width=5cm] at (-8.05,-.5) {\fontsize{9pt}{9pt}\selectfont \color[HTML]{090088} (c) DRCI-MLCP based nudging};

             \node[inner sep=0pt , rectangle,fill=white, minimum height=20pt,minimum width=2cm] at (-4.7,-1.5) {\fontsize{9pt}{9pt}\selectfont \color[HTML]{000000} PPMCC=0.654762};
     
     \node[inner sep=0pt , rectangle,fill=white, minimum height=15pt,minimum width=5cm] at (0,-.5) {\fontsize{9pt}{9pt}\selectfont \color[HTML]{090088} (d) no nudging};
     
      \node[inner sep=0pt , rectangle,fill=white, minimum height=20pt,minimum width=2cm] at (3.6,-1.5) {\fontsize{9pt}{9pt}\selectfont \color[HTML]{000000} PPMCC=0.183036};

\end{tikzpicture}}

\caption{Correlation between the FSA-HealthScoring based system predicted rank and the user assigned rank in all four recommendation scenarios.}
    \label{fig:rankpredErrFSA_SCORING1}
\end{figure}

Here, determined the FSA-HealthScore based system assigned ranks in similar approach as WHO-HealthScore based system assigned ranks. The user-assigned ranks are the same as in the  previous section. Figure \ref{fig:rankpredErrFSA_SCORING1} illustrates the correlation between FSA-HealthScore based system-assigned ranking and user-assigned ranking.

  The high positive PPMCC in recommendation scenarios, under the impression of smart-nudges, such as   DRCI-MLCP nudge(+0.654762),  FSA-ColorCoding nudge (+0.638889), and  WHO-BubbleSlider nudge (+0.504008), suggest that all three nudging scenarios produced a significant positive effect on users' preference towards healthier recipes. The DRCI-MLCP nudging strategy, with the highest positive PPMCC of +0.654762, is proven to be the most effective in motivating users to choose healthier recipes. In contrast, the no-nudging scenario, with the lowest PPMCC of +0.183036, demonstrates no positive or negative trend between the recipe healthiness and user preference.

\subsubsection{ RCM2: Normalized Distance-based Performance Measure (NDPM)}

 NDPM estimates performance based on the relative position of items in the RecList instead of focusing on their exact order \cite{NDPMExamplezhang2016fast,NDPMExampleparadarami2017hybrid}.. For example, suppose an FRS predicts recipe \(r_t\) to be more liked by a user \(u_a\) than a recipe \(r_c\); and \(u_a\) assigns a higher rating to \(r_t\) than \(r_c\). NDPM will determines the prediction as a success, regardless of the number of other recipes \(u_a\)  has rated higher than \(r_t\) and lower than \(r_c\). A perfect NPDM score of 0 suggests that the system correctly predicts the user-rating based order of each recipe pair within the RecList. In contrast, the lowest NPDM score of 1 demonstrates the system failed to predict the user assigned order for each recipe pair, in the RecList.

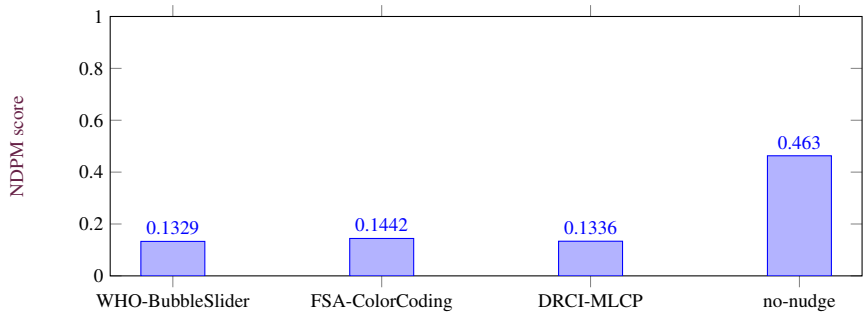
\begin{figure}[!h]
\centering
\small
\scalebox{1}{
\begin{tikzpicture}
    \begin{axis}[
        ybar,
        ymin=0,
        ymax = 1,
        width=.7\textwidth,
        height=5cm,
        bar width=24pt,
        ylabel={\color{dragonberry} NDPM score},
        nodes near coords,
         every node near coord/.append style={/pgf/number format/.cd,fixed,precision=4},
        symbolic x coords={WHO-BubbleSlider, FSA-ColorCoding, DRCI-MLCP, no-nudge},
        xtick = data,
        legend style={at={(0.02,0.02)},anchor=south west}
    ]
    \addplot coordinates {(WHO-BubbleSlider, 0.1329) (FSA-ColorCoding, 0.1442) (DRCI-MLCP, 0.1336) (no-nudge, 0.4630)};
      \legend{}
    \end{axis}
\end{tikzpicture}
}
\caption{NPDM generated by four recommendation scenarios.}
\label{fig:NDPMscore}
\end{figure}

 As the nudging strategies aim to encourage healthier recipes, the system determines the order of two recipes based on their healthiness scores. During the PPMCC based evaluation all four scenarios produced similar results while comparing against both WHO-HealthScore and FSA-HealthScore. Hence,  here on we will consider WHO-HealthScore as the basis of system assigned ranks.  Figure \ref{fig:NDPMscore} illustrates the NPDM based performance evaluation of all four recommendation scenarios. The results show the WHO-BubbleSlider nudging, with the lowest NPDM score of (+0.1329), performs best to predict users' relative preferences for recipes within the RecList. That is, WHO-BubbleSlider performed strongly in encouraging users to like the healthier option between two recipes. The DRCI-MLCP and  FSA-ColorCoding also received very low NDPM of 0.1336 and  0.1442, respectively. The low NPDM score of all three nudging strategies concludes that there remains a bias towards healthier recipes in recommendation scenarios with smart-nudging contents. The most inadequate performance from the no-nudge scenario, with an NDPM score of (+0.4630), indicates that the lack of nudging contents provides low encouragement towards healthy options.

\subsection{Did the Health Aware Smart Nudges Influenced users' Decision Making?}

Nudging aims to attract users' attention towards specific items in an environment or scope where all items are personalized to the user's preference. In this section, we evaluate, \emph{whether any of the proposed smart-nudges produce an impact on users' behavior and decision making}.

\subsubsection{Cumulative First Click Rate (CFCR)}
 
 The First Click Rate (FCR), a category of  Click Through Rate (CTR) matrices, estimates the performance in guiding users' attention towards certain items.  The FCR metric focuses on which item or item-type receives the highest number of \textit{first clicks} from users. FCR is a more appropriate metric of evaluating  \textit{persuasive recommendations}. The CFCR is the cumulative some of FCR over a range of positions, or a set of items, or items with certain features. The CFCR ranges from 0 to 1. The higher the CFCR, the more effective an RS is in persuading users towards the anticipated \textit{behavioral change}. In this work, we determined CFCR as the cumulative sum of FRC on recipes with higher WHO-HealthScore, such as 4, 5, 6, 7, and 8.  CFCR asses the impact generated by the nutri-aware badges. Here, we examine which recommendation scenarios can sluice most \textit{first clicks} towards healthier recipes.  Figure \ref{fig:CFCRgraph} illustrates the CFCR for all four recommendation scenarios. 

    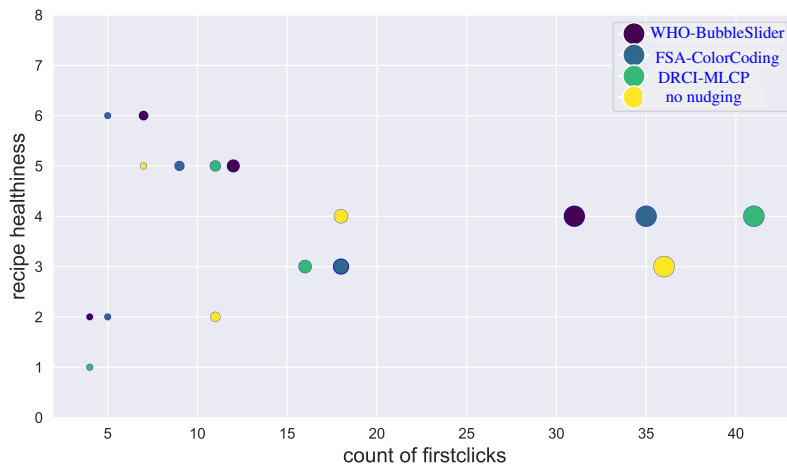
\begin{figure}[!h]
    \centering
    \small
     \scalebox{.9}{ 

    \begin{tikzpicture}
        \node[] (CDCR) at (0,0) {
     \includegraphics[width=.72\textwidth,height=!]{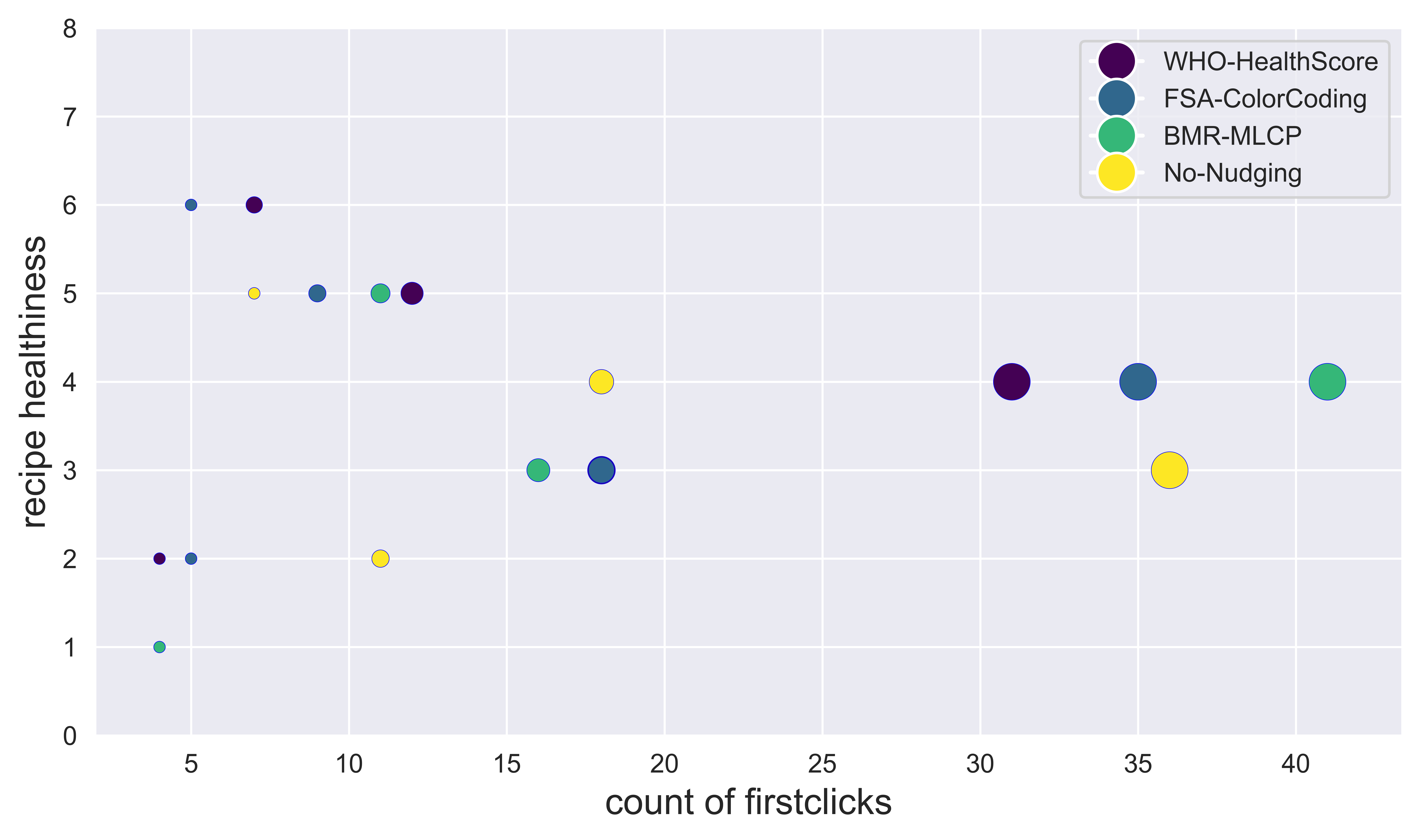}};
     
         \node[inner sep=0pt , rectangle,fill=leveldorGray, minimum height=8pt, minimum width=1.8cm] at (4.6,3) {\fontsize{7pt}{7pt}\selectfont \color{blue} WHO-BubbleSlider };
         
         \node[inner sep=0pt , rectangle,fill=leveldorGray, minimum height=9pt, minimum width=1.8cm] at (4.6,2.62) {\fontsize{7pt}{7pt}\selectfont \color{blue} FSA-ColorCoding };

         \node[inner sep=0pt , rectangle,fill=leveldorGray, minimum height=8pt, minimum width=1.8cm] at (4.4,2.05) {\fontsize{7pt}{7pt}\selectfont \color{blue} no nudging };
            
        \node[inner sep=0pt , rectangle,fill=leveldorGray,  minimum height=8pt, minimum width=1.6cm] at (4.4,2.35) {\fontsize{7pt}{7pt}\selectfont \color{blue} DRCI-MLCP };

    \end{tikzpicture}}
    \caption{Under different recommendation scenarios how the FCR varies on recipes with different WHO-HealthScores. \label{fig:clickCount}}  
\end{figure}

All three recommendation scenarios with a nudging badge produce a CFCR of \(\geq 0.6901 \) and outperform the no-nudge scenario in leading users' attention to healthier recipes. In the DRCI-MLCP nudging scenario, the highest number of participants, \(> 73\%\), cast their first-click on recipes with high  WHO-HealthScore (4-8). The no-nudge scenario performed the poorest, producing a CFCR of (+0.3521), indicating 64\% of first-clicks were cast on recipes with low  WHO-HealthScore (e.g., 0-3). Figure \ref{fig:clickCount} illustrates,  how the \textit{first clicks} are distributed over recipes with different WHO-HealthScores in all four recommendation scenarios.  Although the WHO-BubbleSlider nudging scenario gained lower CFCR than the DRCI-MLCP scenario, it obtained more first-clicks on recipes with the high WHO-HealthScore 6. As shown in figure \ref{fig:clickCount}, the WHO-BubbleSlider nudging generated 100\%, 28\%, and 100\% more first-clicks on recipes with WHO-HealthScore 6 compared to  FSA-ColorCoding, DRCI-MLCP, and no-nudging scenarios, respectively. CFCR shows the nudging strategies are effective in persuading users to visit healthier recipes first.
 
 \begin{figure}[!h]
\centering
\small
\scalebox{1}{
    \begin{tikzpicture}
    \begin{axis}[
        ybar,
        ymin=0,
        ymax = 1,
        width=.7\textwidth,
        height=5cm,
        bar width=22pt,
        ylabel={ \color{dragonberry}CFCR},
        nodes near coords,
         every node near coord/.append style={/pgf/number format/.cd,fixed,precision=4},
        symbolic x coords={WHO-BubbleSlider, FSA-ColorCoding, DRCI-MLCP, no-nudge},
        xtick = data,
        legend style={at={(0.02,0.02)},anchor=south west}
    ]
    \addplot coordinates {(WHO-BubbleSlider, 0.7042) (FSA-ColorCoding, 0.6901) (DRCI-MLCP, 0.7323) (no-nudge, 0.3521)};
      \legend{}
    \end{axis}
\end{tikzpicture}}
\caption{ Performance of different nudging scenarios and the no-nudge baseline based on CFCR. \label{fig:CFCRgraph}}

    \end{figure}


\newpage
    
\subsubsection{The Gained Utility in each Recommendation Scenario: Cumulative Consumption Rate}

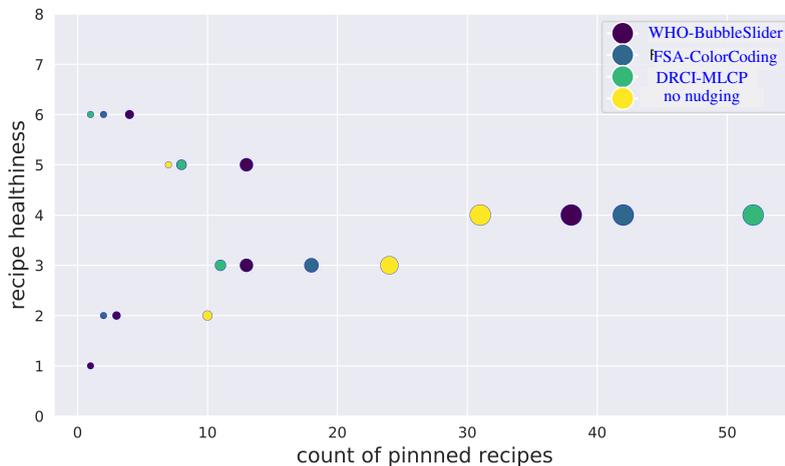
\begin{figure}[!h]
    \centering
    \scalebox{.9}{

        \begin{tikzpicture}
        \node[] (CCR) at (0,0) {
     \includegraphics[width=.72\textwidth,height=!]{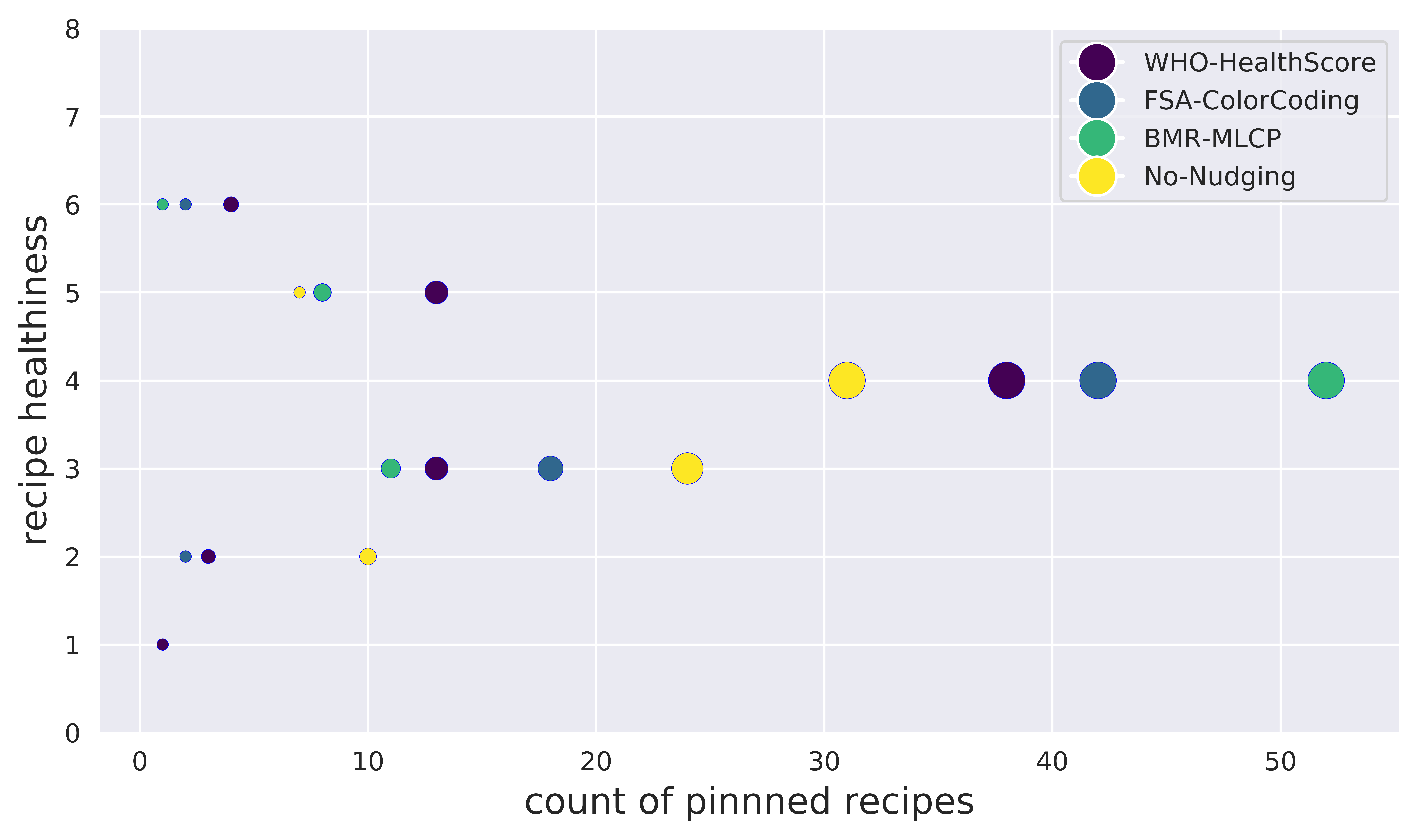}};
     
              \node[inner sep=0pt , rectangle,fill=leveldorGray, minimum height=8pt, minimum width=1.8cm] at (4.6,3) {\fontsize{7pt}{7pt}\selectfont \color{blue} WHO-BubbleSlider };
         
         \node[inner sep=0pt , rectangle,fill=leveldorGray, minimum height=9pt, minimum width=1.8cm] at (4.6,2.62) {\fontsize{7pt}{7pt}\selectfont \color{blue} FSA-ColorCoding };

         \node[inner sep=0pt , rectangle,fill=leveldorGray, minimum height=8pt, minimum width=1.8cm] at (4.4,2.05) {\fontsize{7pt}{7pt}\selectfont \color{blue} no nudging };
            
        \node[inner sep=0pt , rectangle,fill=leveldorGray,  minimum height=8pt, minimum width=1.6cm] at (4.4,2.35) {\fontsize{7pt}{7pt}\selectfont \color{blue} DRCI-MLCP };

    \end{tikzpicture} 
    }\vspace{-.5cm}
    \caption{ The total number of pin gained by different recommendation scenario over WHO-HealthScale.\label{fig:pinCountBasedonWhoScore} }

    \end{figure}

Utility of recommendation strategy corresponds to relevance, usefulness, gained value, and user satisfaction \cite{usersatiesfiction,HerlockerEvaluation}.  Utility can be defined as as an order of preference of consumption \cite{HerlockerEvaluation} and consumption  is a further explicit measure for usefulness of recommendations \cite{ASurveyofAccuracyEvaluationmetricsgunawardana2009survey,Anintelligentsystemforcustomertargetingkim2004intelligent}. In this work we consider \textit{pinning} as \textit{intention to consume}; hence, pinned recipes as \textit{useful recommendation}. During the survey, each user pined one recipe in each recommendation scenario. Figure \ref{fig:pinCountBasedonWhoScore} illustrates,  the distribution of the pinned recipes over the WHO-HealthScale.  Here, we consider pinning a recipe (consumption) as the HIT. The utility of the recommendation scenarios are assessed as the Cumulative HIT Rate (CHITR).  The ratio of \textit{recommendation cases} to \textit{consumed items} gives the HIT Rate of the item. The CHITR is determined as the cumulative sum of  HITs on a range of items.  As this evaluation aims to assess how effective each recommendation scenario is in convincing users to consume healthier recipes, we consider only the recipes, with high Who-HealthScore. The CHITR is determined by pins on recipes with higher WHO-HealthScore (4-8).  For example, in \(n\) recommendation cases, if \(m\) times users pinned recipes with a WHO-HealthScore of 4, then the HIT rate of such recipes is \(\frac{m}{n}\). The CHITR is determined as the sum of HIT rates on recipes with a high WHO-HealthScore, such as 4, 5, 6, 7, and 8.

\begin{figure}[!h]
    \centering
\small
\scalebox{1}{
\begin{tikzpicture}
    \begin{axis}[
        ybar,
        ymin=0,
        ymax = 1,
        width=.7\textwidth,
        height=5cm,
        bar width=22pt,
        ylabel={\color{dragonberry} CHITR},
        nodes near coords,
         every node near coord/.append style={/pgf/number format/.cd,fixed,precision=4},
        symbolic x coords={WHO-HealthScore, FSA-ColorCoding, DRCI-MLCP, no-nudge},
        xtick = data,
        legend style={at={(0.02,0.02)},anchor=south west}
    ]
    \addplot coordinates {(WHO-HealthScore,0.7639) (FSA-ColorCoding, 0.7222) (DRCI-MLCP, 0.8472) (no-nudge, 0.5278 )};
      \legend{}
    \end{axis}
\end{tikzpicture}}
\caption{Performance of different nudging scenarios and the no-nudge baseline based on CHITR. \label{fig:CCRgraph}}
  
    \end{figure}
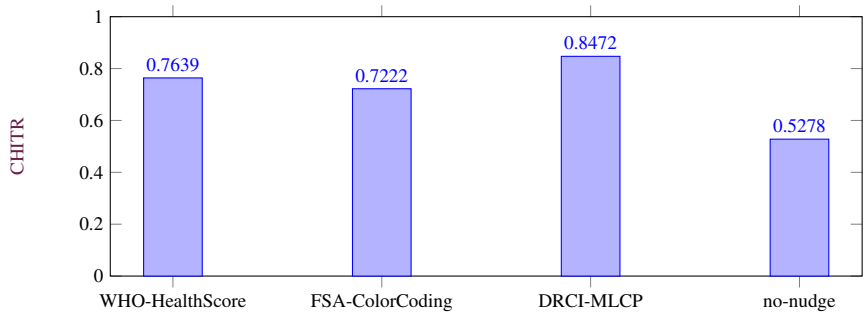

 As shown in figure \ref{fig:CCRgraph}, all three recommendation scenarios with nudging contents produce a CHITR of \(\geq 0.7222 \) consequently outperform the no nudging scenario in pursuing users to consume recipes that are in the healthier half of the WHO-HealthScale (4-8). The DRCI-MLCP nudging scenario with a CHITR of 0.8472 produced the highest incline towards consummations of recipes with higher healthiness scores, as shown in figure \ref{fig:CCRgraph}. Although the WHO-BubbleSlider nudging scenario gained lower CHITR than the DRCI-MLCP scenario, it obtained more pins on recipes with the high WHO-HealthScore (e.g., 5, 6), as shown in figure \ref{fig:pinCountBasedonWhoScore}. The CHITR metric suggests that the DRCI-MLCP nudging produces a significant consumption flow for healthier recipes while WHO-HealthScore leads users to the most healthy recipe in the list.

\subsection{ User Satisfaction: How Good are the Recommendations?\label{sec:usersatisfaction}}

 \pgfplotstableread{ 
Label            verypoor   poor     moderate    good        verygood
no-nudge           72.22	 15.29    9.72	      1.39        1.39
DRCI-MLCP          13.89  	5.56   	  31.95   	   25	        23.60
FSA-ColorCoding 6.95	    12.5   	  20.83	      30.55	        29.16
WHO-HealthScore     6.95    	5.56	      23.61          43.08	        20.83

}\testEffectiveness 

 \pgfplotstableread{ 
Label            verypoor   poor     moderate        good        verygood
no-nudge            55.55	   5.55          9.72	      9.72        19.44
DRCI-MLCP            6.95  	23.61   	23.61   	  20.83	        25
FSA-ColorCoding     1.39	1.39   	    15.27	      27.77	        54.16
WHO-HealthScore     1.39    	6.95	    15.27         30.55	      45.83

}\testUnderstandibility

 \pgfplotstableread{ 
Label              no        maybe    willingtotry    good       strongly
no-nudge           83.33	 5.56       8.3	          1.39        1.39
DRCI-MLCP            12.5  	13.89  	   33.33   	      18.056	  22.22
FSA-ColorCoding     8.3	    12.5   	  13.89	          25	        40.28
WHO-HealthScore     6.95    	6.95	      23.61              43.06        19.44

}\testpersuesive 

 \pgfplotstableread{ 
Label               no        maybe        willingtotry    good       strongly
no-nudge            79.16	   6.95          9.72	       1.39        2.78
DRCI-MLCP            15.5  	   16.67         2.78   	       18.06       47.22
FSA-ColorCoding     12.5	   11.11  	     25.00	       20.83	  30.56
WHO-HealthScore     8.33      6.95	     26.39         31.94	  26.39

}\testLongTermUse

In recent years, user satisfaction score is becoming a more popular metric for evaluating the performance of recommender models and the services offered in an RS \cite{usersatiesfiction}. User satisfaction is a compound concept experienced by the user from the serendipity, unexpectedness, utility, novelty, and usefulness  offered in the recommendations \cite{whatisUserSatisfaction}. It also encorporates the understandably, interactiveness, explanation and adaptiveness of the process of recommendation. In simple words, user satisfaction-focused evaluates the degree of the gain regarding various \textit{experience variables}. To evaluate the smart-nudges on user satisfaction, we decided on four experience variables: \begin{inparaitem} \item[\textbullet]Effectiveness, \item[\textbullet]Ease of Understanding, \item[\textbullet]Persuasiveness, and \item[\textbullet] Suitability for Long-term Use\end{inparaitem}. Based on how satisfied they are, participants evaluated each recommendation scenario, on a  5-star-scale, for each of the four variables, as shown in figure  \ref{fig:performanceTabs1}.


\subsubsection{Effectiveness}

Effectiveness corresponds to the purpose of the RS. Given that the research presented in this work is keen on helping users discover healthy recipes, we defined effectiveness as \emph{how helpful a nudging strategy is while finding a healthy recipe}. Participants assigned ratings based on \emph{ \color{belblue} are there tools that differentiate healthier options from less healthy ones?} and \emph{ \color{belblue}are the healthy options in a RecList aligned to user's taste preference?}.

\noindent\begin{figure}[!h]
\small
\centering
\scalebox{1}{
\begin{tikzpicture}
    \begin{axis}[
            xbar stacked,   
            xmin=0,  
            xmax=105,
            ytick=data,
            width=.55\textwidth,
            height=5cm,
            yticklabels from table={\testEffectiveness}{Label}, 
             legend style={at={(1,1.1)},
      anchor=east,legend columns=5},
]
\addplot [fill=red!90] table [x=verypoor, meta=Label,y expr=\coordindex] {\testEffectiveness};
\addplot [fill=orange!90] table [x=poor, meta=Label,y expr=\coordindex] {\testEffectiveness};   
\addplot [fill=yellow] table [x=moderate, meta=Label,y expr=\coordindex] {\testEffectiveness};
\addplot [fill=lime] table [x=good, meta=Label,y expr=\coordindex] {\testEffectiveness};
\addplot [fill=green] table [x=verygood, meta=Label,y expr=\coordindex] {\testEffectiveness};
\legend{\strut very poor, \strut poor, \strut moderate, \strut good ,\strut excellent}
\end{axis}
\end{tikzpicture}}
  
    \caption{User satisfaction measure: Effectiveness}
    \label{fig:TheperformanceChartEffectiveness}
\end{figure}
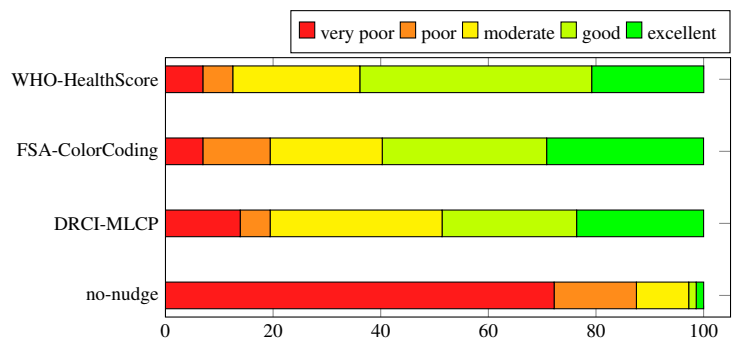

 Figure \ref{fig:TheperformanceChartEffectiveness} illustrates the stacked bar chart \cite{stackedBarChart} of  participants' feedback on how useful they found each recommendation scenario for finding healthy recipes.  Each horizontal bar corresponds to a recommendation scenario, and each color corresponds to a level of performance. The colored sections, stacked against each other within a bar, represent what proportion of the total survey population labeled the scenario as the \textit{performance level} corresponding to their color. Participants found the nudging strategy Who-BubbleSlider to be the most helpful (effective) for finding healthy recipes, as shown in figure \ref{fig:TheperformanceChartEffectiveness}. The WHO-BubbleSlider nudge received \textit{good} or \textit{excellent} feedback from 63.91\% of the participants. The FSA-ColorCoding and the DRCI-MLCP nudging strategies also received  \textit{good} or \textit{excellent} feedback from 59.71\% and 48.6\% of the participants, respectively. The participants rated the no-nudging scenario as the least helpful. 72.22\% of the participant evaluated the no-nudge scenario as \emph{very poor} for effectiveness. The fact possibly influences this feedback, that the no-nudging scenario did not provide any visual guidance towards healthier recipes. The results insist on the necessity of recommendations with health-aware smart-nudges.

\subsubsection{ Ease of Understanding}

Ease of Understanding is determined based on participants' feedback on \textit{the level of effort required from the participant to understand and interpret the visual cues on recipe healthiness.} Participants rated each recommendation scenario based on \emph{ \color{belblue} Are the numeric or color scale consistent over different recipes?} and \emph{ \color{belblue} are the healthiness guidelines conceptually and grammatically correct?} and \emph{ \color{belblue} How much effort dose it requires to figure out what information the visual contents are trying to communicate?}. Figure \ref{fig:TheperformanceChartsEaseofunderstanding} illustrates the user feedback on \textit{ ease of understanding} of each recommendation scenario. The FSA-ColorCoading smart-nudge is determined as the most easy-to-understand nudging strategy, with 81.93\% of the survey population rating the smart-nudge \textit{good} or \textit{excellent} for understandablity. The WHO-BubbleSlider nudging strategy is proven to be the second most easy-to-understand visual influence-contents. 55.56\% of the survey population evaluated the no-nudge scenario as \emph{very poor}. Which makes the no-nudge scenario as the poorest performer regarding understandability.

\noindent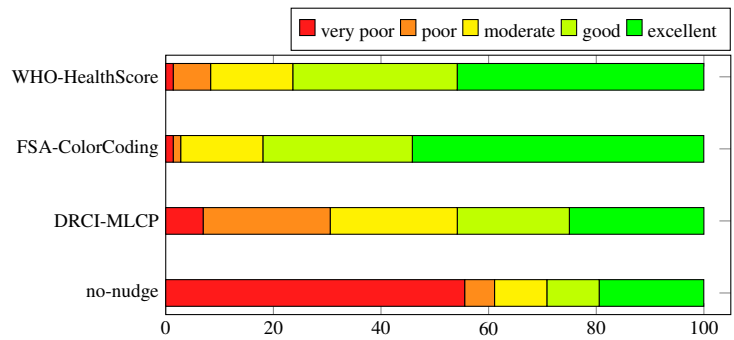
\begin{figure}[!h]
\small
\centering
\scalebox{1}{
\begin{tikzpicture}
    \begin{axis}[
            xbar stacked,   
            xmin=0,
            xmax=105,
            ytick=data,
            width=.55\textwidth,
            height=5cm,
            yticklabels from table={\testUnderstandibility}{Label}, 
             legend style={at={(1,1.1)},
      anchor=east,legend columns=5},
]
\addplot [fill=red!90] table [x=verypoor, meta=Label,y expr=\coordindex] {\testUnderstandibility};
\addplot [fill=orange!90] table [x=poor, meta=Label,y expr=\coordindex] {\testUnderstandibility};   
\addplot [fill=yellow] table [x=moderate, meta=Label,y expr=\coordindex] {\testUnderstandibility};
\addplot [fill=lime] table [x=good, meta=Label,y expr=\coordindex] {\testUnderstandibility};
\addplot [fill=green] table [x=verygood, meta=Label,y expr=\coordindex] {\testUnderstandibility};
\legend{\strut very poor, \strut poor, \strut moderate, \strut good ,\strut excellent}
\end{axis}
\end{tikzpicture}}
    \caption{User satisfaction measure : Understandably \label{fig:TheperformanceChartsEaseofunderstanding} }
    
\end{figure}

\subsubsection{Persuasiveness \label{sec:persuessiveness}}

\begin{figure}[!h]
\centering
\scalebox{1}{
\small
\centering
\begin{tikzpicture}
    \begin{axis}[
            xbar stacked,   
            xmin=0,  
            xmax=105,
            ytick=data,
            width=.55\textwidth,
            height=5cm,
            yticklabels from table={\testpersuesive}{Label}, 
             legend style={at={(1,1.1)},
      anchor=east,legend columns=5},
]
\addplot [fill=red!90] table [x=no, meta=Label,y expr=\coordindex] {\testpersuesive};
\addplot [fill=orange!90] table [x=maybe, meta=Label,y expr=\coordindex] {\testpersuesive};   
\addplot [fill=yellow] table [x=willingtotry, meta=Label,y expr=\coordindex] {\testpersuesive};
\addplot [fill=lime] table [x=good, meta=Label,y expr=\coordindex] {\testpersuesive};
\addplot [fill=green] table [x=strongly, meta=Label,y expr=\coordindex] {\testpersuesive};
\legend{\strut strong no, \strut no, \strut somewhat, \strut strongl ,\strut very strong}
\end{axis}
\end{tikzpicture}
}
    \caption{User satisfaction measure: Persuasiveness }
    \label{fig:TheperformanceChartsPersuasiveness}

\end{figure}
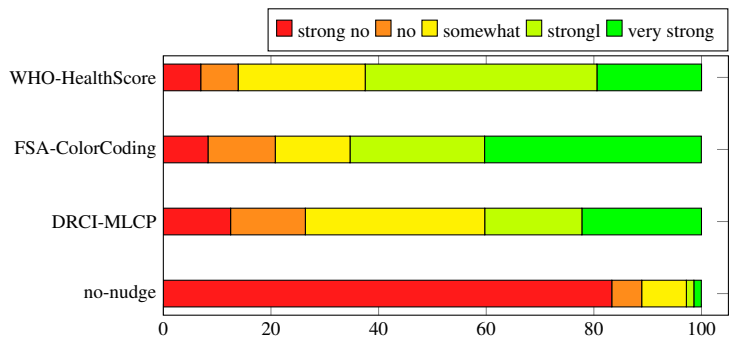

The experience variable \textit{persuasiveness}estimates the degree of encouragement, on consuming healthy recipes, introduced by the healthiness cues. For this variable, participants evaluated each recommendation scenario on a  5-star-scale. The number of stars correspond to the following degrees of gain: \texttt{\small \{1, strong no\}, \{2, no\}, \{3, somewhat\}, \{4, strong\}} and \texttt{\small \{5, very strong \}}.  The nudging strategy FSA-ColorCoding demonstrated as most actively encouraging for healthy recipes, as shown in figure \ref{fig:TheperformanceChartsPersuasiveness}. The FSA-ColorCoding nudge received \textit{strongly} or \textit{very strongly} persuasive feedback from 65.28\% of the participants, followed by the WHO-BubbleSlider and the DRCI-MLCP nudging strategies also receiving at least \textit{strongly} or \textit{very strongly} persuasive feedback from 62.5\% and 40.27\% of the participants, respectively. The participants rated the no-nudging scenario as the least persuasive regarding promoting healthy recipes, and  83.33 \% of the survey population evaluated the no-nudge scenario as \textit{not persuasive at all}. 

\noindent

\subsubsection{Suitability for Long-term Use \label{sec:longterm}} 

\begin{figure}[!h]
\centering
\scalebox{1}{
\small
\centering
\begin{tikzpicture}
    \begin{axis}[
            xbar stacked,   
            xmin=0,
            xmax=105,
            ytick=data,
            width=.55\textwidth,
            height=5cm,
            yticklabels from table={\testLongTermUse}{Label}, 
             legend style={at={(1,1.1)},
      anchor=east,legend columns=5},
]
\addplot [fill=red!90] table [x=no, meta=Label,y expr=\coordindex] {\testLongTermUse};
\addplot [fill=orange!90] table [x=maybe, meta=Label,y expr=\coordindex] {\testLongTermUse};   
\addplot [fill=yellow] table [x=willingtotry, meta=Label,y expr=\coordindex] {\testLongTermUse};
\addplot [fill=lime] table [x=good, meta=Label,y expr=\coordindex] {\testLongTermUse};
\addplot [fill=green] table [x=strongly, meta=Label,y expr=\coordindex] {\testLongTermUse};
\legend{\strut strong no, \strut no, \strut maybe, \strut yes ,\strut strong yes}
\end{axis}
\end{tikzpicture}
}   

    \caption{User satisfaction measure: Suitability for long-term use }
    \label{fig:TheperformanceChartsSuitability}

\end{figure}
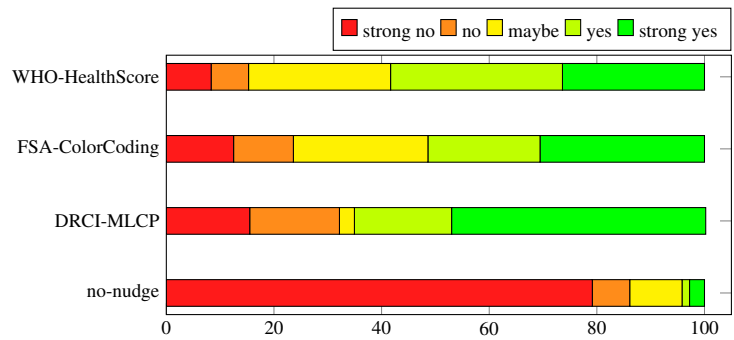

The variable \textit{Suitability for Long-term Use} enables participants to evaluate an RS based on whether the system offers sufficient utility and assistance to support the user as a long-term diet assistance application. This variable represents a composite opinion from the user on overall understandability, persuasiveness, effectiveness, diversity, novelty, and interestingness. For this variable, participants also evaluated each recommendation scenario on a scale same as persuasiveness. According to the participants' feedback, the DRCI-MLCP nudge is the most suitable nudging strategy for long-term use,  as shown in figure \ref{fig:TheperformanceChartsSuitability}. The recommendation scenario with DRCI-MLCP nudge received \textit{yes} or \textit{strong yes} feedback from 65.28\% of the participants, followed by the WHO-BubbleSlider and the FSA-ColorCoding nudging strategies also receiving at least \textit{yes} or \textit{strong yes} feedback from 58.33\% and 51.39\% of the participants, respectively. The participants experienced the no-nudging scenario as the least suitable for long-term use, and 79.16 \% of the survey population evaluated the no-nudge scenario as strongly not suitable for long-term use.

The distribution of the varying feedback based on the four experience variable exhibits that no one nudging scenario outperforms every other regarding all four variables. Instead, each nudging scenario has its advantage and disadvantages. However, section \ref{sec:usersatisfaction} demonstrates that in all the three proposed nudging strategy users experienced a higher degree of assistance and encouragement to find and consume healthier recipes. 

\noindent

\newpage
\section{Discussion  \label{sec:Discussionhealthnudge}}

Under both system ranking approaches, all three recommendation scenarios with smart nudging outperform the no-nudge baseline producing better agreement with the user-assigned rankings. The nudging scenario DRCI-MLCP has proven to be the most effective strategy with the highest PPMCC of (+0.654762) and (+0.769841) in both FSA-HealthScale and WHO-HealthScale base system-rankings respectively. All three nudging methods demonstrated a higher positive correlation to the user-assigned rankings, which provides evidence in support of the hypothesis that \textit{food-healthiness nudges can make users like healthier recipes more}. The DRCI-MLCP nudging performs best in this regard. With the highest CFCR (0.7323) and CHITR (0.8472), the DRCI-MLCP nudging also proved to be best performing strategy in \textit{attracting users' attention to healthy recipes} and \textit{convincing them to consume healthier recipes}, respectively. Users also found the DRCI-MLCP based nudging the most suitable for long-term use, as described in section \ref{sec:longterm}. However, among the three nudging strategies, participants experienced the DRCI-MLCP nudging as the most difficult to understand. 30.56\% of the survey population defined the nudging contents as \textit{difficult} or \textit{very difficult} to understand, as shown in figure \ref{fig:TheperformanceChartsEaseofunderstanding}. These results encourage further research on the design of the DRCI-MLCP nudge.

The CFCR metric suggests the DRCI-MLCP nudging gains the highest exposure for relatively healthier recipes, while the WHO-Bubblesilder directs users to the most healthy recipe in the RecList. At the same time, the WHO-BubbleSlider nudging scenario gained lower CHITR than the DRCI-MLCP scenario but it obtained more pins on recipes with high WHO-HealthScores, as shown in figure \ref{fig:pinCountBasedonWhoScore}. Such results suggest that the WHO-HealthScore strongly favours the most healthy recipes in the Reclist. This nudging-strategy can ensure that the user notices the healthy options available to them. Users also found the WHO-BubbleSlider nudge the most helpful. According to the participants' feedback the WHO-BubbleSlider nudging is the most effective tool in assisting them to differentiate healthier options from less healthy ones, as shown in \ref{fig:TheperformanceChartEffectiveness}. 

Though the FSA-ColorCoding based nudging contents did not outperform WHO-BubbleSlider and DRCI-MLCP based nudging,  FSA-ColorCoding is proven to be the most \textit{easy to understand} and \textit{persuasive} nudging strategy, as shown in figures \ref{fig:TheperformanceChartsEaseofunderstanding} and \ref{fig:TheperformanceChartsPersuasiveness}. The FSA-ColorCoding Disk outperformed both WHO-BubbleSlider and DRCI-MLCP on the count of understandably and persuasiveness, with 81.93\% and 65.28\% (good or above) users responses, respectively. 

The low NPDM score ($\le$0.1442) and high PPMCC ($\ge$0.707679) of all three nudging strategies, as shown in figures \ref{fig:rankpredErrWHOSCORING1} and \ref{fig:NDPMscore}, concludes that there remains a bias towards healthier recipes in recommendation scenarios with smart-nudging contents. The low CFCR (0.3521) of no-nudge scenarios  indicates that healthy recipes are not incentivized in generic personalized recommendations. The CHITR (0.5278) in no-nudge scenarios indicates that there is noticeable possibility that a user may choose a healthy recipe to consume without being influenced by health aware nudges. However, the very high CHITR of all three nudges demonstrates a much greater chance under the influence of healthiness nudges. All three nudging technologies outperformed the no-nudging baseline on effectiveness, understandability, persuasion, and suitability.

\section{Future Work}
One of the most exciting research outcomes of this work is that \textit{intelligent FRS combined with food healthiness knowledge and smart-nudging technology can impact users' food choices}. The results provide evidence supporting the potential of intelligent FRS in educating mass population on healthier food decisions. Following are some of the potential follow-up future research directions that can help improve the FRS strategies and tools.

\begin{itemize}\setlength\itemsep{.05em}
    \item Investigating the impact of the proposed three novel nudging techniques in diverse food decision scope, such as drinks, grocery, restaurant, baby-food, school lunch, diet-plan, and take-aways. 
    \item Designing machine knowledge on the complex relationship between macro-nutrients and the human body and deploying such knowledge to produce tools assisting users with chronic diseases.
    \item Investigating audio as well as visual contents to extend the proposed smart-nudging methods for users with visual disabilities.
    \item Investigating various NLP algorithms to provide a robust explanation of the recommended items and the recipe healthiness cues. 
    
    \item Investigating \textit{gamification} \cite{seaborn2015gamification} technology, such as\textit{ avatar, brooch, class, digital reward, and token}, to keep the user invested in the process of behavioral change,  eating healthy in this case. For example, introducing \textit{brooch} system similar to GitHub \cite{website:github} and GoogleMap's LocalGuide \cite{website:GooglemapLocalGuides} can encourage users to consume healthier options throughout a longer-period.
    \item   Investigating ML techniques to find similar but healthier recipes to a current recipe.  Users often prefer being presented with alternative options over having to perform a new search for different items \cite{alternativenewsRec}. Hence, recommending similar but healthier recipes to the recipe currently being browsed can increase the CHITR of an FRS.

\end{itemize}
\section{Conclusion}

Eating-habits and food-choices are personal and have a significant impact on our long-term \textit{health} and \textit{quality of life}. Eating-habits often develop from past life experiences and current socio-economic state; which makes it challenging to introduce significant changes in one's eating-habits.  However the scope of the recommendation can have a influence on peoples' decision. Recommending food-items, that are aliened with the users' taste, along with persuasive visual motivations towards healthier option can make significant change in\textit{ how a user decide on their  meal}. The work reported in this paper combines the advanced FR model with the theory of choice architecture to develop technology assisting users to get into a healthier eating-habit.  The research outcome demonstrate that the combination of ML and smart-health-nudging can make users keen on healthier food choices. The PPMCCs of all four recommendation scenarios demonstrate that when nudging contents are displayed with personalized recommendations, a user's rating on a recipe positively correlates to the healthiness of the recipe.  In terms of attracting users' attention to the healthier options within the RecList, in comparison to the no-nudge scenario, all three nudging strategies produced a higher CFCR  on healthier recipes. The evidence presented in this paper leads us to conclude that the user experiences more support and encouragement for healthy eating when personalized recommendations are presented with health-aware smart-nudging services. Also, these smart-nudging technologies can successfully introduce changes in users' food-decisions.

\bibliographystyle{plainnat}
\bibliography{main}

\end{document}